\documentclass[12pt]{article}
\pdfoutput=1
\usepackage{bbm}
\usepackage[dvipsnames]{xcolor}
\usepackage{authblk}
\usepackage{latexsym}
\usepackage{epsfig,amssymb,euscript, mathrsfs,cite}
\usepackage{amsmath}
\usepackage{mathrsfs} 
\usepackage{mathtools}
\usepackage{enumerate}
\usepackage{graphicx} 
\usepackage{tikz}
\usepackage{tikz-3dplot}
\usepackage{pgfplots}
\usepackage{caption}
\usepackage{upgreek}
\usetikzlibrary{decorations.pathreplacing,calligraphy}
\usepgfplotslibrary{fillbetween}
\definecolor{MyDarkBlue}{rgb}{0.15,0.15,0.45}
\usepackage[linktocpage=true]{hyperref}
\hypersetup{
colorlinks=true,
citecolor=MyDarkBlue,
linkcolor=MyDarkBlue,
urlcolor=MyDarkBlue,
pdfauthor={},
pdftitle={},
pdfsubject={hep-th}
}
\makeatletter

\newcommand{\Rmnum}[1]{\expandafter\@slowromancap\romannumeral #1@}
\makeatother

\newcommand{\nn}{\notag }
\def\be{\begin{equation}}
\def\ee{\end{equation}}

\newcommand{\del}{\partial}

\newcommand{\ii}{\mathrm{i}}
\newcommand{\ex}{\mathrm{e}}
\newcommand{\diff}{\mathrm{d}}
\newcommand{\dd}{\mathrm{d}}
\newcommand{\R}{\mathbb{R}}
\newcommand{\Z}{\mathbb{Z}}
\newcommand{\vol}{\mathrm{vol}}
\newcommand{\Vol}{\mathrm{Vol}}
\newcommand{\C}{\mathbb{C}}

\newcommand{\cL}{\mathcal{L}}

\newcommand{\I}{\mathrm{I}}
\newcommand{\II}{\mathrm{II}}
\newcommand{\MI}{\mbox{Model I}}
\newcommand{\MII}{\mbox{Model II}}

\newcommand{\bMI}{\mbox{{\bf Model I}}}
\newcommand{\bMII}{\mbox{{\bf Model II}}}

\newcommand{\RF}{\mathcal{F}}

\newcommand{\gauge}{\mathscr{G}}
\newcommand{\tmin}{\mathrm{min}}
\newcommand{\tmax}{\mathrm{max}}
\newcommand{\ggroup}{\mathcal{G}}

\newcommand{\tildeeta}{\eta_{7}}
\newcommand{\tildeJ}{\omega}
\newcommand{\tildesigma}{\psi_{7}}
\newcommand{\tilderho}{\varrho_{7}}

\newcommand{\p}{\mathtt{p}}
\newcommand{\z}{z}
\newcommand{\adots}{\"a}
\newcommand{\xivec}{\mbox{\boldmath$\xi$}}
\newcommand{\zetavec}{\mbox{\boldmath$\zeta$}}

\usetikzlibrary{arrows,arrows.meta,shapes,shapes.misc,patterns,decorations.pathmorphing,decorations.pathreplacing,positioning,chains,fit}
\pgfdeclarepatternformonly{coarsedots}{\pgfqpoint{-1pt}{-1pt}}{\pgfqpoint{5pt}{5pt}}{\pgfqpoint{12pt}{12pt}}{
    \pgfpathcircle{\pgfqpoint{.5pt}{.5pt}}{.5pt}
    \pgfusepath{fill}}
\tikzset{gauge/.style={circle, draw=black!100, thick, minimum size=2mm},  gaugeD/.style={rounded rectangle, draw=black!100,double,thick,minimum size=5mm},  empty/.style={rounded rectangle, draw=white!100, thick, minimum size=5mm}, flavor/.style={rectangle, draw=black!100, thick, minimum size=5mm},flavorD/.style={rectangle, draw=black!100, double,thick, minimum size=5mm}}

\numberwithin{equation}{section}       

\newcommand{\ab}[1]{\textcolor{orange}{{\bf [ab: #1]}}}

\begin{document}

\begin{titlepage}

\vskip 1cm

\begin{center}


{\Large \bf Matrix models \\ 
\vskip 0.3cm
 from black hole geometries}

\vskip 1cm
{Andrea Boido, Alice L\"uscher and James Sparks}

\vskip 1cm

\textit{Mathematical Institute, University of Oxford,\\
Andrew Wiles Building, Radcliffe Observatory Quarter,\\
Woodstock Road, Oxford, OX2 6GG, U.K.\\}

\vskip 0.2 cm

\end{center}

\vskip 0.5 cm

\begin{abstract}
\noindent Supersymmetric, magnetically charged (and possibly accelerating) 
black holes in AdS$_4$ that uplift on Sasaki-Einstein manifolds $Y_7$ to M-theory 
have a dual matrix model description. The matrix model in turn arises by 
localization of the 3d $\mathcal{N}=2$ SCFTs, dual to the AdS$_4$ vacuum, 
on the black hole horizon geometry, which is  a Riemann surface $\Sigma_g$ (or a spindle $\Sigma$). We identify the imaginary part $t$ of the continuously distributed eigenvalues in the matrix model, and their density function $\rho(t)$, with natural geometric quantities associated with the M-theory circle action $U(1)_M$ on the near-horizon geometry AdS$_2\times Y_9$, the internal space $Y_9$ being a $Y_7$ fibration over $\Sigma_g$ (or $\Sigma$). Moreover, we argue that the points where $\rho'(t)$ is discontinuous match with the classical action of BPS probe M2-branes wrapping AdS$_2$ and the M-theory circle. We illustrate our findings with the ABJM and ADHM theories, whose duals have $Y_7 = S^7/\mathbb{Z}_k$, and some of their flavoured variants corresponding to other toric $Y_7$.

\end{abstract}

\end{titlepage}

\pagestyle{plain}
\setcounter{page}{1}
\newcounter{bean}
\baselineskip18pt

\tableofcontents

\newpage

\section{Introduction and overview}\label{sec:intro}

One of the successes of string theory and M-theory has been to give a 
microstate counting of black hole entropy. While the 
seminal work of \cite{Strominger:1996sh} applied to supersymmetric 
black holes in asymptotically flat spacetime, more recently 
a different approach was initiated in \cite{Benini:2015eyy, Benini:2016rke} 
using holography
for asymptotically AdS black holes. This relies on the fact 
that the  partition function of the  
dual field theory may be evaluated exactly using supersymmetric
localization, reducing the computation of the entropy to a matrix model calculation. 
The aim of the present work is to extend this matching further, identifying 
large $N$ matrix model quantities in this setting to counterparts in the gravity duals.

\subsection{Introduction and background}

We consider a general class of supersymmetric, magnetically charged AdS$_4$ 
black holes in M-theory, where the internal space $Y_7$ is 
a Sasaki-Einstein manifold. Before introducing the black hole the 
vacuum of the theory is the Freund-Rubin solution AdS$_4\times Y_7$. 
Dirac quantization imposes
\begin{align}\label{intro:flux}
N = \frac{1}{(2\pi\ell_p)^6}\int_{Y_7} \star_{11} G\in \mathbb{N}\, ,
\end{align}
where $G$ is the M-theory four-form, $\star_{11}$ is the Hodge dual, and $\ell_p$ denotes the 
Planck length. The integer $N$ fixes the overall scale, and may also be interpreted as the number of M2-branes 
sourcing the AdS solution, as we describe further below.

Via the Kaluza-Klein mechanism, if the  
isometry group of $Y_7$ contains a maximal torus $U(1)^s$ 
of rank $s$, then there are associated massless $U(1)$ gauge fields
$A_i$, $i=1,\ldots,s$, 
in the four-dimensional spacetime. One can then consider introducing 
a supersymmetric black hole into this vacuum, carrying conserved charges 
under the gauge fields. Taking the  black hole horizon 
to be a Riemann surface\footnote{Later in the paper we will also consider 
the extension to supersymmetric \emph{accelerating} black holes  \cite{Ferrero:2020twa}, where the horizon is a orbifold surface known as a spindle \cite{Ferrero:2020laf}.}  $\Sigma_g$ 
of genus $g$, we consider black holes with magnetic (but no electric) fluxes
\begin{align}\label{intro:pi}
p_i = \frac{1}{2\pi}\int_{\Sigma_g}\diff A_i \in \Z\, .
\end{align}
In the full M-theory geometry the $p_i$
are Chern classes which describe how the internal space $Y_7$ 
is fibred, as an associated principal $U(1)^s$ bundle, 
over the horizon $\Sigma_g$. The total space $Y_9$ is then
a fibre bundle
\begin{align}\label{intro:fibre}
Y_7 \, \hookrightarrow\, Y_9 \, \rightarrow\, \Sigma_g\, .
\end{align}

Such black hole solutions are in general difficult to construct 
explicitly, not least since in general we lack four-dimensional consistent truncations 
that keep the massless gauge fields $A_i$. Exceptions to this 
are the $U(1)^4\subset SO(8)$ STU gauged supergravity model that arises from reduction on $Y_7=S^7$, 
with the corresponding black holes and holographic microstate counting 
discussed in \cite{Benini:2015eyy}; and the consistent truncation 
of \cite{Gauntlett:2007ma}  to minimal gauged supergravity that includes only the R-symmetry gauge field 
(the graviphoton) for the $U(1)_R$ isometry of $Y_7$, where the 
corresponding ``universal twist'' black holes and microstate counting were discussed in 
\cite{Azzurli:2017kxo}. 

Even with a consistent truncation it 
is not guaranteed to be able to  solve the Einstein equations in closed form, and we shall instead 
follow the approach introduced in  \cite{Couzens:2018wnk}, 
and further developed in \cite{Gauntlett:2019roi, Hosseini:2019ddy, Kim:2019umc}. 
Here rather than studying the full supersymmetric 
extremal black holes, we instead focus on their near-horizon geometry.  Such near-horizon solutions take the form  AdS$_2\times Y_9$, 
where the internal space $Y_9$ has the fibred form 
\eqref{intro:fibre}. Supersymmetric M-theory solutions of this type, 
with no internal four-form flux $G$ on $Y_9$,  were 
classified in \cite{Kim:2006qu}, and require that $Y_9$ is a ``GK geometry'' 
\cite{Gauntlett:2007ts}. We shall review this in more detail in 
section~\ref{sec:gravity}, but for the purposes of this introduction 
we now briefly summarize some relevant features. 

Firstly, $Y_9$ necessarily admits an R-symmetry Killing vector field 
$\xi$, meaning that the preserved Killing spinor of the supergravity solution is charged
under $\xi$. 
 If we introduce a normalized basis of Killing vectors 
$\partial_{\varphi_i}$, $i=1,\ldots,s$, generating the $U(1)^s$ isometry 
of the fibre $Y_7$ in \eqref{intro:fibre}, then we may generically write
\begin{align}\label{intro:xi}
\xi = \sum_{i=1}^s b_i\, \partial_{\varphi_i}\, .
\end{align}
The variables $b_i\in\R $ then specify how $U(1)_R\subset U(1)^s$. 
Secondly, the geometry transverse to the vector field $\xi$ in $Y_9$ 
is K\"ahler, and in the fibre $Y_7$ in particular one can introduce a transverse K\"ahler 
class $[\omega]\in H^2_B(\mathcal{F}_\xi)$, where the latter is 
the basic cohomology for the foliation $\mathcal{F}_\xi$ defined by 
the vector field $\xi$. As in \cite{Hosseini:2019ddy, Boido:2022iye, Boido:2022mbe} 
we then study a restricted class of \emph{flavour twist solutions}, where 
by definition we assume 
\begin{align}\label{intro:flavourtwist}
[\omega] = \Lambda [\varrho] \in H^2_B(\mathcal{F}_\xi)\, .
\end{align}
Here $\Lambda\in\R$ is a constant (which is ultimately fixed by flux quantization in M-theory, as in \eqref{intro:flux}) and $\varrho$ denotes the transverse Ricci form 
for the K\"ahler metric. 
The condition~\eqref{intro:flavourtwist} may be motivated in a number 
of ways. It is in some sense imposing an Einstein equation on the 
transverse K\"ahler metric, at a cohomology level, but in general 
one can certainly turn on more general K\"ahler class parameters. 
These parameters are expected, since in addition to the massless gauge 
fields $A_i$ that result from isometries of $Y_7$, the Kaluza-Klein mechanism 
will also lead to ``baryonic'' $U(1)$ gauge fields, associated with 
five-cycles in $Y_7$, with the number of these 
being $\dim H_5(Y_7,\R)$. When the latter is zero 
(such as for $Y_7=S^7$ discussed above)  the condition 
\eqref{intro:flavourtwist} is automatically true, and the same condition 
also holds for the universal twist black holes, also discussed above. 
On the other hand, when \eqref{intro:flavourtwist} does \emph{not} 
hold it is currently not understood how to match the black hole entropy to a localization calculation, due to associated flat directions for the baryonic symmetries at large $N$ \cite{Hosseini:2019ddy}. In fact, in this paper we shall find a new 
puzzle with interpreting supergravity solutions which do not 
satisfy the condition \eqref{intro:flavourtwist}.

With these ingredients to hand, we can now state how the black hole 
entropy is computed using this formalism \cite{Gauntlett:2019roi, Hosseini:2019ddy}. 
First, we introduce the gravitational free energy
\begin{align}\label{intro:Fholo}
	F_{\mathrm{grav}}[\xi] =\sqrt{\frac{2\pi^6}{27\Vol_S(Y_7)[\xi]}} \, N^{3/2}\, .
\end{align}
Here $\Vol_S(Y_7)[\xi]$ is the Sasakian volume 
of $Y_7$ \cite{Martelli:2005tp, Martelli:2006yb}, and as the notation suggests this may be computed 
just from the topology of $Y_7$ and fixing a choice of 
R-symmetry vector~$\xi$. $F_{\mathrm{grav}}$ is the free energy (the holographically 
renormalized action) of the AdS$_4\times Y_7$ vacuum discussed at the start of this section. The black hole entropy is then given by the elegant formula
\begin{equation}\label{intro:entropy2}
	\mathcal{S}[\xi] =  4 \sum_{i=1}^{s} p_i\, \frac{\partial}{\partial b_i}\, F_{\text{grav}}[\xi]\,\Big|_{b_1 = 1} \: ,
\end{equation}
where $p_i$ are the magnetic charges \eqref{intro:pi}, 
and we have chosen a basis for $\partial_{\varphi_i}$ generating 
$U(1)^s$ such that the Killing spinor is charged only under 
$\partial_{\varphi_1}$, with charge $\tfrac{1}{2}$. 
Supersymmetry is then preserved for such solutions via 
a topological twist, which fixes $p_1=2g-2$, in terms of the genus 
$g$ of the horizon $\Sigma_g$, and 
the condition 
$b_1=1$ in \eqref{intro:entropy2} simply normalizes the 
R-symmetry correctly. 
Crucially, following \cite{Couzens:2018wnk}, 
 in \cite{Gauntlett:2019roi, Hosseini:2019ddy} it is shown that 
solutions to the equation of motion \emph{extremize} 
the entropy function in \eqref{intro:entropy2} 
over the choice of $\xi$, parametrized in terms of $b_i$ via \eqref{intro:xi}.
The extremal value of $\mathcal{S}$ is then the Bekenstein-Hawking 
entropy of the corresponding black hole solution.

The dual field theory calculations, which reproduce the entropy 
in \eqref{intro:entropy2} for certain classes of models, 
were carried out in \cite{Hosseini:2016tor, Hosseini:2016ume}, 
with an interesting alternative approach recently given in 
\cite{Hosseini:2022vho}. As we review in section~\ref{sec:fieldth}, 
the field theories are generically described by 
3d $\mathcal{N}=2$ Chern-Simons-matter theories. 
These arise as the low-energy worldvolume theories on $N$ 
M2-branes at the Calabi-Yau four-fold singularity $X_4=C(Y_7)$, 
and flow to superconformal field theories (SCFTs) in the IR, dual 
to the AdS$_4\times Y_7$ vacuum. This is itself a well-established 
story. 

Introducing a supersymmetric black hole changes 
the spacetime geometry, and correspondingly
the Euclidean conformal boundary on which the field theory is defined is then $\Sigma_g\times S^1$. 
The partition function $Z$ of the Chern-Simons-matter theory 
on the latter background is called the ``topologically twisted index'', 
and this localizes to a matrix model as we review in 
section~\ref{sec:fieldth}. In general the gauge group consists of $\gauge$ 
copies of $U(N)$, labelled by $I=1,\ldots,\gauge$, and one 
can introduce the complexified Cartan-valued variables 
$u_I=\beta (A^I_3+\ii\sigma^I)$. Here $\beta$ is the radius 
of the $S^1$, $A^I_3$ is the component of the $I$-th gauge 
field along this circle direction, while $\sigma^I$ is 
the scalar field in the corresponding vector multiplet.
The localized partition function involves an integral 
over the matrix variables $u_I$, which in the large $N$ limit 
take a continuum form
\begin{align}\label{intro:eigenvalues}
u_I(t) = \ii  N^{1/2}t + v_I(t)\, ,
\end{align}
where the imaginary part $N^{1/2} t$ will play a crucial 
role in what follows. The continuum limit has an associated eigenvalue density $\rho(t)$, and the black hole entropy is identified with 
$\mathrm{Re} \log Z$. 

\subsection{Overview of results}

In this paper we extend the above matching further, and show that many quantities in the large $N$ matrix model 
can be identified directly in the gravitational duals, described by the 
GK geometry on $Y_9$. 

Firstly, a given field theory dual involves specifying a choice 
of the M-theory circle $S^1_M$. This is because the Chern-Simons-matter theories 
are typically engineered from brane configurations in type IIA string theory, 
which involve reducing on $S^1_M$. This can be specified by 
picking a $U(1)_M$ action on the fibre internal space $Y_7$, 
generated by a corresponding Killing vector field $\zeta_M$. 
Analogously to the construction in \cite{Farquet:2013cwa}, 
we then introduce an \emph{M-theory Hamiltonian function} $h_M$, also called \textit{moment map} in the following,
which satisfies the defining equation
\begin{align}
\diff h_M = -\zeta_M \lrcorner\, \omega\, .
\end{align}
Recall here that $\omega$ is the transverse K\"ahler form for the fibre $Y_7$. 
In fact there is a natural choice for the integration constant in $h_M$, 
with an alternative definition of $h_M$ given later in the paper. 
Being a continuous function on a compact space $Y_7$, the image
\begin{align}\label{intro:tau}
h_M(Y_7) = [\tau_{\mathrm{min}},\tau_{\mathrm{max}}]\subset \R\, 
\end{align}
is a compact interval in $\R$, where we denote the image by 
the variable $\tau$. Our first result is that there is a natural 
identification between this variable in gravity and the large $N$ 
eigenvalue distribution on $\Sigma_g\times S^1$, namely
\begin{align}\label{intro:taut}
\tau = \ell_p^3 N^{1/2} t\, ,
\end{align}
where $\ell_p$ is the Planck length, {\it cf.}\ equation \eqref{intro:eigenvalues}. 
We motivate this identification by following a chain of 
relationships in the set-up we have described, starting from the vacuum moduli space of the Chern-Simons-matter theory. We shall see in examples 
that the eigenvalues in the matrix model are then indeed supported on the $t$-interval corresponding to \eqref{intro:tau}, but there are further checks on this 
identification, as we now describe.

Secondly, in the gravity dual we can introduce a one-parameter 
family of five-dimensional spaces
\begin{align}
Y_\tau = h^{-1}_M(\tau)/U(1)_M\, ,
\end{align}
parametrized by $\tau$. These are ``Sasakian quotients'' of $Y_7$ by
the M-theory circle action, 
where the associated moment map level is $\tau$. We then 
find that the eigenvalue density $\rho(t)$ in the matrix model is given 
by the simple formula
\begin{align}
\rho(t) = C\cdot \left.\frac{\diff}{\diff\Lambda} \Vol(Y_\tau)\right|_{\tau = \ell_p^3N^{1/2} t}\, .
\end{align}
Here the constant $C$ is fixed by requiring that $\rho(t)$ is a normalized density,
and again we shall give a more explicit general formula later in the paper. 
Notice the derivative with respect to the parameter $\Lambda$, 
which was introduced as a K\"ahler class parameter via 
\eqref{intro:flavourtwist}.

Finally, we also study certain probe M2-branes in the 
	near-horizon AdS$_2\times Y_9$ geometries, wrapping 
	the (Euclidean) AdS$_2$ part of the spacetime and the M-theory circle direction (so that they are tangent to $\zeta_M$). These correspond to fundamental strings upon reducing to type IIA, and, when supersymmetric, should map to BPS Wilson loops in the fundamental representation of the gauge group wrapping the Euclidean time circle in the dual field theory \cite{Maldacena:1998im}.
	We find that such M2-branes are supersymmetric 
	precisely if they wrap the R-symmetry direction, and so are tangent to $\xi$. This can happen only on loci where $\xi$ and $\zeta_M$ are aligned, which are precisely the \emph{critical points} of $h_M$.
	Denoting with $\p_n$ any point in the $n$-th such (connected) locus, the image $h_M(\p_n)=\tau_n$ defines a value $t_n$ in the matrix model via \eqref{intro:taut}. The values $t_n$ are precisely the points at which the derivative of the eigenvalue density $\rho'(t)$ is discontinuous, and we find them to be related to the renormalized action $I_\text{M2}$ of the above BPS M2-branes as
	\begin{equation}
		I_\mathrm{M2}|_{\p_n}  = N^{1/2} t_n\: .
	\end{equation}
	If we order $t_1=t_{\mathrm{min}}<t_2<\cdots<t_{\mathrm{max}}$, the BPS Wilson loop in the fundamental representation is then
	\begin{align}
		\log\,  \langle W_{\mathrm{fund}}\rangle = - I_\mathrm{M2}|_{\p_1} =- N^{1/2} t_\mathrm{min}\: ,
	\end{align}
	which is the BPS M2-brane with least action. 

This is a very satisfying picture of how the 
large $N$ matrix model, that arises from localizing the
Chern-Simons-matter field theories on $\Sigma_g\times S^1$,  
is related to the gravitational dual. We should emphasize 
that some of the ingredients we have introduced above 
have analogues also in the original AdS$_4\times Y_7$ vacua
 -- for example, the eigenvalue density on $S^3$ 
was interpreted in \cite{Gulotta:2011aa} in terms of counting 
operators in the chiral ring of the gauge theory in flat spacetime. 
The fact that we find similar formulas for the black hole
solutions is non-trivial. Firstly, although the large $N$ matrix 
models on $S^3$ and $\Sigma_g\times S^1$ 
are closely related (as pointed out in \cite{Hosseini:2016tor}), 
before taking the large $N$ limit they are very different. 
This is emphasized by the alternative approach to the 
topologically twisted index and its refinement given in \cite{Hosseini:2022vho}, 
which we will briefly discuss momentarily and in section~\ref{sec:spindle}. Similarly, 
the fact that certain structures in the Sasaki-Einstein geometry of $Y_7$ 
have analogues in GK geometry for $Y_9$ which are fibrations, 
as in \eqref{intro:fibre}, only follows from the quite extensive 
work on developing GK geometry that has been cited above.  
In particular there are some key differences with previous literature, 
for example with the discussion of BPS M2-branes. 
Another important aspect of the results described above 
is that they are formally independent of the choice of 
magnetic charges $p_i$, which have no analogue 
in the AdS$_4\times Y_7$ vacuum. The matching we 
describe in fact holds ``off-shell'', for any choice of R-symmetry vector 
$\xi$; in particular we do not have to pick the superconformal 
R-symmetry which extremizes the entropy \eqref{intro:entropy2} 
(and thus depends implicitly on the magnetic charges $p_i$) in 
order for the above statements to hold, and {\it a priori} 
it is not clear that this had to be the case. 

Many of the structures we have introduced above
also extend to \emph{accelerating} black holes, 
using the results of \cite{Boido:2022iye, Boido:2022mbe}, as 
we discuss in section~\ref{sec:spindle}. 
Here the black hole horizon is replaced by a spindle $\Sigma=\mathbb{WCP}^1_{[m_+,m_-]}$. This is topologically a two-sphere, but with 
conical deficit angles $2\pi(1-1/m_\pm)$ at the poles.
Crucially there is not yet any large $N$ matrix model result to 
compare to, although this is expected to arise from 
a limit of the spindle index introduced in \cite{Inglese:2023wky}. 
However, setting $m_+=m_-=1$ gives the two-sphere, 
but where unlike in \eqref{intro:xi} the R-symmetry vector  $\xi$ on $Y_9$
is now allowed to mix with rotations of the horizon $S^2$. 
We find that the analogous quantities computed in GK 
geometry in this setting, for example BPS M2-brane actions 
which sit at the poles of the $S^2$, precisely match 
the \emph{refined} twisted index of \cite{Hosseini:2022vho}
where an additional chemical potential 
is introduced for rotations of the $S^2$, 
in the background $S^2\times S^1$ on which the field theory is defined!
We present gravity predictions for the large $N$ limit of the spindle index for
$\Sigma\times S^1$.

\subsubsection*{Plan of the paper}

The plan of the rest of the paper is as follows. In section~\ref{sec:fieldth} 
we briefly review how 3d $\mathcal{N}=2$ Chern-Simons-matter theories 
arise on the worldvolumes of $N$ M2-branes at a Calabi-Yau four-fold singularitiy 
$X_4$. Key to this is a choice of M-theory circle and reduction to type IIA string theory, 
with our main examples being the ABJM and ADHM field theories, where in both
cases $X_4=\C^4/\Z_k$. We 
summarize how the partition function of
such Chern-Simons-matter theories on $\Sigma_g\times S^1$ (the ``topologically twisted index'') localizes to a 
matrix model, and take a large $N$ continuum limit. In section~\ref{sec:gravity}, 
which is also largely review, we give an overview of the gravity dual 
solutions, with the near-horizon limit of the supersymmetric extremal 
black holes being described by AdS$_2\times Y_9$ GK geometries in M-theory. 
Here $X_4=C(Y_7)$ is a cone over $Y_7$, and in turn the internal space $Y_9$ is a fibration 
of $Y_7$ over the horizon $\Sigma_g$ \eqref{intro:fibre}.

Having introduced both sides of the duality, in 
section~\ref{sec:matching} we proceed to match quantities on both 
sides in more detail. 
 Motivated by the moduli space discussion in section~\ref{sec:fieldth} 
we introduce the M-theory Hamiltonian function on $Y_9$  and relate this to the 
continuous eigenvalue density in the matrix model. We show in detail 
how this works for our two main ABJM and ADHM examples, and 
also relate this to the matching of BPS M2-brane probes in the gravity background 
and dual Wilson loop VEVs. Section~\ref{sec:spindle} switches gears 
and discusses the generalization to accelerating black holes, 
for which the Riemann surface horizon is replaced by a spindle $\Sigma=\mathbb{WCP}^1_{[m_+,m_-]}$. In this case there is not yet a large $N$ matrix model 
result to compare to, but using \cite{Boido:2022mbe} our  gravity constructions
extend straightforwardly. Remarkably, setting $m_+=m_-=1$ so that $\Sigma=S^2$ our 
results and BPS M2-brane actions in gravity perfectly match the 
recent ``refined twisted index'', introduced in \cite{Hosseini:2022vho}. 
Section~\ref{sec:examples}
illustrates the matching described in section~\ref{sec:matching} 
with further examples, where the calculations utilize appropriate toric geometry
methods. We conclude in section~\ref{sec:discussion} with a 
brief discussion and outlook. 

Some further 
details have been relegated to various appendices: 
Appendix~\ref{app:fieldth} gives more details of the large $N$ matrix models 
with flavours, Appendix 
\ref{app:summary} collects the key results for the two main ABJM and ADHM examples, 
 Appendix~\ref{app:bps} derives the BPS condition for probe M2-branes, 
and finally Appendix~\ref{app:witten} discusses a subtlety in computing the 
M2-brane VEV in gravity, along with its resolution.

\section{M2-brane Chern-Simons theories}\label{sec:fieldth}

\subsection{Geometric engineering}\label{sec:geomengineering}

In this section we review how one can construct 3d $\mathcal{N}=2$ worldvolume theories for M2-branes in backgrounds $\R^{1,2}\times X_4$, 
where $X_4$ is a Calabi-Yau four-fold singularity. This is well-understood 
in cases where $X_4$ may be appropriately viewed as a fibration 
over a Calabi-Yau three-fold $X_3$, and there is a corresponding interpretation in terms of 
D2-branes in type IIA string theory. The fibration is related to 
adding D6-branes and/or turning on Ramond-Ramond two-form flux in
the type IIA setting. We will focus on two simple examples which 
illustrate the general features, where both describe M2-branes in 
quotients of flat spacetime
$\R^{1,2}\times \C^4/\Z_k$. We will then refer to these as primary
examples for illustration throughout the remainder of the paper.

\subsubsection{Type IIA construction}\label{sec:typeIIA}

We start by considering $N$ D2-branes in the background $\R^{1,2}\times X_3\times \R_t$, where $t\in\R$ parametrizes\footnote{Here $t$ is not to be confused with the time direction in $\R^{1,2}$, which 
will play no particular role in this paper.} the real line  and
 $X_3$ is a Calabi-Yau three-fold cone singularity. 
The latter has conical metric $g_{X_3} = \diff r^2 + r^2 g_{Y_5}$, 
where $r\geq 0$ is a radial coordinate for the cone and $(Y_5,g_{Y_5})$ is a Sasaki-Einstein 
five-manifold. There are by now many classes of such $X_3$, 
including infinite families with both explicit metrics and also general existence 
results~\cite{Sparks:2010sn}. While $X_3$
could be arbitrary, we will illustrate our results with two particularly simple 
examples, which we refer to as $\MI$ and $\MII$:
\begin{align}\label{X3defs}
\MI & : \ X_3^{\I} =\C^3\, ,\nonumber\\ 
\MII & : \  X_3^{\II} =  \{\z_1 \z_2=\z_3 \z_4\}\subset \C^4\, .
\end{align} 
Here $X_3^{\II}$ is the well-known conifold singularity, with $Y_5^{\II}=T^{1,1}\cong S^2 \times S^3$,
 while the background for $\MI$ 
is simply flat spacetime, with $Y_5^{\I} = S^5$ with its round metric. The D2-branes 
are placed at the conical singularity $r=0$ of $X_3$ (which is the origin of $\C^3$, $\C^4$, respectively, in the two models \eqref{X3defs}), and the origin $t=0$ of $\R_t$.

The low-energy effective theory on the D2-branes in the $\R^{1,2}$ directions is generically a three-dimensional 
$\mathcal{N}=2$ theory. When the singularity $X_3$ admits a resolution 
to a smooth Calabi-Yau, this effective theory is 
expected to be described by a quiver gauge theory with superpotential $W$.
For example, this will be the case if $X_3$ is a toric Calabi-Yau singularity, 
where the gauge group may be taken to be $\ggroup=U(N)^\gauge$ where $\gauge$ 
is the Euler number of the Calabi-Yau resolution. In this picture, each copy of $U(N)$ arises as the gauge group on a fractional D-brane, wrapping
various collapsed cycles at the singularity, and the bifundamental matter fields in the quiver
are massless strings between these D-branes. For our two examples we have:
\vskip 0.1cm
\begin{quote}
\bMI$_0$: $\mathcal{N}=8$ $U(N)$ SYM. In $\mathcal{N}=2$ 
language this is a $U(N)$ vector multiplet, together with 
three adjoint chiral fields $\Phi_i$, $i=1,2,3$, and superpotential $W=\mathrm{Tr}\, ( \Phi_3[\Phi_1,\Phi_2])$.
\end{quote}
\begin{quote}
\bMII$_0$: Klebanov-Witten theory \cite{Klebanov:1998hh}. This is a $U(N)^2$ quiver gauge theory, 
with bifundemental chiral fields $A_i$, $B_i$ ($i=1,2$), transforming in the  $(\mathbf{N},\overline{\mathbf{N}})$, $(\overline{\mathbf{N}},\mathbf{N})$ representations, respectively,  and superpotential $W=\mathrm{Tr}\, ( A_1B_1A_2B_2 - A_1B_2A_2B_1)$.
\end{quote}
In particular for $\MII_0$ the resolved Calabi-Yau is the resolved conifold, 
which replaces the isolated singular point $\{r=0\}=\{\z_1=\z_2=\z_3=\z_4=0\}\in X_3^{\II}$ by a 
copy of $\mathbb{CP}^1=S^2$. 

To obtain the final models that we are interested in, we add two further ingredients:

\subsubsection*{D6-branes}

\begin{figure}
\begin{center}
\begin{tikzpicture}[scale=1.2,every node/.style={scale=1.2},font=\scriptsize]
    \node[gauge] (t0) at (0,0) {$\, N\,$};
    \node[flavor] (t1) at (3,0) {$k$};
    \draw[->] (t0) edge [out=35,in=180-35,loop,looseness=1] node[anchor=north,yshift=13pt] {$Q_j$} (t1);
    \draw[<-] (t0) edge [out=-35,in=180+35,loop,looseness=1] 
    node[anchor=south,yshift=-16pt] {$\tilde Q_j$}(t1);
    \draw[<<<-] (t0) edge [out=180-35,in=180+35,loop,looseness=9.5] node[anchor=north,xshift=12pt,yshift=27pt] {$\Phi_{1,2,3}$} (t0);
\end{tikzpicture}
\end{center}
\vspace{-.5cm}
    \caption{Quiver diagram for $\MI$: the ADHM theory.}
    \label{fig:mABJM}
\end{figure}

We may add D6-branes to the above 
D2-brane set-up, without breaking any supersymmetry, provided 
these wrap divisors (complex codimension one submanifolds) in $X_3$.  
From the low-energy D2-brane worldvolume perspective, this 
adds chiral matter fields in the fundamental/anti-fundamental representations, 
corresponding to massless strings stretching between the 
D2-branes and D6-branes \cite{Benini:2009qs, Jafferis:2009th}. In particular, we define:
 \vskip 0.1cm
\begin{quote}
\bMI: ADHM theory. We add $k$ pairs of fundamental/anti-fundamental 
chiral fields $(Q_j,\tilde{Q}_j)$, $j=1,\ldots,k$ to  \MI$_0$ [$\mathcal{N}=8$ $U(N)$ SYM] above, with total superpotential 
\begin{align}\label{WI}
W = \mathrm{Tr}\bigg( \Phi_3[\Phi_1,\Phi_2] + \sum_{j=1}^k\tilde{Q}_j \Phi_3 Q_j\bigg)\, .
\end{align}
Physically we have added $k$ D6-branes to the original background, 
located at $\{z_3=0\}\subset \C^3$, and at the origin $t=0$ of $\R_t$. 
The quiver for this theory is shown in Figure \ref{fig:mABJM}. 
\end{quote}

\subsubsection*{Chern-Simons terms}

 In $2+1$ dimensions 
we may instead add an $\mathcal{N}=2$ Chern-Simons interaction for 
each $U(N)$ gauge group factor, proportional to the Chern-Simons levels
$k_I\in\Z$, $I=1,\ldots, \gauge$. As explained in \cite{Aganagic:2009zk}, 
such Chern-Simons interactions may be induced (via the Wess-Zumino couplings on D-branes) by turning 
on Ramond-Ramond (RR) fluxes in the original D2-brane background, 
where the flux threads the various (collapsed) cycles in $X_3$.  
The sum $\sum_{I=1}^\gauge k_I = 2\pi\ell_s \RF_0$, where $\ell_s$ is the string 
length and
$\RF_0$ is the Romans mass \cite{Gaiotto:2009mv}. Thus provided $\RF_0=0$ the resulting 
backgrounds uplift to M-theory. For $\MII_0$ above this leads to:
\begin{figure}
	\begin{center}
		\begin{tikzpicture}[scale=1.2,every node/.style={scale=1.2},font=\scriptsize]
			\node[gauge] (t0) at (0,0) {$N_{+k}$};
			\node[gauge] (t1) at (3,0) {$N_{-k}$};
			\draw[->>] (t0) edge [out=35,in=180-35,loop,looseness=1] node[anchor=north,yshift=13pt] {$A_{1,2}$} (t1);
			\draw[<<-] (t0) edge [out=-35,in=180+35,loop,looseness=1] 
			node[anchor=south,yshift=-16pt] {$B_{1,2}$}(t1);
		\end{tikzpicture}
	\end{center}
	\vspace{-.5cm}
	\caption{Quiver diagram for $\MII$: the ABJM theory.}
	\label{fig:ABJM}
\end{figure}
\vskip 0.1cm
\begin{quote}
\bMII: ABJM theory \cite{Aharony:2008ug}. This is the same \MII$_0$ D2-brane analogue of the Klebanov-Witten theory above, 
with superpotential
\begin{align}\label{WII}
W=\mathrm{Tr}\, ( A_1B_1A_2B_2 - A_1B_2A_2B_1)\, ,
\end{align}
 but with Chern-Simons levels $(k_1,k_2)=(k,-k)$, ensuring $\RF_0=0$. 
This  Chern-Simons coupling is induced by turning on $k$ units of RR two-form flux 
through the (collapsed) $\mathbb{CP}^1$ at the conifold singularity.  
The quiver for this theory is shown in Figure \ref{fig:ABJM}. 
\end{quote}

More generally, given any $X_3$ (with a Calabi-Yau resolution) one could add D6-branes, introducing 
fundamental matter in the field theory, and also turn 
on RR two-form flux, inducing Chern-Simons couplings in the field theory. 
Both types of deformation precisely 
fibre the M-theory circle $S^1_M$ 
over the original type IIA spacetime. 
The uplifted M-theory solution is then $\R^{1,2}\times X_4$, 
where $\mathcal{N}=2$ supersymmetry  fixes $X_4$ to be a
Calabi-Yau four-fold, with the M-theory circle then necessarily complexified 
into  $\C^*_M\equiv \R_t\times S^1_M$, and we may identify $X_3=X_4/\C^*_M$. 
Both of our models uplift to $\R^{1,2}\times \C^4/\Z_k$, 
with different $\Z_k$ quotients, as we describe 
 in more detail in section~\ref{sec:Mthcircle}.

\subsubsection{Moduli space of vacua}\label{sec:VMS}

The class of 3d $\mathcal{N}=2$ field theories we have introduced 
have moduli spaces of vacua, and in this section we briefly comment 
on some aspects of this which will be relevant for the rest of the paper. 
A more complete discussion may be found in the original references
 \cite{Martelli:2008si, Benini:2009qs, Jafferis:2009th, Aganagic:2009zk}.

The Calabi-Yau geometry $X_4$ probed by the M2-branes 
appears directly in their vacuum moduli spaces. The reason for this is simple: 
 for a single brane that probes 
a space $X$ as a pointlike object, and is free to move anywhere on $X$, 
by definition $X$  should appear as (a component of) the 
vacuum moduli space of the worldvolume field theory on that brane.  
Similarly, for $N$ branes the vacuum moduli space 
should contain 
a copy of $\mathrm{Sym}^N X$, the symmetric product 
of $N$ copies of $X$, corresponding to the moduli space 
of $N$ indistinguishable pointlike objects probing $X$.

Let us describe this for a single D2-brane in $\R^{1,2}\times \C^3\times \R_t$, 
setting $N=1$ in $\MI_0$ described at the start of section \ref{sec:typeIIA}.
In $\mathcal{N}=2$ language we have a $U(1)$ vector multiplet, 
which in addition to the gauge field contains a real scalar $\sigma$, 
together with three (uncharged) complex scalar fields $\Phi_i$, $i=1,2,3$. 
The superpotential $W$ is zero for the Abelian theory, and thus the 
vacua are simply described by the vacuum expectation 
values (VEVs) of the $\Phi_i$, which precisely parametrize $X_3^\I = \C^3$, 
together with the VEV of the real scalar $\sigma$, which 
parametrizes $\R_t$. As already mentioned above, 
in three dimensions we may also dualize the $U(1)$ gauge field 
into a periodic scalar field: in the M2-brane lift of the D2-brane, 
the VEV of this periodic scalar precisely parametrizes the M-theory circle 
$S^1_M$. Moreover, after dualizing in this way the $\mathcal{N}=2$ vector multiplet 
becomes a complex scalar field, living in $\C^*_M$. 
Extending to the non-Abelian $U(N)$ case, 
all of these fields become $N\times N$ matrices, transforming in the adjoint 
representation. 
Imposing the F-term relations $\diff W=0$ from the superpotential $W=\mathrm{Tr}\, ( \Phi_3[\Phi_1,\Phi_2])$
implies that the $\Phi_i$ commute, and are thus simultaneously 
diagonalizable, parametrized by their $N$ eigenvalues $\Phi_i^\alpha$, $\alpha=1,\ldots,N$. Similarly, $\sigma$ is now a Hermitian $N\times N$ matrix, 
which is again parameterized by its $N$ real eigenvalues $\sigma^\alpha$. 
Altogether these parametrize $\mathrm{Sym}^N (\C^3\times \R_t)=(\C^3\times \R_t)^N/S_N$, 
with the eigenvalues permuted by the Weyl group $\mathrm{Weyl}(U(N))=S_N$, which
is the permutation group of $N$ objects. 

The M2-brane worldvolume theories of interest deform the 
original D2-brane theories by adding either D6-branes (fundamental/anti-fundamental 
matter) or turning on RR flux (Chern-Simons couplings). 
There is a similar, but more involved, discussion of their vacuum moduli spaces, which 
may be found in 
the original references \cite{Martelli:2008si, Benini:2009qs, Jafferis:2009th, Aganagic:2009zk}. The main point we wish to emphasize here is that
the branch of the vacuum moduli space 
which reproduces $\mathrm{Sym}^N X_4$ has \cite{Martelli:2008si}
\begin{align}\label{sigmas}
\sigma_1 = \sigma_2 = \cdots =\sigma_\gauge\, ,
\end{align}
where $\sigma_I$ is the vector multiplet scalar for the $I$-th gauge group factor
in $\ggroup=U(N)^\gauge$. 
The $N$ real eigenvalues $\sigma_I^\alpha$ of any one of the $\sigma_I$, where $\alpha=1,\ldots, N$, 
may again be thought of as parametrizing the positions of the D2/M2-branes 
along the $\R_t$ direction.  We shall return to this point again
in the next subsection, and also when we consider the holographic 
duals.

In 3d gauge theories 
one can introduce pointlike \emph{monopole operators}, which by definition create 
magnetic flux for a gauge field through an $S^2_{\mathrm{pt}}$ surrounding 
that point. For M2-brane  theories with gauge group $\ggroup=U(N)^\gauge$ 
an important role 
is played by the diagonal monopole operators $T^{(n)}$, which by definition 
create  flux
\begin{align}
	\int_{S^2_{\mathrm{pt}}} \mathrm{Tr}\, F_I = n\in \Z\, , \quad I=1,\ldots, \gauge\, .
\end{align} 
Here $F_I$ is the field strength for the $I$'th gauge field, and it is convenient to further define $T\equiv T^{(1)}$, $\tilde{T}\equiv T^{(-1)}$. In a Chern-Simons theory such monopole 
operators also carry electric charges under the gauge group $\ggroup$. 
An alternative way to introduce $T,\tilde{T}$ is to first notice that in 
a quiver gauge theory nothing is charged under
the diagonal $U(1)_{\mathrm{diag}}\subset U(N)^\gauge$. This  Abelian gauge field may then be 
dualized to a periodic scalar field $\uptau$, with period $2\pi/\gauge$ \cite{Martelli:2008si}. 
The diagonal monopole operators may then be promoted to complex chiral
fields by identifying $T=|T|\, \ex^{\ii \gauge \uptau}$, $\tilde{T}=|\tilde{T}|\, \ex^{-\ii \gauge\uptau}$ \cite{Benini:2009qs}. 

For quiver gauge theories without D6-branes, and hence without fundamental/anti-fundamental matter, such as \MII, one can describe 
the vacuum moduli space without introducing $T$, $\tilde{T}$;  
or rather one can introduce them, but they are essentially then immediately 
redundant, satisfying the trivial relation $T\tilde{T}=1$.  
However, for theories that contain fundamental/anti-fundamental 
matter, the charges of $T$, $\tilde{T}$ under both gauge and global 
symmetries in the quantum theory obtain anomalous contributions 
from fermion zero modes in that matter \cite{Benini:2009qs, Jafferis:2009th}. 
The classical vacuum moduli space is correspondingly 
corrected in the quantum theory, and one necessarily needs to 
introduce $T$, $\tilde{T}$ to describe this 
quantum-corrected moduli space. This leads to:

\begin{quote}
	\bMI:  The monopole operators satisfy the 
	quantum F-term relation 
	\begin{align}\label{TtildeT}
		T\tilde{T} = \Phi_3^k\, .
	\end{align}
	This will be important, both for discussing the matrix model in section 
	\ref{sec:matrixmodel}, and also for comparing to the gravity 
	dual in section \ref{sec:matching}. 
\end{quote}

\subsubsection{M-theory circle}\label{sec:Mthcircle}

As mentioned at the end of section \ref{sec:typeIIA}, 
adding RR two-form flux and/or D6-branes to a type IIA 
spacetime fibres the M-theory circle $S^1_M$, resulting in a
total 11d spacetime of the form $\R^{1,2}\times X_4$. 
Equivalently, one can start with the latter spacetime and a specific choice of M-theory circle, 
and then reduce to type IIA. 
In practice we may fix a Calabi-Yau four-fold $X_4$ and
pick a $U(1)\equiv U(1)_M$ action on $X_4$. The flow of the associated Killing vector $\zeta_M$ generates the M-theory circle action.
Moreover, as explained in the last subsection the field theory moduli space for the Abelian theory precisely realizes $X_4$. 

For our models of interest recall  that 
 $X_4=\C^4/\Z_k$, where we denote standard complex coordinates on $\C^4$ by $z_a=|z_a|\ex^{\ii\phi_a}$, and the $a$-th copy of $\C$ is rotated by $\del_{\phi_a}$ with weight one. The $\Z_k$ quotient is precisely along  $\Z_k\subset U(1)_M$, but where 
the M-theory circle action depends on the model. 
 Let us see this explicitly:
\begin{quote}
	\bMI:  
	Recall that this theory  adds the fundamental/anti-fundamental 
	fields $Q_j$, $\tilde{Q}_j$, $j=1,\ldots,k$, to the $\mathcal{N}=8$ Maxwell
	theory.  
	The branch of the \emph{classical} vacuum moduli space where $Q_j=0=\tilde{Q}_j$ 
	is identical to that for a D2-brane in flat spacetime, namely 
	$\C^*_M\times \C^3$, which has the wrong topology.\footnote{There 
		will be an additional ``Higgs branch'', where $Q_j$, $\tilde{Q}_j$ 
		may obtain vacuum expectation values, but from the F-term 
		equations $\diff W=0$ this can only happen where $\Phi_3=0$, which 
		geometrically is then the D2-branes and D6-brane are coincident. This Higgs 
		branch does not play any role in our discussion.}
	However, as shown in \cite{Benini:2009qs}, the quantum-corrected moduli space 
	for this theory may be parametrized by the three complex scalars $\Phi_1, \Phi_2, \Phi_3$, together
	with the diagonal monopole operators  $T,\tilde{T}$, which 
	recall satisfy the additional
	quantum constraint \eqref{TtildeT}. 
	The latter  equation defines an $A_{k-1}=\C^2/\Z_k\subset \C^3$ 
	singularity, where the $\C^3$ here has coordinates $(T,\tilde{T},\Phi_3)$. 
	We immediately see that the quantum vacuum moduli space 
	is  $X_4^{\I}=\C^2/\Z_k\times \C^2$, with $(\Phi_1,\Phi_2)$ 
	parametrizing the second factor of~$\C^2$. \\
	We may then identify 
	\begin{align}\label{zimodelI}
		T=z_1^k\, , \quad \tilde{T}=z_2^k\, , \quad  
		\Phi_1=z_3\, , \quad  \Phi_2=z_4\, , 
	\end{align}
	in terms of the standard coordinates $z_a$ on $\C^4$. The operators above have charges
	\begin{align}\label{U1MI}
		U(1)_M^\I : (1,-1,0,0)\, ,
	\end{align}
	under the M-theory circle action $U(1)_M$. The $\Z_k$ quotient is 
precisely along $\Z_k^\I\subset U(1)_M^\I$, and we 
introduce the associated generating 
vector field 
 $\zeta_M = \frac{1}{k} \,( \del_{\phi_1}-\del_{\phi_2})$, whose orbits are precisely 
	the M-theory circles $S^1_M$ of the quotient. On the other hand, the fixed point set of the circle action $\{z_1=z_2=0\}$ corresponds to the D6-brane locus in the type IIA spacetime. The original type IIA spacetime 
$X_3^\I=\C^3$ arises as the (Geometric Invariant Theory) quotient
	\begin{equation}
		\C^4/(\C^*_M)^\I = X_3^\I = \C^3\, ,
	\end{equation}
where recall that $(\C^*_M)^\I$ is the complexification of $U(1)_M^\I$.
\end{quote}
\begin{quote}
	\bMII:  
	There are no quantum corrections in this theory, 
	and we may hence describe the vacuum moduli space without 
	introducing $T$, $\tilde{T}$. In the Abelian $N=1$
	theory the superpotential $W$ is zero, and the vacuum moduli 
	space is hence parametrized by the four scalar fields 
	\begin{align}\label{zimodelII}
		A_1 = z_1\, , \quad B_1 = z_2\, , \quad A_2 = z_3\, , \quad B_2 = z_4\, .
	\end{align}
	Nothing is charged under the $U(1)_{\mathrm{diag}}\subset U(1)^2$ diagonal gauge group, while we may identify the remaining 
	$U(1)$ (to be concrete, the second $U(1)$ in $U(1)^2$) 
	with the M-theory circle action. Thus the coordinates $z_a$ on $\C^4$ have charges
	\begin{align}\label{U1MII}
		U(1)_M^{\II}: (1, -1,1,-1)\, .
	\end{align}
The vacuum moduli space is then 
	$X_4^{\II}=\C^4/\Z_k^{\II}$, 
	with the $\Z_k^\II\subset U(1)_M^\II$ quotient arising due to a residual discrete gauge symmetry \cite{Aharony:2008ug}, and we correspondingly introduce the 
vector field $
	\zeta_M = \frac{1}{k} \,(\del_{\phi_1}-\del_{\phi_2}+\del_{\phi_3}-\del_{\phi_4})$.
	 In this example the circle action is free, and hence there are no D6-branes. However Kaluza-Klein reduction along $U(1)_M^\II$ gives rise to RR fluxes corresponding to Chern Simons coupling in the field theory.
The original type IIA spacetime is the quotient
	\begin{equation}
		\C^4/(\C^*_M)^{\II} = X_3^{\II} = \{\z_1 \z_2 = \z_3 \z_4\}\subset\C^4\, .
	\end{equation}
\end{quote}

\subsection{Topologically twisted matrix model}\label{sec:matrixmodel}

The class of 3d $\mathcal{N}=2$ Chern-Simons-matter theories 
we have introduced may be put on $\Sigma_g\times S^1$ with a topological twist along $\Sigma_g$, preserving supersymmetry, where $\Sigma_g$ is a Riemann surface of genus $g$.\footnote{From the type IIA perspective, this is achieved by wrapping the D2-branes on $\Sigma_g$.} 
The path integral of such a theory is called the topologically twisted index. In the following we review the localization of the latter, which leads to an effective matrix model description whose properties can be extracted in the large $N$ limit. 
Our discussion 
follows the original references \cite{Benini:2015noa, Benini:2016hjo, Benini:2015eyy}.

\subsubsection{Chemical potentials versus R-charges}

The topologically twisted index is a function of  background magnetic fluxes $\mathfrak{n}_A$ and fugacities $y_A$ for the flavour symmetries of the theory. 
The fugacities $y_A=\ex^{\ii\pi\Delta_A}$ are equivalently expressed in terms of the chemical potentials $\Delta_A$ and these are the variables we use in the following.\footnote{Note that we take a different convention from \cite{Benini:2015eyy}, namely $\Delta^\textrm{there}=\pi\Delta$.} Another important fugacity is that for the topological symmetry, namely  $\upxi=\ex^{\ii\pi\Delta_m}$.

On the other hand, the partition function on $S^3$ instead depends on trial R-charges, which are also denoted $\Delta_A$ in the literature. We shall use $\Delta_A^{S^3}$ to distinguish them where necessary. The chemical potentials and the R-charges appear in an identical fashion in the computation of the large $N$ partition functions on $\Sigma_g\times S^1$ and $S^3$ respectively, which leads to the formal identification
$\Delta_A=\Delta_A^{S^3}$ in \cite{Hosseini:2016tor}. The latter reference also noticed that for the topological symmetry these parameters are related up to a sign 
$\Delta_m=-\Delta_m^{S^3}$. These identifications are crucial for this work, as holographically the parameters matched to geometry are the R-charges, and then by extension the chemical potentials. Additionally, it is also expected 
that the chemical potentials should be R-charges for the 1d supersymmetric 
theory that appears on flowing to the IR, after compactification on 
$\Sigma_g$. 
We will make some of these statements more precise in section~\ref{sec:1stmatching}. In line with this comment, we introduce the formal variables $\Delta_T$ and $\Delta_{\tilde T}$ which we identify with the R-charges of the diagonal  monopole operators, such that
\begin{equation}\label{Deltam}
	\Delta_m=-\Delta_m^{S^3}=-\frac{1}{2}(\Delta_T^{S^3}-\Delta_{\tilde T}^{S^3})=-\frac{1}{2}(\Delta_T-\Delta_{\tilde T})\,.
\end{equation}
The second equality is the definition of the ``bare" monopole charge $\Delta_m^{S^3}$ \cite{Jafferis:2011zi}. 
Recall that the monopole operators $T$ and $\tilde T$ also obey an F-term relation, which translates into a constraint on their R-charges, and by extension to our variables. 
Together with \eqref{Deltam} this gives two constraints on $\Delta_T$ and $\Delta_{\tilde T}$, and it is consistent to phrase any result in terms of these rather than $\Delta_m$. 
As we shall see, many of the results we shall obtain 
take a more simple, universal form when expressed in terms of $\Delta_{T,\tilde T}$.

Finally, we note that the above variables are not all independent, but rather obey constraints coming from the superpotential.
First, the invariance of the superpotential under global symmetries imposes $\prod_A y^A=1$ in each term of the superpotential, such that\footnote{Looking only at the constraint on the fugacities it appears that we could have $\sum_A \Delta_A =2\ell$, $\ell\in\Z$ in general. However it as been argued in \cite{Hosseini:2016tor} that any $\ell\neq 1$ either does not give a consistent solution, or gives a solution related to the $\ell=1$ solution by discrete symmetries.}
\begin{equation}\label{Wconstraints}
	\sum_{A\in W} \Delta_A = 2\, , 
\end{equation}
where the sum runs over the fields appearing in each monomial of the superpotential.
On the other hand the superpotential has charge 2 under the R-symmetry $\Delta_W = 2$, giving rise to the constraint $\sum_A \Delta_A^{S^3} =2$. This guarantees that the identification $\Delta_A=\Delta_A^{S^3}$ is sensible.
Second, there is also a constraint on the fluxes arising because we are imposing a topological twist to preserve supersymmetry on the 
Riemann surface~$\Sigma_g$
\begin{equation}\label{Wconstraints2}
	 \sum_{A\in W} \mathfrak{n}_A=2g-2\, .
\end{equation}
Note that we take a different sign convention from \cite{Benini:2015eyy} for the fluxes, in order to match the geometry conventions of \cite{Boido:2022mbe} later on.

\subsubsection{Localization on $\Sigma_g\times S^1$}\label{sec:localize}

The index can be evaluated \textit{exactly} using supersymmetric localization \cite{Pestun:2007rz}. That is, for BPS observables 
the path integral is exactly equal to a (finite-dimensional) 
integral over certain supersymmetric field configurations, dressed by 
a one-loop determinant that captures fluctuations.
Recall that $\mathcal{N}=2$ theories have vector multiplets with fields $(A_\mu,\lambda,\sigma, D)$, where $A_\mu$ is the gauge field with curvature $F_{\mu\nu}$, $\lambda$ is an adjoint-valued Dirac fermion, while $\sigma, D$ are adjoint-valued scalars. 
The BPS configurations for an $\mathcal{N}=2$ vector multiplet on $\Sigma_g\times S^1$, denoting the $\Sigma_g$ directions with orthonormal frame directions $\ex^1$, $\ex^2$ and the $S^1$ with $\ex^3$, are such that $D=\ii F_{12}$, $F_{13}=F_{23}=0$, while $\sigma$ and $A_3$ take constant values along $S^1$.
The BPS moduli space is then characterized by the Cartan-valued variables
\begin{equation}\label{uBPS}
	u = \beta(A_3 + \ii \sigma) \, , \qquad  \mathfrak{m}= \frac{1}{2\pi}\int_{\Sigma_g} F \, ,
\end{equation}
where $\beta$ is the radius of the $S^1$ and the gauge magnetic fluxes $\mathfrak{m}$ take values in the co-root lattice of the gauge group.\footnote{This BPS locus is to be contrasted with the one on $S^3$ where only $\sigma=-D$ is non-zero, and in particular  there are no gauge magnetic fluxes.}
For each $U(N)$ gauge group, $u$ takes the form of an $N\times N$ matrix, which can be diagonalized by a gauge transformation. Then, for the
$I$'th gauge group factor,
\begin{equation}
	u_I = \mathrm{diag}(u_I^1,\dots,u_I^N) \,,
\end{equation}
such that the full BPS moduli space is parametrized by the $\gauge$ sets of $N$ eigenvalues $u_I^\alpha$, $\alpha=1,\ldots,N$, $I=1,\ldots,\gauge$. 

The localized partition function can then be recast into a contour integral over $u_I$ and a sum over $\mathfrak{m}$.
\begin{equation}\label{locZ}
	Z(\Delta,\mathfrak{n})=\sum_\mathfrak{m}\oint_\mathcal{C} Z_{\mathrm{int}}(u,\mathfrak{m};\Delta,\mathfrak{n})\,.
\end{equation}
We are not interested in the explicit form of the integrand at this stage. What we want to highlight is that even though the moduli space depends on the gauge content of the model only, the path integral itself also includes the matter content. Therefore, as mentioned before, it is a function of the chemical potentials and the flavour magnetic fluxes of the theory. 
Deforming the contour and performing the sum in \eqref{locZ}, the twisted index reduces to a contour integral with simple poles at $	\ex^{\ii B_I^\alpha}-1=0$ where $\ex^{\ii B_I^\alpha}$ depends on the field content; explicit rules to build this function can be found in \cite{Hosseini:2016tor}. The index is therefore given by a sum of residues evaluated at the solutions of the so-called Bethe ansatz equations (BAEs) 
\begin{equation}
	\ex^{\ii B_I^\alpha}=1 \implies \quad  B_I^\alpha + 2\pi n^\alpha_I = 0\, ,
\end{equation}
where the variables $n_I^\alpha$ parameterize the phase ambiguities, and are chosen in order to cancel the long range forces \cite{Benini:2015eyy}. Furthermore, the BAEs can be obtained as critical points of a function known as the Bethe potential $\mathcal{U}$,
\begin{equation}\label{BAE}
 \mathrm{BAEs:}\qquad\frac{\del\, \mathcal{U}}{\del u^\alpha_I} = 0\,.
\end{equation}
In summary, this reformulates the problem of evaluating the index to the one of extremizing the Bethe potential.  

\subsubsection{The large $N$ limit}\label{sec:largeN}

In the large $N$ limit, \eqref{BAE} can be solved numerically by taking the following ansatz for the eigenvalues, which is partly based on some analytical arguments and on the numerical simulations in \cite{Benini:2015eyy},
\begin{align}\label{uansatz}
	u_I^\alpha = \ii t^\alpha N^{1/2} + v_I^\alpha + o(N^0)\, ,
\end{align}
where $t^\alpha$, $v_I^\alpha$ are real and taken to be $\mathcal{O}(N^0)$ 
in the large $N$ limit. 
Notice that \eqref{uansatz} may equivalently be written 
as \begin{align}\label{Imus}
	\mathrm{Im}\, u_1=\cdots = \mathrm{Im}\, u_\gauge =  \mathrm{diag}(t^1,\cdots,t^N) N^{1/2}\, .
\end{align}
Note this condition implies that $\sigma_1=\cdots=\sigma_\gauge$, 
which is \emph{precisely} the same condition that arises on the vacuum moduli space -- see equation \eqref{sigmas}. This is a non-trivial statement: \eqref{Imus} 
describes the large $N$ eigenvalue distribution 
of the field theory localized on $\Sigma_g\times S^1$, while 
\eqref{sigmas} is a statement about the moduli space 
of the same field theory on flat spacetime $\R^{1,2}$. 
Recall that in the latter case we interpreted $\sigma_1=\cdots=\sigma_\gauge$ 
as the position of the corresponding D2-branes in $\R_t$. We will return 
to this comment in section \ref{sec:matching}, when we come to 
interpret the large $N$ saddle point distribution directly in the gravity dual.

The numerical simulations also show that the imaginary parts 
of the eigenvalues $t^\alpha$ become dense at large $N$, and are distributed over a finite interval $[t_\tmin,t_\tmax]$. We may 
order the eigenvalues using Weyl transformations so that $t^\alpha$ are increasing, 
and introduce the corresponding (discrete) density function
\begin{align}
	\rho(t) = \frac{1}{N}\sum_{\alpha=1}^N \delta(t-t^\alpha)\, , \qquad \int_{t_\tmin}^{t_\tmax}\dd t \,\rho(t)=1\, .
\end{align}
In the $N\rightarrow \infty$ limit the eigenvalues $t^\alpha$ become a continuous variable $t$, with continuous eigenvalue density $\rho(t)$. In particular \eqref{uansatz} becomes a continuous function  
\begin{equation}
	u_I(t)=\ii t N^{1/2}+v_I(t)\,.
\end{equation}
In fact, the Bethe potential can be formulated in terms of $\rho(t)$ and $v_I(t)$, and the solutions of the BAEs are found by extremizing the potential with respect to these functions.
Again there are rules, which can be found in \cite{Hosseini:2016tor} and that we review in Appendix~\ref{app:fieldth}, to build $\mathcal{U}$ in terms of these variables directly from the field content of the theory. 
The normalization of $\rho$ is imposed via the introduction of a Lagrange multiplier $\mu$, and the quantity to extremize is actually
\begin{equation}\label{lagrangemult}
	\frac{\mathcal{U}}{\ii N^{3/2}}-\pi\mu\left(\int\dd t\,\rho(t)-1\right)\,.
\end{equation}
We now present the results of this extremization, which may be found in \cite{Hosseini:2016ume} and \cite{Benini:2015eyy} respectively:
\begin{quote}
	\begin{figure}
		\centering
		\begin{tikzpicture}
			\draw[-stealth] (-1,0) -- (9.5,0) node[anchor=north] {$t$};
			\draw	(0,0) node[anchor=north] {$t_1$}
			(5,0) node[anchor=north] {$t_2$}
			(8,0) node[anchor=north] {$t_3$};
			\draw[-stealth] (5,0) -- (5,2.5) node[anchor=west] {$\rho(t)$};
			\draw[thick, blue] (0,0) -- (5,2.)--(8,0);
		\end{tikzpicture}
		\caption{Schematic behaviour of the piecewise linear distribution of eigenvalues in the large $N$ limit for the $\MI$ ADHM theory.}
		\label{fig:rhoADHM} 
	\end{figure}
	\bMI: The chemical potentials satisfy the constraints 
	\begin{align}\label{RmodelI}
		\sum_{i=1}^3 \Delta_{\Phi_i} = 2\, , \quad \Delta_{T}+\Delta_{\tilde{T}} = k \Delta_{\Phi_3}\, ,
	\end{align}
	leaving three independent. The first  constraint
	is simply the superpotential one, where the $W$ is given in \eqref{WI}, 
	while the second constraint follows from \eqref{TtildeT}.
	The saddle point eigenvalue density is
	\begin{align}\label{rhoADHM}
		\rho(t) = \begin{cases} \displaystyle\frac{\mu+\Delta_{\tilde{T}}t}{\pi^2\Delta_{\Phi_1}\Delta_{\Phi_2}\Delta_{\Phi_3}}\, , & t_\tmin<t<0\, ,\\[10pt]
			\displaystyle\frac{\mu-\Delta_{{T}}t}{\pi^2\Delta_{\Phi_1}\Delta_{\Phi_2}\Delta_{\Phi_3}}\, , & 0\leq t<t_\tmax\, ,
		\end{cases}
	\end{align}
	where the endpoints are given by
	\begin{align}\label{intervalADHM}
		t_{\tmin} = -\frac{\mu}{\Delta_{\tilde{T}}}\, , \qquad
		t_{\tmax} = \frac{\mu}{\Delta_T}\, ,
	\end{align}
	and the extremal value of the Lagrange multiplier is 
	\begin{align}\label{muI}
		\mu  = \pi\sqrt{\frac{2}{k}\Delta_{\Phi_1}\Delta_{\Phi_2}\Delta_T\Delta_{\tilde{T}}}\, .
	\end{align}
	In particular for generic chemical potentials, $\rho(t)$ is linear in two regions (see Figure~\ref{fig:rhoADHM}). 
\end{quote}

\begin{quote}
	\begin{figure}
		\centering
		\begin{tikzpicture}
			\draw[-stealth] (-1,0) -- (9.5,0) node[anchor=north] {$t$};
			\draw	(0,0) node[anchor=north] {$t_1$}
			(4,0) node[anchor=north] {$t_2$}
			(7,0) node[anchor=north] {$t_3$}
			(8,0) node[anchor=north] {$t_4$};
			\draw[-stealth] (5,0) -- (5,2.5) node[anchor=west] {$\rho(t)$};
			\draw[dotted] (4,0) -- (4,2);
			\draw[dotted] (7,0) -- (7,1.5);
			\draw[thick, blue] (0,0) -- (4,2.) -- (7,1.5)--(8,0);
		\end{tikzpicture}
		\caption{Schematic behaviour of the piecewise linear distribution of eigenvalues in the large $N$ limit for the $\MII$ ABJM theory.}
		\label{fig:rhoABJM} 
	\end{figure}
	\bMII: The chemical potentials $\Delta_{A_i}$, $\Delta_{B_i}$, $i=1,2$, 
	satisfy the single superpotential constraint
	\begin{align}
		\Delta_{A_1}+\Delta_{B_1}+\Delta_{A_2}+\Delta_{B_2} = 2 \, ,
	\end{align}
	where $W$ is given by \eqref{WII}. 
	Without loss of generality, 
	we may assume that 
	$\Delta_{A_1}\leq\Delta_{A_2}$, $\Delta_{B_1}\leq\Delta_{B_2}$. 
	One finds the following eigenvalue density
	\begin{align}\label{rhoABJM}
		\rho(t) = \begin{cases} \displaystyle\frac{\mu +  k \Delta_{B_1}t}{\pi^2(\Delta_{A_1}+\Delta_{B_1})(\Delta_{A_2}+\Delta_{B_1})(\Delta_{B_2}-\Delta_{B_1})}\, , & t_1<t<t_2\, ,\\[12pt]
			\frac{2\mu + k (\Delta_{B_1}\Delta_{B_2}-\Delta_{A_1}\Delta_{A_2})t}{\pi^2(\Delta_{A_1}+\Delta_{B_1})(\Delta_{A_1}+\Delta_{B_2})(\Delta_{A_2}+\Delta_{B_1})(\Delta_{A_2}+\Delta_{B_2})}\, , &  t_2<t<t_3\, , \\[10pt]
			\displaystyle\frac{\mu - k \Delta_{A_1}t}{\pi^2(\Delta_{A_1}+\Delta_{B_1})(\Delta_{A_1}+\Delta_{B_2})(\Delta_{A_2}-\Delta_{A_1})}\, , & t_3<t<t_4\, .
		\end{cases}
	\end{align}
which is linear in the three regions separated by 
\begin{align}\label{intervalABJM}
	 t_1= -\frac{\mu}{k\Delta_{B_1}}\, , \quad 
	t_2 = -\frac{\mu}{k\Delta_{B_2}}\, , \quad t_3 = \frac{\mu}{k\Delta_{A_2}}\, , \quad \quad  t_4= \frac{\mu}{ k\Delta_{A_1}}\, ,
\end{align}
where $t_\tmin=t_1<t_2<t_3<t_4=t_\tmax$, as depicted in Figure \ref{fig:rhoABJM}. The extremal value of the Lagrange multiplier is
\begin{align}\label{muII}
	\mu = \pi\sqrt{2 k\Delta_{A_1}\Delta_{A_2}\Delta_{B_1}\Delta_{B_2}}\, .
\end{align}
\end{quote}

The free energy $F=-\log Z$ can also be computed directly from the Bethe potential evaluated on the BAEs solutions \cite{Hosseini:2016tor} 
\begin{align}\label{freeenergy}
	\log Z_{\Sigma_g\times S^1}(\Delta,\mathfrak{n}) = 
	\frac{1}{\pi}\sum_{A}\, \left[ \mathfrak{n}_A  \frac{\partial \,\mathcal{ \bar U}(\Delta)}{\partial \Delta_A} \right] \, ,\\
	\mathcal{\bar U}(\Delta) \equiv -\ii\,\mathcal{U}|_\mathrm{BAEs}  = \frac{2\pi\mu}{3}N^{3/2}\, ,
\end{align} 
where the sum runs over all the fields in the quiver, as well as the monopole operators. Note that some fields can be eliminated using the constraints \eqref{Wconstraints} and \eqref{Wconstraints2}.
For our examples, this recovers some known results:
\begin{quote}
	\bMI
	\begin{equation}\label{FtwistedADHM}
\hspace{-1.2cm}	\log Z^\mathrm{ADHM}_{\Sigma_g\times S^1} = \frac{\pi}{3}\sqrt{\frac{2}{k} \Delta_{\Phi_1}\Delta_{\Phi_2}\Delta_{T}\Delta_{\tilde T}}\left(\frac{\mathfrak{n}_{\Phi_1}}{\Delta_{\Phi_1}}+\frac{\mathfrak{n}_{\Phi_2}}{\Delta_{\Phi_2}}+\frac{\mathfrak{n}_{T}}{\Delta_{T}}+\frac{\mathfrak{n}_{\tilde T}}{\Delta_{\tilde T}}\right)N^{3/2}\, ,
	\end{equation}
	and
	\begin{equation}\label{FS3ADHM}
		\frac{2}{\pi}\,\mathcal{\bar U}(\Delta) = \frac{4\pi}{3}\sqrt{\frac{2}{k}\Delta_{\Phi_1}\Delta_{\Phi_2}\Delta_{T}\Delta_{\tilde T}}N^{3/2}=F^\mathrm{ADHM}_{S^3}\, .
	\end{equation}
\end{quote}
\begin{quote}
\bMII
\begin{equation}\label{FtwistedABJM}
	\hspace{-1.2cm}\log Z^\mathrm{ABJM}_{\Sigma_g\times S^1} = \frac{\pi}{3}\sqrt{2 k\Delta_{A_1}\Delta_{A_2}\Delta_{B_1}\Delta_{B_2}}\left(\frac{\mathfrak{n}_{A_1}}{\Delta_{A_1}}+\frac{\mathfrak{n}_{A_2}}{\Delta_{A_2}}+\frac{\mathfrak{n}_{B_1}}{\Delta_{B_1}}+\frac{\mathfrak{n}_{B_2}}{\Delta_{B_2}}\right)N^{3/2}\, ,
\end{equation}
and
\begin{equation}\label{FS3ABJM}
	\frac{2}{\pi}\mathcal{\bar U}(\Delta) = \frac{4\pi}{3}\sqrt{2 k\Delta_{A_1}\Delta_{A_2}\Delta_{B_1}\Delta_{B_2}}N^{3/2}=F^\mathrm{ABJM}_{S^3}\, .
\end{equation}
\end{quote}
There are two things to notice here. First, the similarity of the expressions between the two models. This is easily understood from looking at the dual geometries which we will do in section~\ref{sec:1stmatching}. Second, that $\mathcal{\bar U}$ corresponds to the expression of the free energy on $S^3$ upon identifying the chemical potential with the R-charges, which has already been noted in \cite{Hosseini:2016tor}. There is no {\it a priori} reason for these quantities to be related from a purely field theoretic point of view. For some related 
discussion in the gravity dual, see  \cite{Gauntlett:2019roi}. 

Finally we shall also be interested in the expectation values of supersymmetric Wilson loop
operators.
Recall that a Wilson loop in a representation $\mathcal{R}$ is defined as 
\begin{equation}
	W_\mathcal{R} = \mathrm{Tr}_\mathcal{R}\mathcal{P}\exp{\oint \dd \tau \left(\ii A_\mu \dot x^\mu +\sigma|\dot x|\right)} \, .
\end{equation}
In order for the Wilson loop to be supersymmetric, it needs to wrap the $S^1$ direction \cite{Benini:2015noa}. It then reduces to $W_\mathcal{R}=\mathrm{Tr}_\mathcal{R} \exp(\ii u)$ with $u$ defined in \eqref{uBPS}.
The expectation value is obtained by inserting this loop operator into the path integral, and normalizing by a factor of $1/Z$, so that $\langle 1 \rangle =1$.
In the large $N$ limit, the factors of the partition function simply cancel. Therefore, in the fundamental representation, 
\begin{equation}
	\langle W_\mathrm{fund} \rangle  =\sum_{I=1}^\gauge\sum_{\alpha=1}^N e^{\ii u_I^\alpha}
	=N\sum_{I=1}^\gauge\int_{t_\tmin}^{t_\tmax} \dd t \rho(t) \ex^{-N^{1/2}t+\ii v_I(t)}\, ,
\end{equation}
such that at leading order in $N$
\begin{equation}\label{logW}
	\log\, \langle W_\mathrm{fund}\rangle = -N^{1/2}t_\tmin \, .
\end{equation} 
Compare this with the discussion in  \cite{Farquet:2013cwa}, 
on $S^3$, which is similar. 

\section{Gravity duals}\label{sec:gravity}

In this section we introduce and then further study the geometry of the M-theory backgrounds dual to the 3d field theories introduced in section~\ref{sec:fieldth}. Firstly, we briefly introduce the vacuum AdS$_4 \times Y_7$ solutions, with $Y_7$ being a 7d Sasaki-Einstein manifold, which are dual to the field theories placed on $S^3$ (the conformal boundary 
of global Euclidean AdS$_4$). Next, we consider inserting a black hole with horizon $\Sigma_g$ into such a  background. This is the gravity setting relevant for reproducing the topologically twisted index at large $N$ discussed in section \ref{sec:matrixmodel}, as the conformal boundary with the insertion of a black hole is $\Sigma_g\times S^1$ rather than $S^3$.

The near-horizon geometry of such extremal black holes takes the form of AdS$_2 \times Y_9$ with $Y_9$ a GK manifold \cite{Kim:2006qu,Gauntlett:2007ts}. More specifically $Y_9$ here is the total space of a fibration with base $\Sigma_g$ and fibre $Y_7$, as in 
\eqref{intro:fibre}.  
We present the main properties of such fibred $Y_9$ geometries and summarize the geometric extremization procedure to get the entropy and the R-charges in gravity.
Finally, we illustrate concretely all the above constructions by looking at case where $Y_7$ is a 7-sphere (or a $\Z_k$ quotient thereof), which corresponds to the gravity dual of the Models I and II previously introduced.

\subsection{AdS$_4$ backgrounds in M-theory}\label{sec:ads4}

To set the stage, we briefly illustrate the near-horizon limit of the 11d geometries corresponding to the 3d SCFTs on $S^3$ we introduced in section~\ref{sec:fieldth}.
This will also allow us to introduce various notation and geometric 
concepts that will be needed for the black hole generalization, where 
the  internal manifold $Y_7$ will fiber over the horizon $\Sigma_g$ in the later subsections.

The starting point consists of $\R^{1,2}\times X_4$ backgrounds in M-theory where $X_4$ is a Calabi-Yau four-fold cone. There are many classes of constructions of such Calabi-Yau's, including explicit metrics and existence results, and some of these will be considered later in section \ref{sec:examples}. However, for the sake of clarity, we mainly focus on the simplest possible example, namely $X_4=\C^4$ (or a $\Z_k$ quotient
thereof), which leads to the Model I and II introduced previously in section~\ref{sec:fieldth}. 
Following \cite{Martelli:2005tp} we go ``off-shell'' and consider $X_4$ equipped with a K\"ahler, but 
not necessarily Ricci-flat, cone metric
\begin{align}\label{gX4}
	\diff s^2_{X_4} = \diff r^2 + r^2\, \diff s^2_{Y_7}\, .
\end{align}
When the cone \eqref{gX4} is K\"ahler the induced metric $\diff s^2_{Y_7}$  on $Y_7=\{r=1\}\subset X_4$
is by definition Sasakian and therefore comes equipped with 
a unit length Killing vector field $\xi$ called \emph{Reeb vector field}. In particular the metric on $Y_7$ can be written as
\begin{equation}\label{Y7split}
	\diff s^2_{Y_7} = \tildeeta^2 + \diff s_{\text{6d}}^2 (\tildeJ_S) \: ,
\end{equation}
where $\tildeeta$ is the one-form dual to $\xi$, i.e.\ $\xi \lrcorner\, \tildeeta = 1$ (and $\xi \lrcorner\, \diff \tildeeta = 0$). 
The metric \eqref{Y7split} itself 
is transversely Kähler, with Kähler two-form $\tildeJ_S$ and $\diff\tildeeta = 2 \tildeJ_S$.\footnote{
	The superscript in $\tildeJ_S$ is meant to highlight that this Kähler two-form is such that the metric \eqref{Y7split} is Sasakian. Later on we will lift this assumption and consider a generic Kähler two-form $\tildeJ$.} The transverse Kähler form on the base of the cone can be related with the Kähler form $\mathcal{J}$ on the cone itself in the following way
\begin{equation}\label{kahler_relation}
	\mathcal{J} \equiv \frac{1}{2} \, \diff(r^2 \tildeeta) = r\,\diff r \wedge \tildeeta + r^2\,\omega_S \quad \implies \mathcal{J}\,\big|_{r=1} = \omega_S\:.
\end{equation}

The holomorphic $(4,0)$-form $\Psi_{(4,0)}$ on the cone $X_4 = C(Y_7)$ has charge 4 under $\xi$ i.e.
\begin{equation}\label{Psi}
	\cL_\xi \Psi_{(4,0)} = 4 \ii  \Psi_{(4,0)} \:.
\end{equation}
We assume that $Y_7$ has a $U(1)^s$ isometry with $1 \leq s \leq 4$, and we call the corresponding generating vector fields $\partial_{\varphi_i}$, $i=1,\dots,s$, with the coordinates $\varphi_i$ having period $2\pi$. The case $s = 4$ corresponds to a toric cone and will be considered in more detail in section~\ref{sec:toric}. Having introduced the coordinates $\varphi_i$, we may parameterize the Reeb vector field $\xi$ as
\begin{equation}\label{Reebvec}
	\xi = \sum_{i=1}^s b_i\, \partial_{\varphi_i} \:.
\end{equation}
In particular, we choose the basis in \eqref{Reebvec} in such a way that $\Psi_{(4,0)}$ has charge 1 under $\partial_{\varphi_1}$ and is uncharged under all other $\partial_{\varphi_i}$, so that \eqref{Psi} is achieved by setting $b_1 = 4$.

Placing $N$ M2-branes at the singular point $\{r=0\}\in X_4$, including 
their backreaction and then taking the near-horizon limit, results 
in the associated AdS$_4$ backgrounds 
\begin{align}\label{AdS4}
	\diff s^2_{11} & = L^2\left(\tfrac{1}{4}\,\diff s^2_{\mathrm{AdS}_4}+ \diff s^2_{Y_7} \right)\, ,\nonumber\\
	G&  = \tfrac{3}{8}L^3 \vol_{\mathrm{AdS}_4}\, .
\end{align}
Here $g_{\mathrm{AdS}_4}$ is the metric on a unit  radius AdS$_4$, with $\vol_{\mathrm{AdS}_4}$ 
the corresponding volume form, while $G$ is the M-theory four-form. 
The constant length scale 
$L$ is fixed by flux quantization via
\begin{align}
	L^6=\frac{(2\pi\ell_p)^6N}{6\Vol_S(Y_7)}\, ,
\end{align}
where $\ell_p$ is the eleven-dimensional Planck length, $\Vol_S(Y_7)$ is the volume of the Sasakian metric on $Y_7$, and the integer flux $N$ given by \eqref{intro:flux}
is the number of M2-branes. 

The holographically renormalized gravitational 
free energy of such backgrounds is 
\begin{align}\label{Fholo}
	F_{\mathrm{grav}}[\xi] = \frac{\pi}{2G_4} =  \sqrt{\frac{2\pi^6}{27\Vol_S(Y_7)[\xi]}} \, N^{3/2}\, ,
\end{align}
where $G_4$ is the four-dimensional effective Newton constant. 
Here $\Vol_S(Y_7)[\xi]$ highlights the fact that the Sasakian volume of
$Y_7$ is a function of the Reeb vector field $\xi$ \cite{Martelli:2005tp, Martelli:2006yb}. The eleven-dimensional metric and four-form 
in \eqref{AdS4} give an $\mathcal{N}=2$ supersymmetric solution to eleven-dimensional 
supergravity precisely when the Sasakian metric $\diff s^2_{Y_7}$ is also Einstein, 
or equivalently when the K\"ahler cone metric \eqref{gX4} is also Ricci-flat, {\it i.e.}\ Calabi-Yau. In \cite{Martelli:2005tp, Martelli:2006yb} it was shown that 
such metrics precisely minimize the volume $\mathrm{Vol}(Y_7)[\xi]$, 
as a function of $\xi$ subject to the charge constraint \eqref{Psi} which 
fixes its normalization. It immediately follows 
that the extremal value of $\xi$ then maximizes the holographic free energy $F_{\mathrm{grav}}$ in \eqref{Fholo}.

\subsection{Near-horizon geometry of AdS$_4$ black holes}\label{sec:GK}

From a geometric engineering perspective, taking an M2-brane worldvolume field theory and putting it on a Riemann surface $\Sigma_g$ amounts to wrapping the M2-branes at the tip of the CY cone $X_4$ over $\Sigma_g$. When taking the near-horizon limit this set-up results in a particular class of AdS$_2 \times Y_9$ backgrounds where $Y_9$ is a GK manifold \cite{Kim:2006qu, Gauntlett:2007ts} taking the particular form of a fibration over $\Sigma_g$ with fibres $Y_7$ as in \eqref{Y7split}, although with Kähler form $\tildeJ$ in general different from the Sasakian Kähler form $\tildeJ_S$. These are the near-horizon limits of supersymmetric, asymptotically AdS$_4$,  magnetically charged black holes, 
that we began our discussion with at the start of the paper. 

\subsubsection{GK geometry and geometric extremization}\label{sec:geoextr}
We start by describing the M-theory background arising from a general GK manifold $Y_9$. We consider solutions of the form
\begin{align}
	\diff s_{11}^2 &=  \ex^{-2B/3}\, (\diff s^2_{\text{AdS}_2} + \diff s^2_{Y_9}) \:, \label{11dmetric} \\
	G &=  \vol_{\text{AdS}_2} \wedge F \:, \label{G4_gk}
\end{align}
where $\diff s^2_{\text{AdS}_2} $ is the metric on AdS$_2$ with unit radius, and $B$ and $F$ are respectively a function and a closed two-form on a compact manifold $Y_9$. 
Via Dirac quantization the seven-form flux $\star\, G$ needs to be quantized over a set of codimension 2 submanifolds $\Sigma_\Upsilon \subset Y_9$, which constitute a basis for the free part of $H_7(Y_9,\Z)$:
\begin{equation}\label{flux_quant}
	\frac{1}{(2\pi \ell_p)^6}\,\int_{\Sigma_\Upsilon} \star\, G = \mathcal{N}_\Upsilon \in \Z \:.
\end{equation}

Imposing supersymmetry, {\it i.e.}\ the existence of Killing spinors, it can be shown \cite{Gauntlett:2007ts} that the 9d Riemannian manifold $Y_9$ is equipped with a unit norm Killing vector field $\xi$, now called the R-symmetry vector,\footnote{The reason we used the same symbol to denote the R-symmetry vector and the Reeb vector in section \ref{sec:ads4} will become clear momentarily.} whose dual one-form we call $\eta$. The metric on $Y_9$ is adapted to the corresponding foliation $\mathcal{F}_\xi$, taking the form
\begin{equation}\label{GKmetric}
	\diff s^2_{Y_9} = \eta^2 + \ex^B\,\diff s_{\text{8d}}^2(J) \: .
\end{equation}
Here $\diff s_{\text{8d}}^2(J)$ is the 8d transverse metric, which is Kähler with Kähler two-form $J$, Ricci two-form $\varrho = \diff P$, and a positive Ricci scalar $R>0$. Note that in local coordinates we may write
\begin{equation}
	\xi = \partial_z \:, \qquad\qquad \eta = \diff z + P \:,
\end{equation}
so that $\diff \eta = \varrho$. The transverse Kähler form $J$ fully determines this class of geometries as
\begin{equation} \label{F_gk}
	\ex^B = \frac{R}{2} \:, \qquad\qquad F = -J + \diff\left(\ex^{-B} \eta\right).
\end{equation}

The cone $C(Y_9)$, with metric $\diff s^2_{C(Y_9)} = \diff r^2 +r^2\,\diff s_{Y_9}^2$, has a canonically defined integrable complex structure and it admits a closed nowhere-vanishing holomorphic (5,0)-form $\Psi_{(5,0)}$, hence it is a Calabi-Yau in this sense \cite{Gauntlett:2007ts}. Furthermore, $\Psi_{(5,0)}$ has definite charge under the $U(1)$ action generated by $\xi$:
\begin{equation}\label{hol_charge}
	\mathcal{L}_\xi \Psi_{(5,0)} = \ii \Psi_{(5,0)} \:.
\end{equation}

The geometries described are supersymmetric and satisfy the 11d equations of motion provided that the transverse Kähler metric satisfies the equation
\begin{equation}\label{PDE}
	\square R = \frac{1}{2}\,R^2 - R_{ab}R^{ab}\:,
\end{equation}
where $R_{ab}$ is the Ricci tensor and $\square$ the Laplacian. 
However, we will again go ``off-shell'' following \cite{Couzens:2018wnk} and we will not impose the PDE \eqref{PDE} on the Kähler two-form. Instead we will follow an alternative approach, which is the analogue of the constrained extremization mentioned at the end of section \ref{sec:ads4}. This consists of two steps: 
rather than imposing \eqref{PDE}, one instead imposes its \emph{integral} 
over $Y_9$ to hold, namely 
\begin{align}
	\int_{Y_9} \eta  \wedge \varrho^2 \wedge \frac{J^2}{2} &= 0 \label{constraint}\:.
\end{align}
It then makes sense to impose the flux quantization conditions \eqref{flux_quant}, which read
\begin{align}
	\int_{\Sigma_\Upsilon} \eta  \wedge \varrho \wedge \frac{J^2}{2} &= (2\pi \ell_p)^6\,\mathcal{N}_\Upsilon \:. \label{flux_quant2}
\end{align}
It can be proved that these integrals depend only on the (basic) cohomology classes of $J$ and $\varrho$. 
The second step then involves varying the \textit{supersymmetric action}
\begin{equation}\label{Ssusy}
	S_{\text{SUSY}} (\xi, [J])= \int_{Y_9} \eta \wedge \varrho \wedge \frac{J^3}{3!} \:,
\end{equation}
subject to the constraints \eqref{flux_quant2} and \eqref{constraint}. 
Solving these for the cohomology class of $J$, 
the result is a function depending on the R-symmetry vector $\xi$ only, which is the analogue of \eqref{Fholo} in the AdS$_4$ case. This function is proportional to the ``trial entropy" of the black hole
\begin{equation}\label{entropy}
	\mathcal{S} = \frac{4\pi}{(2\pi)^8 \ell_p^9}\, S_{\text{SUSY}}\:.
\end{equation}
Extremizing $\mathcal{S}$ over the R-symmetry vector $\xi$ and imposing its positivity yields the on-shell entropy.

\subsubsection{Fibered $Y_9$}\label{sec:fibration}
As anticipated we want to put a black hole in the AdS$_4$ background and study its near-horizon limit. This amounts to consider AdS$_2 \times Y_9$ backgrounds as in \eqref{11dmetric} where $Y_9$ takes the fibered form
\begin{equation}\label{fibration}
	Y_7 \hookrightarrow Y_9 \rightarrow \Sigma_g \:,
\end{equation}
where $Y_7$ is a manifold as in \eqref{Y7split} with an {\it a priori} generic Kähler two-form $\tildeJ$ instead of the Sasakian one. The R-symmetry vector $\xi$ is assumed to be tangent to $Y_7$, hence it can be parametrized in the same way as the Reeb vector of the fibre \eqref{Reebvec}.
Physically the fibration \eqref{fibration} may be achieved by turning on $s$ $U(1)$ gauge fields $A_i$ supported on $\Sigma_g$ with integer magnetic fluxes
\begin{equation}\label{twist_fluxes}
	\frac{1}{2\pi}\int_{\Sigma_g} \diff A_i = p_i \in \Z \:,
\end{equation}
which are nothing but the magnetic charges of the black hole.
Mathematically, this amounts to constructing a principal $U(1)^s$ bundle over $\Sigma_g$ with Chern numbers $p_i$. The holomorphic (5,0)-form $\Psi_{(5,0)}$ is obtained as a wedge product of the canonical (1,0)-form on $\Sigma_g$ and the (4,0)-form $\Psi_{(4,0)}$, and differently from the AdS$_4$ case it has charge 1 under the $U(1)$ action generated by $\xi$, hence $b_1 = 1$. In order for $\Psi_{(5,0)}$ to be globally well-defined form, in our basis we need to set $p_1 = 2g-2$ \cite{Gauntlett:2018dpc}.
In the following we will refer to the integers $p_i$ as \emph{flavour twisting parameters}. 

Denoting $\tildesigma$ as the one-form such that locally $\tildeJ = \diff\tildesigma$, concretely the fibration amounts to replace
\begin{align}
	\tildeeta &\quad\rightarrow\quad \tildeeta^\text{t} \equiv \, \tildeeta\, \big|_{\diff \varphi_i \to \diff \varphi_i + A_i} \:, \label{eta_t}\\[6pt]
	\tildesigma &\quad\rightarrow\quad \tildesigma^\text{t} \equiv \, \tildesigma\,\big|_{\diff \varphi_i \to \diff \varphi_i + A_i} \label{sigma_t}
\end{align}
where the superscript $\text{t}$ stands for ``twisted". Then one can show that for $Y_9$ as in \eqref{fibration} we have
\begin{align}
	\eta &= \tildeeta^\text{t} + \text{basic} \:, \label{totaleta}\\[6pt]
	J &= \diff\tildesigma^\text{t} + A\,\vol_{\Sigma_g} + \text{basic exact} \:, \label{kahlerform}
\end{align}
where we take the normalization to be $\int_{\Sigma_g}\vol_{\Sigma_g}=1$, $A$ is parametrizing the Kähler class of the Riemann surface, and ``basic" here is with respect to the R-symmetry foliation $\mathcal{F}_\xi$. Note that the equation $\diff \eta = \varrho$ at the cohomology level becomes
\begin{equation}
	[\diff \tildeeta^\text{t}] = [\varrho] \in H_B^2(\mathcal{F}_\xi) \:. 
\end{equation}

To summarize: we started from a manifold $Y_7$ with $U(1)^s$ isometry, within which we picked an R-symmetry vector field $\xi$, parametrized in terms of $b_i$ as in \eqref{Reebvec}. Then we built the fibration \eqref{fibration} specified by some twisting parameters $p_i \in \Z$. The resulting manifold is a foliation defined by the above choice of $\xi$, which furthermore depends on the Kähler form of the transverse metric \eqref{kahlerform}, which is unspecified up to this point. 

\subsubsection{Flavour twist}\label{sec:flavtwist}
As discussed in the introduction, 
in the remainder of the paper we will be interested in the so called ``flavour twist", which amounts to imposing that the Kähler class of the fibre is proportional to its Ricci form class, {\it i.e.}\ 
\begin{equation}\label{flavour_twist}
	[\tildeJ] = \Lambda\, [\tilderho] \:.
\end{equation}
The full Kähler class $[J]$ is then parameterized by two parameters only: $A$ and $\Lambda$. This is the case which is under control from the holographic perspective, where a holographic 
matching of the free energy has been achieved  \cite{Hosseini:2019ddy, Gauntlett:2019roi}. Note that for manifolds $Y_7$ with no baryonic symmetries, {\it  i.e.}\ $H^2(Y_7, \R) = 0$ (such as $S^7$ and $\Z_k$ quotients thereof, as in our main examples), the condition \eqref{flavour_twist} is automatically true. 

With this assumption, the geometric extremization procedure summarized in section~\ref{sec:geoextr} simplifies, but we will not go through the steps here.\footnote{See \cite{Hosseini:2019ddy} (and note that what we call ``flavour twist" is dubbed ``mesonic twist" in \cite{Hosseini:2019ddy}).} In the present work all we will need is an expression for the  Kähler class parameter $\Lambda$ which can be obtained by picking the 7-cycle given by the fibre $Y_7$ itself in \eqref{flux_quant2}.
Indeed, the resulting flux is equivalent to \eqref{intro:flux} and gives the number of M2-branes $N$, and it is quadratic in $\Lambda$
\begin{equation}\label{flux_quant3}
	\frac{\Lambda^2}{2}\int_{Y_7} \tildeeta \wedge \tilderho^3 = (2\pi \ell_p)^6\,N\:.
\end{equation}
Now note that the integral at the left hand side is proportional to the Sasakian volume of $Y_7$
\begin{equation}\label{sasakvol}
	\Vol_S(Y_7) = \frac{1}{(2b_1)^3}\int_{Y_7} \tildeeta \wedge \frac{\tilderho^3}{3!} \:,
\end{equation}
hence we can solve \eqref{flux_quant3} to obtain\footnote{A thorough discussion of the choice of sign in taking the square root is given in \cite{Boido:2022mbe}.}
\begin{equation}\label{Lambda}
	\Lambda = \frac{(2\pi \ell_p)^3}{\sqrt{24b_1^3 \, \Vol_S(Y_7)}}N^{1/2}\:.
\end{equation}
Then the remaining equations in \eqref{constraint} and \eqref{flux_quant2}  fix $A$, and put  constraints on the remaining quantized fluxes $\mathcal{N}_\Upsilon$ in terms of the R-symmetry vector $b_i$ and the flavour twisting parameters $p_i$, and eventually one obtains the trial black hole entropy with a flavour twist
\begin{equation}\label{entropy2}
	\mathcal{S}[\xi] =  4 \sum_{i=1}^{s} p_i\, \frac{\partial}{\partial b_i}\, F_{\text{grav}}[\xi]\,\Big|_{b_1 = 1} \:,
\end{equation}
where $ F_{\text{grav}}[\xi]$ is the trial free energy \eqref{Fholo} of the vacuum AdS$_4$ solution we started with.\footnote{Note that the free energy appearing in \eqref{entropy2} is intended to be before setting $b_1 = 4$.}

It is useful to introduce further observables: the geometric R-charges. These are integrals associated to 5d $U(1)^s$-invariant supersymmetric submanifolds in the fibres $S_a \subset Y_7$ and they are dual to the R-charges of baryonic operators associated with M5-branes wrapping them. In particular, they are defined as
\begin{equation}\label{Rcharges}
	R_a \equiv \frac{1}{N}\, \frac{4\pi }{(2\pi\ell_p)^6 }\,\int_{S_a} \tildeeta\wedge \frac{\tildeJ^2}{2} \:.
\end{equation}
Analogously to what we did above, we can introduce the Sasakian volume of these submanifolds as
\begin{equation}
	\Vol_S(S_a) = \frac{1}{(2b_1)^2}\int_{S_a} \tildeeta \wedge \frac{\tilderho^2}{2} \:,
\end{equation}
so that in the flavour twist case the geometric R-charges reduce to the ratio between Sasakian volumes
\begin{equation}\label{Rcharges2}
	R_a  = \frac{2\pi}{3b_1}\,\frac{\Vol_S(S_a)}{\Vol_S(Y_7)}\:,
\end{equation}
with $R_a > 0$ \cite{Boido:2022mbe}.

Finally, other relevant quantities associated to these submanifolds are the quantized fluxes $M_a$ that are obtained by considering the integral in \eqref{flux_quant2} on fibrations $S_a \hookrightarrow \Sigma_a \rightarrow \Sigma_g$. Note that these are different from the $\mathcal{N}_\Upsilon$ in \eqref{flux_quant2} and in general they are non-vanishing even if $H_5(Y_7,\R) = 0$ (see \cite{Boido:2022mbe} for more details on this). For the flavour twist they are related to the geometric R-charges $R_a$ by
\begin{equation}\label{fluxes}
	M_a = \frac{N}{2} \sum_{i=1}^{4} p_i\, \frac{\partial}{\partial b_i} \, (b_1 R_a) \:.
\end{equation}

\subsection{Example: $Y_7 = S^7$}\label{sec:S7example}

In this section we illustrate the simplest example of the above geometry, corresponding to the fibre being a 7-sphere.

The cone $X_4 = C(S^7)$ in this case is simply $\C^4$, hence we can introduce four complex coordinates $z_a = |z_a|\ex^{\ii \phi_a}$, $a=1,2,3,4$, with $\phi_a$ having period $2\pi$. The holomorphic form $\Psi_{(4,0)}$ is the standard one on $\C^4$ 
\begin{equation}
	\Psi_{(4,0)} = \diff z_1 \wedge \diff z_2 \wedge \diff z_3 \wedge \diff z_4 \:.
\end{equation}
The standard K\"ahler form for the flat metric on $\C^4$ is
\begin{equation}\label{kahlerformC4}
	\mathcal{J} = \frac{\ii}{2} \, \sum_{a=1}^{4} \diff z_a \wedge \diff \bar{z}_a = \sum_{a=1}^{4} \diff(\tfrac{1}{2}|z_a|^2) \wedge \diff \phi_a = \sum_{a=1}^{4} \diff y_a \wedge \diff \phi_a\:,
\end{equation}
where in the last step we introduced the coordinates $y_a = \frac{1}{2} \, |z_a|^2 \geq 0$. Note that the last form in \eqref{kahlerformC4} is still valid when we lift the Einstein condition on $\C^4$, though the coordinates $y_a$ are defined differently.

The number of independent $U(1)$ isometries is clearly $s=4$, hence there has to exist a linear change of variables relating the vector fields $\partial_{\phi_a}$ rotating the coordinates $z_a$ with weight one with the general basis introduced in \eqref{Reebvec}
\begin{equation}\label{vectorfieldS7}
	\partial_{\phi_a} = \sum_{i=1}^{4} v_{ai}\, \partial_{\varphi_i}\:,
\end{equation}
where $v_{ai}$ are 16 constants.
Recalling that we require $\Psi_{(4,0)}$ to have charge one under $\partial_{\varphi_1}$ and to be uncharged under the remaining $\partial_{\varphi_i}$ we can read off
\begin{equation}\label{toricdata_S7}
	v_1 = (1,0,0,0) \:, \quad v_2 = (1,1,0,0) \:, \quad v_3 = (1,0,1,0) \:, \quad v_4 = (1,0,0,1) \:.
\end{equation}

The Sasakian volume as a function of the R-symmetry vector is given by
\begin{equation}\label{volS7}
	\Vol_S(S^7) = \frac{\pi^4}{3b_2\,b_3\,b_4\,(b_1-b_2-b_3-b_4)}\: .
\end{equation}
Setting $z_a=0$ selects a copy of $\C^3\subset\C^4$ which is the cone over the submanifold $S_a$. The corresponding Sasakian volumes are given by
\begin{equation}\label{volsubS7}
	\Vol_S(S_1)=\frac{\pi^3}{b_2\,b_3\,b_4}, \qquad \Vol_S(S_a)=\frac{\pi^3\,b_a}{b_2\,b_3\,b_4\,(b_1-b_2-b_3-b_4)}\,, \quad a=2,3,4 \:.
\end{equation}

From the quantities above we can directly read off the final results for the gravitational free energy of the AdS$_4$ solution \eqref{Fholo}, and the black hole entropy \eqref{entropy2}, the geometric R-charges \eqref{Rcharges2}, and the fluxes \eqref{fluxes} through the submanifolds $\Sigma_a \subset Y_9$ of the AdS$_2$ solution:
\begin{align}
	F_{\mathrm{grav}} &=  \frac{\pi}{3}\sqrt{2b_2\,b_3\,b_4\,(b_1-b_2-b_3-b_4)} \, N^{3/2}\: ,\nn \\[4pt]
	\mathcal{S} &= \frac{4\pi}{3}\,N^{3/2} \sum_{i=1}^{4} p_i\, \frac{\partial}{\partial b_i}\, \sqrt{2b_2\,b_3\,b_4\,(b_1-b_2-b_3-b_4)} \:, \nn \\[4pt]
	R_1&=\frac{2}{b_1}\left(b_1-b_2-b_3-b_4\right), \qquad \ \  \, R_a=\frac{2 b_a}{b_1}\,, \quad a=2,3,4 \:, \label{RchargesS7}\nn \\[4pt]
	M_1 &= N (p_1 - p_2 - p_3 - p_4 )\:, \qquad\;\; M_a=N\,p_a\,, \quad a=2,3,4 \:.
\end{align}
One should then set $b_1 = 1$ and extremize $\mathcal{S}$ in order to obtain the on-shell quantities. Note that we have
\begin{equation}
	\sum_{a=1}^{4} R_a = 2 \:, \qquad \qquad \sum_{a=1}^{4} M_a = (2g-2) N \:.
\end{equation}
It is also convenient to give an expression for the R-symmetry vector in terms of the original basis $\del_{\phi_a}$. One may check that
\begin{equation}\label{xiS7}
	\xi = \frac{b_1}{2} \sum_{a=1}^{4} R_a \, \del_{\phi_a} \:.
\end{equation}
Using this expression to compute $\xi \lrcorner\, \mathcal{J}$ with both \eqref{kahlerformC4} and \eqref{kahler_relation} we find that the radius in the coordinates $\{y_a\}$ is
\begin{equation}\label{radiusC4}
	r^2 = \sum_{a=1}^{4} R_a\,y_a \:.
\end{equation}

Later we will be interested in $\Z_k$ quotients of the 7-sphere.
This will be reviewed for the two main models of interest in 
 section \ref{sec:1stmatching}. However, at the level of the various 
quantities obtained via integration introduced above, the quotient simply amounts to
adding a factor of  $k \in \Z$ in the denominator of the volumes \eqref{volS7} and \eqref{volsubS7}. Moreover, for later purposes it is useful to invert the relations \eqref{RchargesS7} and express $F_{\text{grav}}$ and $\mathcal{S}$ in terms of the geometric R-charges, so that eventually we have
\begin{align}\label{FgravS7}
	F_\mathrm{grav}&=\frac{4\pi}{3}\sqrt{2 k R_1 R_2 R_3 R_4}\, N^{3/2}\:, \\[4pt]
	\label{entropyS7}
	\mathcal{S}&=\frac{\pi}{3}\sqrt{2 k R_1 R_2 R_3 R_4}\,\sum_{a=1}^4 \frac{(M_a/N)}{R_a} N^{3/2}\:,
\end{align}
where we have already set $b_1= 4$ in the former and $b_1=1$ in the latter. Finally, let us also give an expression for the K\adots hler class parameter \eqref{Lambda} in the flavour twist as we will need it many times later
\begin{equation}\label{LambdaC4}
	\Lambda = \frac{\pi \ell_p^3\, N^{1/2}}{2} \, \sqrt{2k R_1 R_2 R_3 R_4} \:.
\end{equation}

\section{Matrix model geometry}\label{sec:matching}

The topologically twisted matrix models described in section~\ref{sec:fieldth} are dual to the black hole near-horizon geometries presented in section~\ref{sec:gravity}. In this section we make a precise identification of various quantities on both sides of this duality. We start in section \ref{sec:1stmatching} by reviewing the matching between chemical potentials and R-charges, and the free energy with the entropy. In section \ref{sec:mm} then we introduce the M-theory Hamiltonian function, which allows us to give a further geometrical interpretation of the matrix model, or more particularly its eigenvalue density discussed in section \ref{sec:density}. The relation to probe M2-branes wrapping the M-theory circle, 
and their Wilson loop duals, is described in section \ref{sec:M2match}. 
Along the way we describe the general matching for any Calabi-Yau four-fold cone $X_4$, but illustrate with our two main examples where $X_4=\C^4/\Z_k$; further examples are discussed in section~\ref{sec:examples}.

\subsection{Holography of free energies and R-charges}\label{sec:1stmatching}

As already discussed in section \ref{sec:Mthcircle}, given any M-theory 
solution we may in principle make different choices of M-theory circle 
$U(1)\equiv U(1)_M$ acting on the spacetime, and reduce 
in different ways to type IIA. Starting with a given AdS$_2\times Y_9$
background, with $Y_9$ a fibration of $Y_7$ over $\Sigma_g$ as introduced in the previous section, different choices of M-theory circle acting on $Y_7$ lead 
to different  Chern-Simons-matter theories on $\Sigma_g \times S^1$. 
Our principal examples of this are Models $\I$ and $\II$, which 
are dual to M-theory geometries with  $Y_7=S^7/\Z_k$ and $U(1)_M$ actions given in \eqref{U1MI} and \eqref{U1MII} respectively, where the $\Z_k$ quotient is taken along the corresponding M-theory circle. 

Having fixed a particular GK geometry and choice of M-theory circle, and hence 
a choice of UV field theory on $\Sigma_g \times S^1$, we would like 
to identify variables on each side of the duality. 
From a geometric engineering point of view,  baryonic operators arise from M5-branes wrapping five-dimensional submanifolds $S_A\subset Y_7$. The R-charge of the corresponding field is then given in term of the Sasakian volume of $S_A$ in the same way we defined the geometrical R-charges in \eqref{Rcharges2};\footnote{The distinction between $S_A$ and $S_a$ is purely motivated by the fact that in some examples the choice we make in gravity for the latter submanifolds does not directly correspond to the $S_A$ associated to physical operators but rather to linear combinations thereof. Correspondingly, the same will be true for any associated quantity \textit{e.g.}\ $R_A$ vs $R_a$.} this 
was first studied in the D3-brane holographic context in \cite{Berenstein:2002ke}.
 Moreover, recalling the identification between chemical potentials and R-charges, we get the following chain of identifications
\begin{align}\label{Deltaxi}
	\Delta_A = \Delta_A^{S^3} 
	=R_A\, .
\end{align}
Here $A$ labels any field of the theory, including monopole operators whose chemical potentials are only formally defined.  
Similarly, the field theory fluxes $\mathfrak{n}_A$ through $\Sigma_g$ are related in the same way to the geometrical fluxes:
\begin{align}\label{nPhi}
\mathfrak{n}_A = \frac{M(S_A)}{N}\, ,
\end{align}
where $M(S_A)$ is the quantized flux through $S_A \hookrightarrow \Sigma_A \rightarrow \Sigma_g$, which is computed via \eqref{fluxes}.

Let us illustrate this for our two models:

\begin{quote}
	\bMI: 
	The identifications \eqref{zimodelI} give the relation between the fields in this model and the divisors on $\C^4/\Z_k$. In this case, the submanifolds $S_A$ directly correspond to (multiples of) $S_a$ (recall the latter are defined by $\{z_a = 0\}$).
	Substituting this into \eqref{Deltaxi} gives the identifications
	\begin{align}\label{dicoADHM}
		\Delta_{T} = kR_1\, , \quad \Delta_{\tilde{T}} = kR_2\, , \quad 
		\Delta_{\Phi_1} = R_3\, , \quad \Delta_{\Phi_2} = R_4\, ,
	\end{align}
	\begin{align}\label{dicoADHMflux}
	\mathfrak{n}_{T} = k\,\frac{M_1}{N}\, , \quad \mathfrak{n}_{\tilde{T}} = k\,\frac{M_2}{N}\, , \quad 
	\mathfrak{n}_{\Phi_1} = \frac{M_3}{N}\, , \quad \mathfrak{n}_{\Phi_2} = \frac{M_4}{N}\, . 
	\end{align}
\end{quote}

\begin{quote}
	\bMII:  We may again use \eqref{Deltaxi} together with 
	the identifications \eqref{zimodelII} to deduce the mapping
	\begin{align}\label{dicoABJM}
		\Delta_{A_1} = R_1\, , \quad 
		\Delta_{B_1} = R_2\, , \quad 
		\Delta_{A_2} = R_3\, , \quad 
		\Delta_{B_2} = R_4\, ,
	\end{align}
\begin{align}\label{dicoABJMflux}
	\mathfrak{n}_{A_1} = \frac{M_1}{N}\, , \quad \mathfrak{n}_{B_1} = \frac{M_2}{N}\, , \quad 
	\mathfrak{n}_{A_2} = \frac{M_3}{N}\, , \quad \mathfrak{n}_{B_2} = \frac{M_4}{N}\, . 
\end{align}
\end{quote}

Having made these identifications allows one to then match other physical quantities
on both sides of the correspondence. In particular minus the free energy of the field theory on  $\Sigma_g\times S^1$
reproduces the entropy $\mathcal{S}$:
\begin{equation}\label{FSmatch}
	\log Z_{\Sigma_g\times S^1}[\Delta,\mathfrak{n}]= \mathcal{S}[R,M]\, .
\end{equation}
The matching \eqref{FSmatch} has been verified in multiple examples \cite{Benini:2015eyy,Hosseini:2019ddy,Gauntlett:2019roi}, and 
is particularly 
straightforward to see in our main examples. For Model I, using the dictionary \eqref{dicoADHM}--\eqref{dicoADHMflux}, $\log Z$ in \eqref{FtwistedADHM} matches  the entropy \eqref{entropyS7} on $S^7$.
Likewise for Model II, the dictionary \eqref{dicoABJM}--\eqref{dicoABJMflux} relates  \eqref{FtwistedABJM} with  \eqref{entropyS7}. 
In a similar fashion, a well-known matching is that of the free energy on $S^3$ with $F_\mathrm{grav}$, which can be observed explicitly for both models as \eqref{FS3ABJM} and \eqref{FS3ADHM} match with \eqref{FgravS7}. Interestingly $F_{S^3}$ and $F_\mathrm{grav}$ are proportional to $\mu$ and $\Lambda$ respectively, such that a direct consequence of their matching is  
\begin{equation}\label{Lambdamu}
	\Lambda = \frac{\ell_p^3  N^{1/2}}{2}\mu\,,
\end{equation}
which is indeed what we observe by comparing \eqref{LambdaC4} with \eqref{muI} and \eqref{muII}.

\subsection{M-theory circle moment map}\label{sec:mm}

Our aim in this section is to provide a geometric interpretation of 
the matrix model described in section \ref{sec:largeN}, or more 
precisely the large $N$ saddle point distribution. To this end, we would first like 
to identify the variable $t$ from a geometric perspective in M-theory. 

Recall that, in the  large $N$ limit, the variable $t$ corresponds to the continuum limit of the imaginary parts of the eigenvalues in \eqref{uansatz}. On the other hand, 
before taking the large $N$ limit the eigenvalues $t^\alpha$, $\alpha = 1, \dots, N$, are simply defined as the values of the scalars $\sigma^\alpha$ (up to an overall proportionality constant).  Physically the 
VEVs of the scalars 
$\sigma^\alpha$ also parametrize the position of the $N$ D2-branes in the $\R_t$ direction, 
as described in section~\ref{sec:VMS}. Via this chain of identifications, 
it is natural to conclude that the matrix model variable $t$, for the 
large $N$ eigenvalue distribution, should be related to the coordinate parametrizing $\R_t$ in the type IIA construction. The aim of the rest of this section will be to make this statement  precise.

Let us start by considering the uplifted geometry in M-theory without the branes' backreaction. Recall from section \ref{sec:Mthcircle} that the Killing vector $\zeta_M$ by definition generates the $U(1)_M$ isometry of the M-theory circle action. This $U(1)_M$ action is Hamiltonian on $X_4$, namely there exists a function $\mu_M \colon X_4 \to \R$, called the moment map,
such that
\begin{equation}\label{mmap}
	\diff \mu_M = -\, \zeta_M \lrcorner\, \mathcal{J} \:,
\end{equation}
where $\mathcal{J}$ is the original K\adots hler form on $X_4$. 
The crucial point is that the image of $\mu_M$ is precisely identified with the $\R_t$ direction. In fact, from a physical perspective, the D-term equations that fix the VEVs of $\sigma^\alpha$ ({\it i.e.}\ the position of the M2-branes along $\R_t$), 
 correspond geometrically to fixing level sets of the moment map \cite{Martelli:2008si}!

Let us illustrate this in the two examples with $X_4 = \C^4 / \Z_k$. We will focus on the Abelian case ($N=1$), where note that the K\adots hler form is given in \eqref{kahlerformC4}.

\begin{quote}
	\bMI: From \eqref{U1MI} we recall that the M-theory circle is given by $\zeta_M^\I =\frac{1}{k}\,( \partial_{\phi_1} - \partial_{\phi_2})$. Hence, using the definition \eqref{mmap}, the moment map in this case is simply
	\begin{equation}\label{mmapI}
		\mu_M^\I(z_a) = \frac{1}{2k}\left(|z_1|^2 - |z_2|^2\right).
	\end{equation}
	On the other hand the D-term equation for $k=1$ reads \cite{Benini:2009qs} 
	\begin{equation}
		|T|^2 - |\tilde{T}|^2 = \frac{\sigma}{2\pi} \:,
	\end{equation}
	and hence from the identifications in \eqref{zimodelI} we can see that the D-term equation for $k=1$ can be expressed as\footnote{
Recall that $\C^2/\Z_k\subset \C^4/\Z_k=X_4$ is realized via the embedded equation 
$T\tilde{T}=\Phi_3^k$ in $\C^3$ \eqref{TtildeT}. This means that for $k>1$ 
we cannot identify $T$, $\tilde{T}$ with the coordinates $z_1$, $z_2$ 
on the covering space $\C^2$ of $\C^2/\Z_k$, and correspondingly the classical 
metric on the field theory moduli space is different from the 
spacetime metric. However, these comments are only meant
to motivate the precise identifications later.}
	\begin{equation}
		\mu_M^\I(z_a) = \frac{\sigma}{\pi} \:.
	\end{equation}
\end{quote}

\begin{quote}
	\bMII: Here, the M-theory circle is in \eqref{U1MII} and reads $\zeta_M^{\II} = \frac{1}{k}\,(\partial_{\phi_1} - \partial_{\phi_2} + \partial_{\phi_3} - \partial_{\phi_4})$. From \eqref{mmap} we then read the moment map
	\begin{equation}\label{mmapII}
		\mu_M^{\II} (z_a)= \frac{1}{2k}\left(|z_1|^2 - |z_2|^2 +|z_3|^2 - |z_4|^2 \right).
	\end{equation}
	The D-term equation for this theory is \cite{Aharony:2008ug} 
	\begin{equation}
		\frac{1}{k} \left(|A_1|^2 - |B_1|^2 + |A_2|^2 -|B_2|^2\right) = \frac{\sigma}{2\pi} \:.
	\end{equation}
	From the identifications \eqref{zimodelII} it follows that the D-term equation is again
	\begin{equation}
		\mu_M^{\II}(z_a) = \frac{\sigma}{\pi} \:.
	\end{equation}
\end{quote}

Our claim is that this picture is ``preserved", in a sense that we will specify momentarily, when taking the large $N$ limit in field theory and the near-horizon limit in gravity.

As explained in section \ref{sec:largeN}, in field theory the eigenvalues $u_I^\alpha$ in the large $N$ limit become infinite in number, and continuously distributed, with their imaginary part parametrized by the real variable $t$ which takes values in a finite interval $[t_\text{min}, t_\text{max}]$. Moreover, this interval is divided into a certain number of subintervals, on each of which the eigenvalue density $\rho(t)$ is linear; 
in particular $\rho'(t)$ is discontinuous at the endpoints of each subinterval. 
The gravity duals are described by 
the AdS$_2 \times Y_9$ backgrounds in M-theory described in section \ref{sec:GK}. In this context, the $U(1)_M$ action on the fibre $Y_7$ is still Hamiltonian, and hence we can define the moment map $h_M$ (now regarded as a function on $Y_7$) via
\begin{equation}
	\diff h_M = -\, \zeta_M \lrcorner\, \omega \: .
\end{equation}
Recall here that $\omega$ is the transverse K\adots hler form on $Y_7$, and in the flavour twist case this is given by \eqref{flavour_twist}. From \eqref{kahler_relation} and the fact that $\diff \tildeeta =\tilderho$ it follows that, for the flavour twist case
\begin{equation}\label{hM}
	h_M = 2\Lambda\, \mu_M \,\big|_{r=1} = \Lambda\, \zeta_M \lrcorner\, \tildeeta \:,
\end{equation}
where $\mu_M$ is the moment map on the cone $X_4$ introduced above in \eqref{mmap}. 
 Since $h_M$ is a continuous real function defined on a compact space, its image is an interval
\begin{equation}\label{himage}
	h_M (Y_7) = [\tau_\text{min},\tau_\text{max}] \subset \R \:.
\end{equation}
The function $h_M$ is defined on $Y_7$, rather than directly on 
a  K\"ahler manifold (which is instead here a transverse K\"ahler structure), but 
nevertheless it is straightforward to import some standard results from the 
symplectic geometry literature (see {\it e.g.}\ \cite{KIRWAN1998135}). 
Similar reasoning was used in \cite{Farquet:2013cwa}. 
First, the interval in \eqref{himage} is divided into 
subintervals whose boundaries are given by the images of the critical points 
of the moment map, so $\{\diff h_M=0\}$. In turn these are precisely the loci 
where $\zeta_M$ is aligned with the Reeb vector field $\xi$, so that 
along such a locus $\zeta_M \lrcorner\, \tildeeta$ is a constant. 
This includes the special case where $\zeta_M$ is zero, which 
are fixed point sets signalling the presence of D6-branes. We shall return to what distinguishes these subintervals geometrically in the next subsection, while the relation of these points with physical observables will be discussed in section~\ref{sec:M2match}.

We conclude that there must be a holographic map between the field theory variable $t$ and the geometry variable $\tau$ parametrizing the image of the moment map. In particular, from the examples we are going to consider shortly and later in section \ref{sec:examples}, we find the 
precise identification~is
\begin{equation}\label{tau_to_t}
	\tau =  \ell_p^3  N^{1/2} \, t \:.
\end{equation}

Let us once again illustrate how this works concretely for the examples with $X_4 = \C^4/\Z_k$ {\it i.e.}\ $Y_7 = S^7/\Z_k$. Recall the radial coordinate on the cone is given by \eqref{radiusC4}
\begin{equation}\label{RmmapS7}
	r^2 = \sum_{a=1}^4 R_a\,y_a \:,
\end{equation} 
with $y_a \geq 0$.
Note that when $R_a=\frac{1}{2}$ for all $a=1, 2,3,4$ we recover the usual radius for a round 7-sphere. Setting $r^2 = 1$ constrains the possible values of $y_a$, in particular 
\begin{equation}
	0 \leq	y_a \leq \frac{1}{R_a} \:.
\end{equation} 

\begin{quote}
	\bMI: The moment map on $S^7/\Z_k$ is given by
	\begin{equation}\label{mmapI_2}
		h_M^\text{I} = \frac{2\Lambda}{k} \,(y_1 - y_2)\,\big|_{r^2=1} \:,
	\end{equation}
	whose image is clearly
	\begin{equation}
		h_M^\text{I} (S^7/Z_k) = \left[-\frac{2\Lambda}{kR_2},\,\frac{2\Lambda}{k R_1}\right] \:.
	\end{equation}
	Using \eqref{LambdaC4} for the K\adots hler class parameter, this precisely maps to the interval $[t_{\text{min}},t_\text{max}]$ with extrema \eqref{intervalADHM} via \eqref{tau_to_t} and the identification among the R-charges \eqref{dicoADHM}.
	Moreover, the only other fixed point locus of the $U(1)_M$ action on $S^7/\Z_k$ is $\{z_1 = z_2 = 0\}$ which reproduces the point $t=0$, arising from the D6-brane insertion.
\end{quote}

\begin{quote}
	\bMII: Here we solve the constraint $r^2 = 1$ for one of the coordinates and substitute it back into the moment map. In order to obtain $\tau_\text{min}$ it is convenient to solve for $y_4$, so that
	\begin{align}
\hspace{-0.4cm}		h_M^\text{II} (y_a) & = \frac{2\Lambda}{k}\bigg[\left(1+ \frac{R_1}{R_4}\right)y_1 - \left(1- \frac{R_2}{R_4}\right)y_2  + \left(1+ \frac{R_3}{R_4}\right)y_3\ - \frac{1}{R_4}\bigg] .
	\end{align}
	In field theory we assumed a particular ordering for the chemical potentials which must be mirrored here: $R_1 \leq R_3$ and $R_2 \leq R_4$. As a consequence the coefficient of $y_2$ in the expression above is negative, and it is then clear that the minimum of $h_M^\text{II} (y_a)$ is achieved when
	\begin{equation}
		y_1 = y_3 = 0 \:, \quad y_2 = \frac{1}{R_2} \quad \implies \quad  \tau_1 \equiv\tau_\text{min}  = -\frac{2\Lambda}{kR_2} \:.
	\end{equation}
	Similarly, solving for $y_3$ it is easy to obtain the maximum
	\begin{equation}
		\tau_4 \equiv \tau_\text{max} = \frac{2\Lambda}{kR_1} \:.
	\end{equation}
	Notice that, as explained in general above, $\tau_\text{min}$ and $\tau_\text{max}$ are images of loci where  $\zeta_M$ is aligned with the R-symmetry vector $\xi$, 
specifically here where  $\{z_1 = z_3 = z_4 = 0\}$ and $\{z_2 = z_3 = z_4 =0\}$. Indeed, there are other two such loci/image points:
	\begin{align}
		\{z_1 = z_2 = z_3 =0\}& \quad \implies \quad \tau_2 = - \frac{2\Lambda}{kR_4} \:,\\[4pt]
		\{z_1 = z_2 = z_4 =0\}& \quad \implies \quad \tau_3 =  \frac{2\Lambda}{kR_3} \:.
	\end{align}
	By virtue of \eqref{Lambdamu}, \eqref{dicoABJM}, and \eqref{tau_to_t}, $\tau_1$, $\tau_2$, $\tau_3$, $\tau_4$ are precisely matched with \eqref{intervalABJM}.
\end{quote}

To conclude this section, let us notice that here neither the 3d space where the field theory lives, nor the fact that $Y_7$ is fibered over $\Sigma_g$, played a crucial role. Therefore, everything we have said holds, with minor adjustments, for both 3d theories localized on $S^3$ and on $\Sigma_g \times S^1$.

\subsection{Eigenvalue density from geometry}\label{sec:density}
In the previous section we identified the variable $t$ in field theory as the value of the moment map associated to the M-theory circle action.
We can actually do better and provide a prescription to compute the full eigenvalue density function of the matrix model $\rho(t)$ from the geometry of the near-horizon dual. Despite focusing on the AdS$_2 \times Y_9$ geometries, applying our approach to AdS$_4 \times Y_7$ is rather straightforward. Note that for a certain set of the latter geometries with toric $Y_7$ some results concerning this very same matter were already found in the past -- see appendix C of \cite{Gulotta:2011aa} and section 3.6 of \cite{Farquet:2013cwa}. However, here we will give a much more complete picture which works off-shell and in principle also for non-toric $Y_7$, and more specifically also for the near-horizon black hole 
GK geometries.

We start by using the moment map $h_M$ defined in \eqref{hM} to build a 5d space out of  the fibre $Y_7$, by restricting to the level set $h_M^{-1}(\tau) \subset Y_7$ and then quotienting by the M-theory circle action
\begin{equation}\label{Ytau}
	Y_\tau \equiv h_M^{-1}(\tau)/U(1)_M \:.
\end{equation}
In the special case when $Y_7$ is quasi-regular, namely that $b_i \in \mathbb{Q}$ for all $i=1,\dots, s$, so that the orbits of the R-symmetry vector close, $Y_\tau$ has a particular structure which derives from the following mathematical construction. We first introduce the 6d  K\adots hler base space $K_6$ obtained by quotienting
\begin{equation}
	K_6 \equiv Y_7 / U(1)_\xi \:.
\end{equation}
Note that this quotient would be very badly behaved  if we did not assume $b_i \in \mathbb{Q}$, because the generic orbits of $\xi$ would then be open. In passing, let us highlight that the critical points of $h_M$ which delimit the subintervals in the image $h_M(Y_7)$ correspond precisely to the fixed points of the $U(1)_M$ action on $K_6$; indeed these are either fixed points on $Y_7$ itself, which are related to D6-brane insertions, or points where $\zeta_M$ is aligned with $\xi$ in the ambient space $Y_7$, as in such a situation the former projects to zero in the quotient above.

Given that the moment map $h_M$ is independent of $\xi$, it can be thought as a function on $K_6$. We can then consider introducing a 4d subspace $K_\tau \subset K_6$ as a symplectic quotient with Fayet-Iliopoulos parameter $\tau$ 
\begin{equation}
	K_\tau \equiv K_6 /\!\!/ U(1)_M \:,
\end{equation}
and by construction $K_\tau$ is itself a K\adots hler space with a K\adots hler form $\omega_\tau$ that is inherited from the original $\omega$ on $K_6$. Finally, we restore the $U(1)_\xi$ direction by building the fibration
\begin{equation}
	S^1_\xi \hookrightarrow Y_\tau \rightarrow K_\tau \:.
\end{equation}
The induced K\adots hler class $[\omega_\tau]$ depends linearly on $\tau$, with a different slope in each of the subintervals in the image of $h_M$ delimited by the $U(1)_M$ fixed points \cite{KIRWAN1998135}; therefore it is quite natural to relate it with $\rho(t)$. Our precise claim is that 
\begin{equation}\label{densitygeo}
	\rho(t) \propto \frac{\diff}{\diff\Lambda}\int_{Y_\tau} \eta \wedge \frac{\omega_\tau^2}{2} \,\bigg|_{\tau = \ell_p^3 N^{1/2} \, t} = \frac{\diff}{\diff\Lambda} \,\Vol(Y_\tau)\,\bigg|_{\tau = \ell_p^3  N^{1/2}\, t}\:.
\end{equation}
The normalization constant is then fixed by requiring $\rho(t)$ to integrate to 1.
Finally, in order for the construction above to make sense even when $Y_7$ is irregular ($b_i \notin \mathbb{Q}$), we note that $\Vol(Y_\tau)$ depends continuously on the R-symmetry vector $b_i$, hence its expression can be extended to the general case by continuity.

Next, we will illustrate the above in the two examples with $Y_7 = S^7/\Z_k$, but first let us comment on how to go about computing the volume $\Vol(Y_\tau)$. For toric $Y_7$ it is quite straightforward and we will provide a general recipe later in section \ref{sec:examples}. In general, without further assumptions on the geometry of $Y_7$, it may be difficult to 
produce closed form expressions. We envisage that an approach based on a fixed point formula, such as that in \cite{Wu:1992ty}, could however be effective.

For $Y_7 = S^7/\Z_k$ we use the coordinates $\{y_a, \phi_a\}$ introduced in section \ref{sec:S7example}. Modulo angles, the space $Y_\tau$ is described by imposing both $r^2 = 1$ and $h_M(y_a) = \tau$ which result in the two linear equations describing hyperplanes in $\R^4$. Recalling that $r^2$ is given by \eqref{RmmapS7} and that $y_a \geq 0$, the former equation actually defines a tetrahedron with vertices
\begin{equation}
	p_a = \frac{1}{R_a} \, e_a \:,
\end{equation}
where $e_a$ is the standard orthonormal basis on $\R^4$. Imposing the moment map level set equation then amounts to slicing this tetrahedron with a hyperplane. The intersection of the two will be a 2d polygon $Q_\tau$, particularly either a triangle or a quadrilateral. For the flavour twist, by definition the following cohomological relation holds
\begin{equation}
	[\omega] = \Lambda [\diff \tildeeta] = 2\Lambda [\omega_S] \:.
\end{equation}
From \eqref{kahlerformC4} one can see that the measure induced via $\omega_S$ 
is simply the standard Euclidean measure on  $\R^4$ (times a factor $k^{-1}$ which comes from the $\Z_k$ quotient), and correspondingly $\omega$ induces $(2\Lambda)^4$ times the Euclidean volume form. It follows that 
\begin{equation}\label{volYtau}
	\Vol(Y_\tau) = \frac{(2\pi)^3}{k}\,(2\Lambda)^2\, \Vol_E(Q_\tau) \, ,
\end{equation}
where $\Vol_E$ denotes the Euclidean volume, and the factor $(2\pi)^3$ comes from integrating over the angles. 

Let us also introduce a notation for the vectors orthogonal to the two hyperplanes, which are nothing but $\xi$ and $\zeta_M$ in coordinates. The former is 
\begin{equation}
	\xivec = \frac{1}{2}\,(R_1, R_2, R_3, R_4) \:,
\end{equation}
while the latter will depend upon the model we are considering and we will denote it by $\zetavec$. It will also be useful to introduce a name for the component of $\zetavec$ orthogonal to $\xivec$, namely
\begin{equation}\label{zetaperp}
	\zetavec_\perp \equiv \zetavec - \frac{\zetavec \cdot \xivec}{|\xivec|^2} \, \xivec \:.
\end{equation}
In both \MI \ and \MII, it turns out the properly normalized $\rho(t)$ is given in gravity by
\begin{equation}\label{normalized_rho}
	\rho(t) = \frac{1}{32  \pi^5 \ell_p^3  N^{1/2}\,|\xivec||\zetavec_\perp|}\,\frac{\diff}{\diff \Lambda}\, \Vol(Y_\tau)\,\bigg|_{\tau = \ell_p^3  N^{1/2}\, t} \:.
\end{equation}

\begin{quote}
	
	\begin{figure}
		\centering
		\begin{tikzpicture}
			\filldraw[color=red, fill=red!10, very thick] (2,.2)--(2.25,2.2)-- (2.75,-1)--(2,.2) node[midway, left] {$Q_\tau$};	
			\filldraw[color=blue, fill=blue!10, very thick] (4,.4)--(4.5,4.45)--(5.5,-2)--(4,.4) node[near end, left] {$Q_\tau$};	
			\filldraw[color=Green, fill=Green!10, very thick] (6,.6)--(6.25,2.6)-- (6.75,-.6)--(6,.6) node[midway, left] {$Q_\tau$};	
			
			\draw[thick,dashed] (0,0) -- (8,.8);
			\draw[thick] (4.5,4.5)--(5.5,-2);
			\draw[thick] (0,0)--(4.5,4.5)--(8,.8)--(5.5,-2)--(0,0);
			\filldraw[black] (0,0) circle (2pt) node[anchor=east]{$p_2$};
			\filldraw[black] (8,.8) circle (2pt) node[anchor=west]{$p_1$};
			\filldraw[black] (4.5,4.5) circle (2pt) node[anchor=south east]{$p_3$};
			\filldraw[black] (5.5,-2) circle (2pt) node[anchor=north east]{$p_4$};	
			
			\filldraw[red] (2,.2) circle (2pt) node[anchor=south east]{$q_1$};
			\filldraw[red] (2.25,2.25) circle (2pt) node[anchor=east]{$q_3$};		
			\filldraw[red] (2.75,-1) circle (2pt) node[anchor=north]{$q_2$};		
			\filldraw[blue] (4,.4) circle (2pt) node[anchor=south east]{$q_1$};
			\filldraw[blue] (4.5,4.5) circle (2pt) node[anchor=south west]{$q_3$};
			\filldraw[blue] (5.5,-2) circle (2pt) node[anchor=north west]{$q_2$};
			\filldraw[Green] (6,.6) circle (2pt) node[anchor=south east]{$q_1$};
			\filldraw[Green] (6.25,2.65) circle (2pt) node[anchor=west]{$q_3$};	
			\filldraw[Green] (6.75,-.6) circle (2pt) node[anchor= west ]{$q_2$};	
		\end{tikzpicture}
		\caption{Tetrahedron describing $S^7$ and its slicing corresponding to \MI \ -- ADHM theory. The red slice is for $\tau_{\text{min}} < \tau <0$, blue slice for $\tau = 0$, and green slice for $0 <\tau < \tau_{\text{max}} $. The perimeter of $Q_\tau$ varies with $\tau$ qualitatively as $\rho$ in Figure~\ref{fig:rhoADHM}.}
		\label{fig:rhoADHMgeom} 
	\end{figure}
	
	\bMI: The hyperplane described by the level set equation of the M-theory circle moment map in \eqref{mmapI_2} satisfies the equation
	\begin{equation}\label{hyperplaneI}
		y_1 - y_2 = \frac{k}{2\Lambda}\, \tau \:,
	\end{equation}
	with $-\frac{2\Lambda}{k R_2} \leq \tau \leq \frac{2\Lambda}{k R_1}$, and the vector $\zetavec$ is
	\begin{equation}
		\zetavec = \frac{1}{k}\,(1,-1,0,0) \:.
	\end{equation}
	Note that for $\tau=0$, both vertices $p_3$ and $p_4$ of the tetrahedron belong to $Q_{\tau=0}$, therefore the hyperplane \eqref{hyperplaneI} is parallel to one of the edges of the tetrahedron. So $Q_\tau$ is always a triangle but we have to compute $\Vol_E(Q_\tau)$ separately for $\tau <0$ and $\tau >0$, which corresponds to the two subintervals of the image of $h_M^\text{I} $ that we identified in the previous section (see Figure~\ref{fig:rhoADHMgeom}). Let us start with $\tau < 0$. By standard linear algebra, one can find that the vertices of $Q_\tau$ are
	\begin{align}
		q_1 &= \left(\frac{2\Lambda + k R_2\tau}{2\Lambda(R_1 + R_2)},\frac{2\Lambda - k R_1\tau}{2\Lambda(R_1 + R_2)},0,0\right), \nonumber\\
		q_2 &= \left(0,-\frac{k\tau}{2\Lambda},\frac{2\Lambda + k R_2\tau}{2\Lambda R_3},0\right), \nonumber\\
		q_3 &= \left(0,-\frac{k\tau}{2\Lambda},0,\frac{2\Lambda + k R_2\tau}{2\Lambda R_4}\right)\,. 
	\end{align}
	Given that $\xivec$ and $\zetavec_\perp$ are both orthogonal to the plane where $Q_\tau$ lies, the volume of $Q_\tau$ can be related to the volume of the 4d parallelotope whose edges are the aforementioned vectors plus two of the edges of $Q_\tau$. Concretely
	\begin{align}
\label{volQsmall}
			\Vol_E(Q_{\tau<0}) &= \frac{1}{2}\,\left|\det\left(\frac{\xivec}{|\xivec|},\frac{\zetavec_\perp}{|\zetavec_\perp|}, q_2 - q_1, q_3 - q_1\right)\right| \nonumber\\
			&= \frac{(2\Lambda+k R_2 \tau)^2}{8\Lambda^2 R_3 R_4 (R_1+ R_2)} \sqrt{(R_1+R_2)^2+2(R_3^2+R_4^2)} \:.
	\end{align}
	Similarly, for $\tau >0$ we get the vertices
	\begin{align}
		q_1 &= \left(\frac{2\Lambda + k R_2\tau}{2\Lambda(R_1 + R_2)},\frac{2\Lambda - k R_1\tau}{2\Lambda(R_1 + R_2)},0,0\right),\nonumber \\
		q_2 &= \left(\frac{k\tau}{2\Lambda},0,\frac{2\Lambda - k R_1\tau}{2\Lambda R_3},0\right),\nonumber \\
		q_3 &= \left(\frac{k\tau}{2\Lambda},0,0,\frac{2\Lambda - k R_1\tau}{2\Lambda R_4}\right)\,, 
	\end{align}
	and therefore the volume
	\begin{equation}\label{volQbig}
		\Vol_E(Q_{\tau>0}) = \frac{(2\Lambda-k R_1 \tau)^2}{8\Lambda^2 R_3 R_4 (R_1+ R_2)} \sqrt{(R_1+R_2)^2+2(R_3^2+R_4^2)} \:.
	\end{equation}
	From \eqref{volQsmall} and \eqref{volQbig} we can directly compute $\rho(t)$ as explained above. Fixing the normalization as in \eqref{normalized_rho}, we find
\begin{equation}
	\rho(t) = \frac{1}{ \ell_p^3  N^{1/2}}\left.\begin{cases}
		\dfrac{2\Lambda + k R_2 \tau}{\pi^2R_3 R_4(R_1+R_2)}\,,\quad -\dfrac{2\Lambda}{k R_2} < \tau <0\\[10pt]
		\dfrac{2\Lambda - k R_1 \tau}{\pi^2R_3 R_4(R_1+R_2)}\,, \quad\: 0 < \tau < \dfrac{2\Lambda}{k R_1}
	\end{cases} \right|_{\tau = \ell_p^3  N^{1/2} \, t},
\end{equation}
	which perfectly matches with the eigenvalue distribution \eqref{rhoADHM} of the ADHM theory.
\end{quote}

\begin{quote}
	
	\begin{figure}
		\centering
		\begin{tikzpicture}
			
			\filldraw[color=red, fill=red!10, very thick] (2,.2)--(2-2*0.2,.2+2*0.7)-- (2+.8*2.7,.2-.8*2.1)--(2,.2) node[near end, below] {$Q_\tau$};	
			\filldraw[color=blue, fill=blue!10, very thick] (4,.4)--(4-4*0.2,.4+4*0.7)--(4-4*0.2+.73*2.5,.4+4*0.7-.73*2.8)--(4+.8*2.7,.4-.8*2.1)--(4,.4) node[near end, below] {$Q_\tau$};	
			\filldraw[color=Green, fill=Green!10, very thick] (6,.6)--(6-4.75*0.2,.6+4.75*0.7)-- (6+.4*2.7,.6-.4*2.1)--(6,.6) node[near end, below] {$Q_\tau$};	
			
			\draw[thick,dashed] (0,0) -- (8,.8);
			\draw[thick] (4.5,4.5)--(5.5,-2);
			\draw[thick] (0,0)--(4.5,4.5)--(8,.8)--(5.5,-2)--(0,0);
			\filldraw[black] (0,0) circle (2pt) node[anchor=east]{$p_2$};
			\filldraw[black] (8,.8) circle (2pt) node[anchor=west]{$p_1$};
			\filldraw[black] (4.5,4.5) circle (2pt) node[anchor=south]{$p_3$};
			\filldraw[black] (5.5,-2) circle (2pt) node[anchor=north]{$p_4$};	
			
			\filldraw[red] (2,.2) circle (2pt) node[anchor=south east]{$q_1$};
			\filldraw[red] (2-2*0.2,.2+2*0.7) circle (2pt) node[anchor=east]{$q_3$};		
			\filldraw[red] (2+.8*2.7+.05,.2-.8*2.1-.05) circle (2pt) node[anchor=north]{$q_2$};		
			\filldraw[blue] (4,.4) circle (2pt) node[anchor=south east]{$q_1$};
			\filldraw[blue] (4-4*0.2,.4+4*0.7) circle (2pt) node[anchor=east]{$q_4$};		
			\filldraw[blue] (4-4*0.2+.73*2.5,.4+4*0.7-.73*2.8) circle (2pt) node[anchor=east]{$q_3$};
			\filldraw[blue] (4+.8*2.7,.4-.8*2.1) circle (2pt) node[anchor=west]{$q_2$};
			\filldraw[Green] (6,.6) circle (2pt) node[anchor=south east]{$q_1$};
			\filldraw[Green] (6-4.75*0.2,.6+4.75*0.7) circle (2pt) node[anchor=west]{$q_3$};	
			\filldraw[Green] (6+.4*2.7,.6-.4*2.1) circle (2pt) node[anchor= west ]{$q_2$};	
		\end{tikzpicture}
		\caption{Tetrahedron describing $S^7$ and its slicing corresponding to \MII \ -- ABJM theory.  The red slice is for $\tau_1 < \tau < \tau_2$, blue slice for $\tau_2 < \tau < \tau_3$, and green slice for $\tau_3 <\tau < \tau_4$. The perimeter of $Q_\tau$ varies with $\tau$ qualitatively as $\rho$ in Figure~\ref{fig:rhoABJM}.}
		\label{fig:rhoABJMgeom} 
	\end{figure}
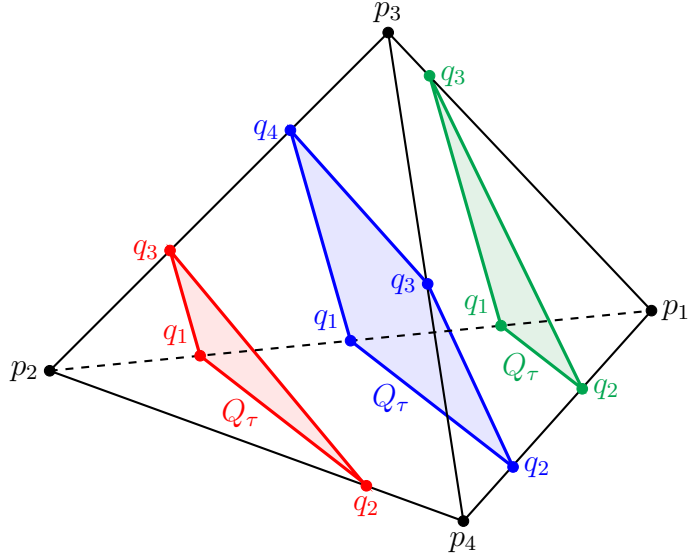
	\bMII: Here the hyperplane corresponding to the moment map level set is described by
	\begin{equation}\label{hyperplaneII}
		y_1 - y_2 + y_3 - y_4 = \frac{k}{2\Lambda} \, \tau \:,
	\end{equation}
	with $-\frac{2\Lambda}{k R_2} \leq \tau \leq \frac{2\Lambda}{k R_1}$ (recall we assumed $R_1 \leq R_3$ and $R_2 \leq R_4$), and the vector $\zetavec$ is
	\begin{equation}
		\zetavec = \frac{1}{k}\,(1,-1,1,-1) \:.
	\end{equation} 
	In this case there is no value of $\tau$ for which both $p_3$ and $p_4$ belong to the hyperplane \eqref{hyperplaneII}, hence the latter hyperplane is not parallel to any edge and the situation is as depicted in Figure~\ref{fig:rhoABJMgeom}. In the intervals $\tau_\text{min} = \tau_1 < \tau < \tau_2$ and $\tau_3 < \tau < \tau_4= \tau_\text{max}$ the polyhedron $Q_\tau$ is a triangle and the computation goes through similarly to \MI. Instead in the interval $\tau_2 < \tau < \tau_3$,  $Q_\tau$ is a quadrilateral whose vertices are
	\begin{align}
		q_1 &= \left(\frac{2\Lambda + k R_2\tau}{2\Lambda(R_1 + R_2)},\frac{2\Lambda - k R_1\tau}{2\Lambda(R_1 + R_2)},0,0\right), \nonumber\\
		q_2 &= \left(\frac{2\Lambda + k R_4\tau}{2\Lambda(R_1 + R_4)},0,0,\frac{2\Lambda - k R_1\tau}{2\Lambda(R_1 + R_4)}\right), \nonumber\\
		q_3 &= \left(0,0,\frac{2\Lambda + k R_4\tau}{2\Lambda(R_3 + R_4)},\frac{2\Lambda - k R_3\tau}{2\Lambda(R_3 + R_4)}\right),\nonumber \\
		q_4 &= \left(0,\frac{2\Lambda - k R_3\tau}{2\Lambda(R_2 + R_3)},\frac{2\Lambda + k R_2\tau}{2\Lambda(R_2 + R_3)},0\right)\,. 
	\end{align}
	The Euclidean volume of $Q_\tau$ can be computed as the sum of the volumes of two triangles, say those with vertices $\{q_1,q_2,q_3\}$ and $\{q_1,q_3,q_4\}$
	\begin{align}
			\Vol_E(Q_{\tau_2 <\tau<\tau_3}) &= \frac{1}{2}\,\left|\det\left(\frac{\xivec}{|\xivec|},\frac{\zetavec_\perp}{|\zetavec_\perp|}, q_2 - q_1, q_3 - q_1\right)\right|\nonumber \\
			&\quad+\frac{1}{2}\,\left|\det\left(\frac{\xivec}{|\xivec|},\frac{\zetavec_\perp}{|\zetavec_\perp|}, q_3 - q_1, q_4 - q_1\right)\right|\,.
		\end{align}
	We do not write the explicit expression as it is long and not particularly insightful. Eventually, fixing the normalization as in \eqref{normalized_rho}, we find
\begin{equation}
	\rho(t) = \frac{1}{ \ell_p^3  N^{1/2}}\left.\begin{cases}
			\frac{2\Lambda + k R_2 \tau}{\pi^2(R_1+R_2)(R_2+R_3)(R_4-R_2)}\,, \,\qquad\quad {\scriptstyle\tau_1 < \tau <\tau_2}\\[4pt] 
			\frac{4\Lambda + k (R_2R_4 - R_1 R_3) \tau}{\pi^2(R_1+R_2)(R_2+R_3)(R_1+R_4)(R_3+R_4)}\,, \: {\scriptstyle\tau_2 < \tau <\tau_3}\\[4pt]
			\frac{2\Lambda - k R_1 \tau}{\pi^2(R_1+R_2)(R_1+R_4)(R_3-R_1)}\,, \qquad\quad\, {\scriptstyle\tau_3 < \tau < \tau_4}
		\end{cases}\hspace{-.2cm} \right|_{\tau = \ell_p^3  N^{1/2}t },
\end{equation}
with 
\begin{equation}
	\tau_1=-\frac{2\Lambda}{kR_2}\,,\quad
	\tau_2=-\frac{2\Lambda}{kR_4}\,,\quad
	\tau_3=\frac{2\Lambda}{kR_3}\,,\quad
	\tau_4=\frac{2\Lambda}{kR_1}\,,\quad
\end{equation}
	which perfectly matches with the eigenvalue distribution \eqref{rhoABJM} of the ABJM theory. 
	
\end{quote}

\subsection{M2-branes and Wilson loops}\label{sec:M2match}

In this section we place probe M2-branes in the black hole backgrounds from section~\ref{sec:gravity}. In particular, we are interested in M2-branes wrapping AdS$_2$ and an additional circle $S^1_{\text{M2}} \subset Y_9$ in the internal manifold. When these branes are BPS and wrap a particular $S^1$ copy, their classical action is related to BPS Wilson loop expectation values in the dual matrix model.

\subsubsection{BPS probe M2-branes}\label{sec:M2}

Since we wish to ultimately match to the Euclidean localized field 
theory, we need 
to Wick rotate our background\footnote{In this instance $t^{(\text{L})}$ is the time coordinate rather than the direction $\R_t$ introduced at the beginning of section \ref{sec:typeIIA}.}
\begin{equation}\label{wick}
	t^{(\text{L})} = - \ii t^{(\text{E})} \:, \qquad \vol^{(\text{L})} = -\ii\, \vol^{(\text{E})} \:, \qquad  \ii I^{(\text{L})} = - I^{(\text{E})} \:,
\end{equation}
where the superscripts $(\text{L})$ and  $(\text{E})$ denote respectively quantities in Lorentzian and Euclidean signature and $I$ here stands for any action. Consequently, the backgrounds of interest become EAdS$_2 \times Y_9$. In the rest of the section we will drop the superscripts and work in Euclidean signature unless otherwise stated.

The classical action of a probe M2-brane with worldvolume $\Sigma_{\text{M2}}$ in Euclidean signature is \cite{Bergshoeff:1987cm}\footnote{We use the conventions of \cite{Gauntlett:2003di}, Wick-rotated as in \eqref{wick}.}
\begin{equation}\label{M2action}
	I_{\text{M2}}=\frac{1}{(2\pi)^2 \ell_p^3}\left[-\Vol\left(\Sigma_{\text{M2}}\right)-\mathrm{i} \int_{\Sigma_{\text{M2}}} C \right] .
\end{equation}
Notice that 
this formula is gauge-invariant under shifts of the three-form gauge field $C \to C + \diff\Omega$, with $\Omega$ a two-form, provided that $\Omega\,|_{\del \Sigma_{\text{M2}}} = 0 $. However, the formula~\eqref{M2action} requires 
careful interpretation, as we will discuss next.
By virtue of \eqref{G4_gk} and \eqref{F_gk}, we can express $C$ as
\begin{equation}\label{C_generalgauge}
	C = -\ii\,\vol_{\text{EAdS}_2} \wedge \mathcal{A} + \diff \Omega \:,
\end{equation}
where $\Omega$ parametrizes the gauge freedom and we have defined
\begin{equation}
	\mathcal{A} = -\psi + \ex^{-B} \eta \:,
\end{equation}
with $\diff \psi = J$. Notice here that $J$ is closed, so we may always \emph{locally} 
write $J=\diff\psi$, with $\psi$ a transverse one-form ({\it i.e.}\ $\xi\,\lrcorner\,\psi=0$), but in general  this will not be globally 
well-defined. Indeed, $[J]\in H^2_B(\mathcal{F}_\xi)$ 
always defines a non-zero basic cohomology class -- if it were 
a basic exact form then using Stokes' theorem the volume of $Y_9$ would necessarily be zero.\footnote{This is a modification of a standard argument in K\"ahler geometry, showing that the K\"ahler class of a compact manifold is always non-zero.} 
Explicitly, recalling \eqref{kahlerform} we can write
\begin{equation}
	\mathcal{A} = -(\tildesigma^\textit{t} + A\, \sigma_{\Sigma_g}) + \ex^{-B} \eta + \mbox{basic exact}\:,
\end{equation}
where both $\tildesigma$ and $\sigma_{\Sigma_g}$, such that $\diff\sigma_{\Sigma_g}=\vol_{\Sigma_g}$, are \textit{local} one-forms only. 
There is no possible choice of gauge fixing that turns the total $C$-field \eqref{C_generalgauge} into a globally defined form.
However, in the following we will be interested in a particular case where $S^1_{\text{M2}}$ is tangent to the fibres $Y_7$. Therefore, if we can turn $C$ into a global form when restricted to the fibres, that would yield a well-defined M2-brane action when substituted into \eqref{M2action}.
Assuming the flavour twist condition \eqref{flavour_twist}, we can do this by choosing
\begin{equation}
	\Omega =  -\ii \,\vol_{\text{EAdS}_2} \wedge \left(- \Lambda z\right)\: ,
\end{equation}
so that \eqref{C_generalgauge} reads (\textit{cf.}\ \eqref{totaleta})
\begin{equation}\label{Calmostglobal}
	C = -\ii\, \vol_{\text{EAdS}_2} \wedge (-\Lambda\eta - A\, \sigma_{\Sigma_g}+ \ex^{-B}\eta + \mbox{basic exact})\:.
\end{equation}
Restricting the above expression to a point on the Riemann surface, the term proportional to $\sigma_{\Sigma_g}$ drops out and everything left is proportional to $\eta$, which itself is a global one-form since the only class in $H^2_B(\mathcal{F}_\xi)$ that is zero when pulled back to a class in $H^2(Y_9)$ is that generated by $[\varrho]$. In fact as discussed in 
\cite{Couzens:2018wnk} for the class of GK geometries we are considering 
in fact $H^2(Y_9)\cong H^2_B(\mathcal{F}_\xi)/[\varrho]$, where 
$\varrho= \diff\eta$ immediately implies that $[\varrho]=0\in H^2(Y_9)$. 

Substituting \eqref{Calmostglobal} into \eqref{M2action} and recalling we are interested in the case where $\Sigma_\mathrm{M2} = \text{EAdS}_2 \times S^1_{\text{M2}}$, we get
\begin{equation}\label{M2action2}
	I_{\text{M2}} = \frac{1}{(2\pi)^2 \ell_p^3} \left[ (-2\pi) \int_{S^1_{\text{M2}}}\left( - \ex^{-2B/3}\,\vol_{S^1_{\text{M2}}}  +\Lambda \eta - \ex^{-B} \eta\right) \right],
\end{equation}
where the factor $(-2\pi)$ appearing in front of the first integral comes from the regularized volume\footnote{Specifically, one can use holographic 
	normalization to define this rigorously.} of EAdS$_2$ \cite{Hung:2011nu}
\begin{equation}
	\Vol(\text{EAdS}_2) = -2\pi \:.
\end{equation}

We are interested in computing the action \eqref{M2action2} for BPS M2-branes. In general, a probe M2-brane will break the supersymmetry of the background, in the sense that acting with a supercharge on the configuration fields of the M2-brane yields a non-vanishing variation. However, the M2-brane action possesses another fermionic symmetry, called $\kappa$-symmetry \cite{Bergshoeff:1987cm},\footnote{This is a symmetry of the Green-Schwarz brane action which includes also fermionic degrees of freedom. However, when evaluating the Green-Schwarz action on a classical bosonic background such a contribution vanishes and one is left with the action \eqref{M2action}.} which we can use to compensate for the supersymmetry variation. In Appendix \ref{app:bps} we discuss in some detail the condition to retain some supersymmetry with the M2-brane insertion, and we find that the circle $S^1_{\text{M2}}$ needs to be aligned with the R-symmetry circle, {\it i.e.}\
\begin{equation} \label{bps_cond}
	\vol_{S^1_\text{M2}} = -\ex^{-B/3}\,\eta\:.
\end{equation}
Then, in the action \eqref{M2action2} the first and third terms inside the first integral cancel to give the final result
\begin{equation}\label{eq.SM2}
	I_{\text{M2}}^{\text{BPS}}
	= - \frac{\Lambda}{2\pi\ell_p^3}\, \int_{S^1_{\text{M2}}}\eta  \:,
\end{equation}
where $\Lambda$ is given explicitly in \eqref{Lambda}, and depends on $\Vol_S(Y_7)^{-1/2}$ and $N^{1/2}$. Note that this integral is a quantity that will differ from model to model and from point to point and will have to be computed in each example. 
For ease of notation, we will drop the BPS superscript in the following, and by $I_{\text{M2}}$ we will always mean the action of a BPS M2-brane. 

Finally, let us make a further comment on 
the $C$-field \eqref{Calmostglobal}. This is valid for solutions in the 
flavour twist class, but outside this class it is not clear 
how to generalize the gauge-fixing we have carried out. More specifically, 
it is not clear how to assign well-defined M2-brane 
actions outside the flavour twist class. Notice that while the action of an M2-brane is invariant 
under gauge transformations of the $C$-field, the 
issue in this case is that we are working with a complex solution, with 
purely imaginary $C$-field, and the saddle point solution 
then in general manifestly depends on the (complex) gauge choice. On the other hand, 
the holographic matching of quantities in gravity 
and localized field theory/matrix models only seems to 
hold for the same flavour twist class, and it is possible that these observations are related.

\subsubsection{Wilson loop dual}\label{sec:Wilsonian}

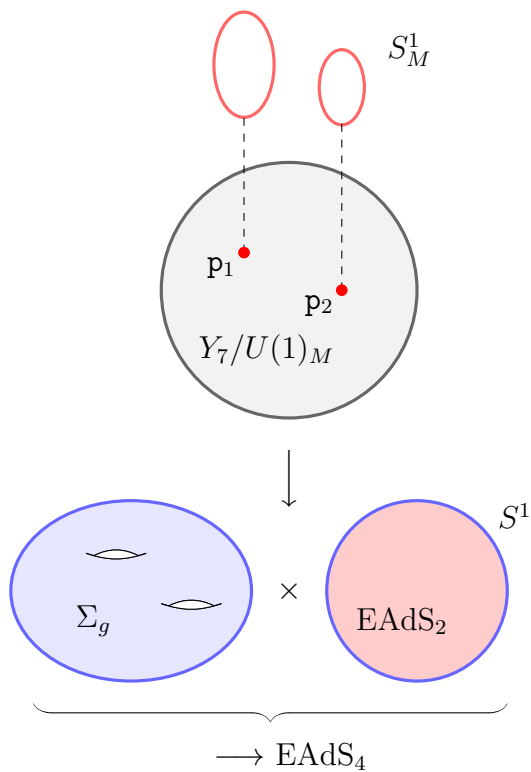
\begin{figure}[h]
	\centering
	\begin{tikzpicture}
		\filldraw[color=blue!60, fill=blue!10, very thick](-2,0) ellipse (1.6 and 1.2);
		\node at (-2.5,-.4)  {$\Sigma_g$};
		\draw [name path = a1] (-2.5,.465) .. controls (-2.3,.57) and (-2.1,.57) .. (-1.9,.465);
		\draw [name path = b1] (-2.6,.5) .. controls (-2.3,.4) and (-2.1,.4) .. (-1.8,.5);
		\tikzfillbetween[of= a1 and b1]{white}
		\draw (-2.5,.465) .. controls (-2.3,.57) and (-2.1,.57) .. (-1.9,.465);
		\draw (-2.6,.5) .. controls (-2.3,.4) and (-2.1,.4) .. (-1.8,.5);
		\draw [name path = a2] (-1.5,-.2) .. controls (-1.3,-.1) and (-1.1,-.1) .. (-0.9,-.2);
		\draw [name path = b2] (-1.6,-.17) .. controls (-1.3,-.27) and (-1.1,-.27) .. (-0.8,-.17);
		\tikzfillbetween[of= a2 and b2]{white}
		\draw (-1.5,-.2) .. controls (-1.3,-.1) and (-1.1,-.1) .. (-0.9,-.2);
		\draw (-1.6,-.17) .. controls (-1.3,-.27) and (-1.1,-.27) .. (-0.8,-.17);
		\filldraw[color=blue!60, fill=red!20, very thick](1.8,0) circle (1.2);
		\node at (1.6,-.4)  {EAdS$_2$};
		\node at (3.1,1.)  {$S^1$};
		\filldraw[color=black!60, fill=black!5, very thick](0.1,4) circle (1.7);
		\node at (-0.2,3.2)  {$Y_7/U(1)_M$};
		\draw[color=red!60, very thick](-.5,7) ellipse (.4 and .7);
		\draw[dashed] (-.5,4.5)--(-.5,6.3); 
		\filldraw [red] (-.5,4.5) circle (2pt);
		\node at (-.8,4.3)  {$\p_1$};
		\draw[color=red!60, very thick](.8,6.7) ellipse (.3 and .5);
		\node at (1.7,7.2)  {$S^1_M$};
		\draw[dashed] (.8,4)--(.8,6.3); 
		\filldraw [red] (.8,4) circle (2pt);
		\node at (.5,3.8)  {$\p_2$};
		\draw [decorate,
		decoration = {calligraphic brace,mirror,amplitude = 7pt}] (-3.3,-1.5) --  (3.,-1.5);
		\node at (0,-2.2)  {$\;\;\longrightarrow$ EAdS$_4$};
		\node at (0.1,0)  {$\times$};    
		\node at (0.1,1.5)  {$\Big\downarrow$};    
		
	\end{tikzpicture}
	\caption{We consider an extremal black hole with Euclidean near horizon geometry EAdS$_2\times \Sigma_g$ in EAdS$_4$. The full geometry is EAdS$_2\times Y_9$ with $Y_9$ a fibration of $Y_7$ over $\Sigma_g$. The probe M2-branes wrap the EAdS$_2$ and a copy of the M-theory circle $S^1_M$ at the points $\p_n$ (red) where $\zeta_M$ is aligned with the R-symmetry vector $\xi$ at $\p_n$. The dual field theory lives on $\Sigma_g\times S^1$ (blue), with $S^1$ being the boundary of EAdS$_2$. This $S^1$ is also the direction wrapped by the Wilson loop, in order for it to be BPS.}
	\label{fig:setup}
\end{figure}

In order for the probe M2-branes to be holographically dual to Wilson loops in the fundamental representation, they should wrap EAdS$_2\times S^1_M$ -- see Figure~\ref{fig:setup}.\footnote{Similar M2-branes were 
considered in \cite{Farquet:2013cwa}, although without any black hole in AdS$_4$ present.}$^,$\footnote{Note that we can use a probe approximation because we are interested in Wilson loops in the fundamental representation which are ``light", meaning that the corresponding brane  has negligible backreaction. The equivalent statement in field theory is that the insertion of the Wilson loop in the path integral does not affect the eigenvalue density itself, to leading order.} Indeed Wilson loops are dual to fundamental strings from the type IIA perspective \cite{Maldacena:1998im}, so we should take $S^1_\mathrm{M2} = S^1_M$.\footnote{Note that, as anticipated, this circle is always tangent to the fibres $Y_7$, hence the expression \eqref{M2action2} is well-defined.}
In the previous subsection we derived that the M2-branes were BPS for $S^1_\mathrm{M2}$ aligned with the R-symmetry circle. Then further imposing that they wrap the M-theory circle, the branes are BPS at the points where $\zeta_M$ is aligned with the R-symmetry vector $\xi$. There are a finite number of such connected loci, which can be determined explicitly for each choice of $Y_7$ together with a $U(1)_M$ action. Let us denote by $\p_n$ a point in the $n$-th such locus. 
Holographically the dual Wilson loop expectation value is determined by the M2-brane action evaluated at $\p_n$, where we should sum (or integrate) over all the possible configurations  
\begin{equation}\label{holoW}
	\langle W_\mathrm{fund}\rangle = \sum_n\int \dd \p_n\,  \ex^{-I_\mathrm{M2}|_{\p_n}}\,.
\end{equation}
Interestingly the value of the action only depends on the locus and not on the specific point within. One should also presumably integrate over the locus, with some measure, 
although the choice of measure is  a subleading effect in $N$, and 
we have correspondingly been schematic in writing \eqref{holoW}.
Another interesting point is that the Wilson loop 
for  the class of supersymmetric black holes we are considering is not 
zero due to integration over  gauge-equivalent configurations, discussed in \cite{Witten:1998zw}. This is a subtle point, but clearly important for the identification that 
follows, and the way that the argument in \cite{Witten:1998zw} is 
evaded in this setting is discussed further in Appendix~\ref{app:witten}.

The ordering of the loci is arbitrary from the geometry point of view, and we order them such that $I_\mathrm{M2}|_{\p_n}$ is increasing with $n$. 
At large $N$ we only consider the dominant contribution in the sum in \eqref{holoW}, which comes from the most negative M2-brane action 
\begin{equation}
	\log\,  \langle W_\mathrm{fund}\rangle =  -I_\mathrm{M2}|_{\p_1}\,.
\end{equation}
Comparing with the expression for the Wilson loop at large $N$ in field theory \eqref{logW}, we have a matching $I_\mathrm{M2}|_{\p_1}=N^{1/2}t_\tmin$. Our claim is that more is true and actually the M2-brane action reproduces all the points $t_n$ at which 
the derivative of the eigenvalue density is discontinuous:
\begin{equation}\label{M2matching}
	I_\mathrm{M2}|_{\p_n} = N^{1/2}t_n\,,
\end{equation}
where recall that $\p_n$ is a point in the $n$-th region where $\zeta_M$ and $\xi$ are aligned.
Let us denote $\beta_n$ the factor of proportionality at such a point, 
\begin{equation}\label{Mlength}
	\zeta_M = \beta_n \xi \quad \implies \int_{S^1_M|_{\p_n}}\eta=-2\pi \beta_n\,,
\end{equation}
as $\xi\lrcorner\,\eta=1$ and the sign is due to the orientation determined in \eqref{bps_cond}. 
Therefore the action of a BPS M2-brane dual to a Wilson loop reads
\begin{equation}\label{eq.SM2fin}
	I_{\text{M2}}|_{\p_n}
	= \frac{\Lambda}{\ell_p^3}\beta_n\:.
\end{equation}

The statement \eqref{M2matching} is actually equivalent to the argument in section~\ref{sec:mm} that the discontinuity points of $\rho'$ are reproduced by the fixed points of the moment map. Indeed recall that $h_M=\Lambda\, \zeta_M\lrcorner\,\eta$, and we already mentioned that the critical points of $h_M$ were those where $\zeta_M$ and $\xi$ are aligned. At a point $\p_n$ where $\zeta_M = \beta_n\, \xi$ we have $\tau_n\equiv h_M|_{\p_n}=\Lambda \beta_n$, such that
\begin{equation}\label{ttaubetaM2}
	t_n = \frac{\tau_n}{\ell_p^3 N^{1/2}}=
	\frac{\Lambda \beta_n}{\ell_p^3 N^{1/2}}=\frac{I_{\text{M2}}|_{\p_n}}{N^{1/2}}.
\end{equation}
Let us next examine this in our favourite examples. Recall that on $S^7/\Z_k$
\begin{equation}
	\xi = \sum_{a=1}^4 \frac{R_a}{2}\,\del_{\phi_a}\,.
\end{equation} 
\begin{quote}
	\bMI:
	For ADHM 
	\begin{equation}
		\zeta_M^\mathrm{I} = \frac{1}{k}(\del_{\phi_1}-\del_{\phi_2})\,,
	\end{equation}
	such that $\zeta_M^\mathrm{I}=\beta_n\xi$ in the regions $\p_1\in(0,z_2,0,0)$, $\p_2\in(0,0,z_3,z_4)$, $\p_3\in(z_1,0,0,0)$, 
with
	\begin{equation}
	 \beta_1 = -\frac{2}{k R_2}\,, \quad \beta_2=0\,,\quad 
	 \beta_3 = \frac{2}{k R_1}\,.
	\end{equation}
This reproduces the points in \eqref{intervalADHM}, recalling the relation \eqref{Lambdamu} between $\Lambda$ and $\mu$.
\end{quote}
\begin{quote}
	\bMII:
	For ABJM 
	\begin{equation}
		\zeta_M^\mathrm{II} = \frac{1}{k}(\del_{\phi_1}-\del_{\phi_2}+\del_{\phi_3}-\del_{\phi_4})\,,
	\end{equation}
	such that $\zeta_M^\mathrm{II}$ and $\xi$ are aligned on the regions where only one of the coordinate is non-zero. Ordering them by increasing $\beta_n$ we denote $\p_1 \in(0,z_2,0,0)$, $\p_2 \in (0,0,0,z_4)$, $\p_3 \in (0,0,z_3,0)$, $\p_4 \in (z_1,0,0,0)$, giving
	\begin{equation}
		\beta_1 = -\frac{2}{k R_2}\,,\quad \beta_2 = -\frac{2}{k R_4}\,, \quad 	\beta_3 = \frac{2}{k R_3}\,,\quad \beta_4 = \frac{2}{k R_1}\,,
	\end{equation}
	which exactly reproduces the subinterval points in \eqref{intervalABJM}. \\
	In the case where some of the R-charges are equal the regions may change topology. In particular when $R_a=1/2$ for all $a$, there are two loci  on which $\zeta_M^\mathrm{II}$ and $\xi$ are aligned, namely $\p_1\in\{z_1=z_3=0\}$, $\p_2\in\{z_2=z_4=0\}$, where 
		\begin{equation}
		\beta_1 = -\frac{4}{k}\,,\quad   \beta_2 = \frac{4}{k}\,,
	\end{equation}
	and $\rho(t)=2\mu/\pi$ is just a constant function between these two points. In our polytope picture this corresponds to the moment map plane being parallel to both the edge between $p_1$ and $p_3$ and that between $p_2$ and $p_4$, so that $Q_\tau$ is always a quadrilateral. In this case, we also have 
	\begin{equation}
		\Lambda = \frac{\pi\ell_p^3 N^{1/2}}{8}\sqrt{2k}= \frac{\pi\ell_p^3 N^{1/2}}{2}\mu\,.
	\end{equation}
This particular case describes the so-called ``universal twist'', 
which is an explicitly known black hole solution of minimal 4d gauged supergravity 
that uplifts on general classes of internal space $Y_7$ \cite{Azzurli:2017kxo}, including 
$S^7/\Z_k$ discussed here.
\end{quote}

\section{Matrix models on the spindle}\label{sec:spindle}
In this section we generalize the results obtained so far
 to the case where the black hole has a further parameter corresponding to the acceleration. The near-horizon geometry remains of the form described in section \ref{sec:geoextr}, with an internal GK manifold $Y_9$, but the Riemann surface in the fibration \eqref{fibration} is replaced by the 2d orbifold $\Sigma \equiv \mathbb{WCP}^1_{[m_+,m_-]}$, where $m_\pm\in\mathbb{N}$ are coprime and determine the acceleration \cite{Boido:2022iye,Boido:2022mbe}. The space $\Sigma$, known as a spindle, is topologically a two-sphere with conical singularities at the two poles parametrized by the positive integers $m_\pm$. Notice that we recover $S^2$ by setting $m_+=m_- = 1$.

We first review the relevant geometric properties of said fibrations, based on the discussion in \cite{Boido:2022mbe}, and then proceed to outline how we expect the large $N$ matrix model arising from localizing the dual 3d $\mathcal{N}=2$ SCFT on $\Sigma \times S^1$ to behave. There are currently no results concerning such large $N$ matrix models from a purely field theoretic approach, though some related work on the ``refined twisted index", with which we make contact here, has been carried out in \cite{Hosseini:2022vho}, 
and may be matched to our gravitational formulas.

\subsection{Accelerating black holes and block formulas}
In \cite{Boido:2022iye} it was conjectured that any supersymmetric, magnetically charged, and accelerating black hole in AdS$_4$, which uplifts to M-theory on a Sasaki-Einstein manifold like \eqref{Y7split}, has a near-horizon geometry of the form \eqref{11dmetric} where $Y_9$ takes the fibred form
\begin{equation}\label{fibration_spindle}
	Y_7 \hookrightarrow Y_9 \rightarrow \Sigma \:.
\end{equation}
Here $\Sigma$ is a spindle parametrized by $m_\pm \in \mathbb{N}$. These fibrations were constructed explicitly in \cite{Boido:2022mbe}, and their main difference with respect to \eqref{fibration} is that the R-symmetry vector $\xi$ is not necessarily orthogonal to $\Sigma$; in general it may have a component tangent to the spindle, which is the black hole horizon. Denoting $\partial_{\varphi_0}$ the generator of the $U(1)$ isometry which rotates $\Sigma$ about its axis, and keeping the notation $\partial_{\varphi_i}$ for a basis of generators for the $U(1)^4$ action on the fibre $Y_7$, we parametrize $\xi$ as
\begin{equation}\label{fullxi}
	\xi = \sum_{\mu = 0}^{s} b_\mu \, \partial_{\varphi_\mu} \:,
\end{equation}
with $b_0 \neq 0 $ in general and $b_1 = 1$ as before. Other than this, the twisting is obtained by turning on $s$ gauge magnetic fluxes analogous to those in \eqref{twist_fluxes}
\begin{equation}
	\frac{1}{2\pi} \int_\Sigma \diff A_i = \frac{p_i}{m_+ m_-}\:, \qquad p_i \in \Z \:.
\end{equation}
The $i=1$ direction is special and we need to set 
\begin{equation}
	p_1 = -\sigma m_+ - m_- \:,
\end{equation}
where $\sigma = \pm 1$ determines whether the mechanism for preserving supersymmetry is the ``twist" or the ``anti-twist", which are both distinct from the topological twist for preserving supersymmetry on a Riemann surface \cite{Ferrero:2021etw}. Note that in the twist case $\sigma = +1$ when setting $m_\pm = 1$ we have $p_1 = -2$, which agrees with what we had in section \ref{sec:fibration} for $g=0$.

Differently from the Riemann surface case, here the topology of the fibres in \eqref{fibration_spindle} do depend on the specific point of the base. We will be particularly interested in the fibres over the two poles of the spindle
\begin{equation}\label{Ypm}
	Y_\pm \equiv Y_7 / \Z_{m_\pm} \:.
\end{equation}
The vector fields $V_+ = \sum_{\mu = 0}^s v_{+\mu}\,\partial_{\varphi_\mu}$ and $V_- = \sigma\sum_{\mu = 0}^s v_{-\mu}\,\partial_{\varphi_\mu}$ which rotate the normal directions to $Y_\pm$ are given by
\begin{equation}\label{vpm}
	v_+ = (m_+, 1, - a_+ p_2, \dots, -a_+ p_s) \:, \qquad v_- = (-\sigma m_-, 1, - \sigma a_- p_2, \dots, -\sigma a_- p_s) 	\:,
\end{equation}
where $a_\pm$ are coprime integers satisfying	$a_- m_+ + a_+ m_- = 1$.\footnote{The integers $a_\pm$ exist by Bezout's lemma but are not unique. However, this is not an issue as they drop out from any on-shell physical quantity.} Thanks to \eqref{vpm}, we can introduce two new vectors fields $\xi_\pm$ as the orthogonal projections of the R-symmetry vector $\xi$ onto $Y_\pm$
\begin{equation}\label{xipm}
	\xi_\pm \equiv \xi \mp \frac{b_0}{m_\pm}\,V_\pm = \sum_{i=1}^{s} b_i^{(\pm)} \, \partial_{\varphi_i} \:,
\end{equation}
where we define the shifted vectors
\begin{equation}\label{bpm}
	b_i^{(+)} \equiv b_i - \frac{b_0}{m_+}\,v_{+i} \:, \qquad\qquad b_i^{(-)} \equiv b_i + \frac{b_0}{\sigma m_-}\,v_{-i} \:.
\end{equation}

By applying a simple lemma in differential geometry, in \cite{Boido:2022mbe} it was shown that in this set-up many geometric quantities that are integrals over $Y_9$ receive contributions only from the two poles of the spindle and hence reduce to integrals over the fibres $Y_\pm$, resulting in the so-called \textit{block formulas}.\footnote{These block formulas in GK geometry have recently been derived using a fixed point formula 
for an equivariant version of the master volume in \cite{Martelli:2023oqk}.}
 In particular, this is the case for all the quantities of physical interests such as the supersymmetric action \eqref{Ssusy}, the constraint \eqref{constraint}, and the quantized fluxes \eqref{flux_quant2}. Moreover, the geometric R-charges defined in \eqref{Rcharges} get doubled as we can define both $R_a^+$ and $R_a^-$ by integrating over $S_a^\pm$ \textit{i.e.}\ the copies of the codimension two submanifold $S_a \subset Y_7$ over the two poles of the spindle. For a more detailed discussion and the explicit expressions we refer the reader to \cite{Boido:2022mbe}, while here we will directly focus on the flavour twist case.

In order to implement the flavour twist condition here we cannot simply impose \eqref{flavour_twist}, as the transverse K\adots hler form of the fibres $\omega$ is not guaranteed to be the same at every point of the spindle -- and it is not in general. 
However, the analogous conditions in this case are instead 
\begin{equation}
	[J|_{Y_\pm}] = \Lambda_\pm [\varrho|_{Y_\pm}] \:, \qquad\qquad \Lambda_\pm \in \R \:.
\end{equation}
The two K\adots hler class parameters are specified by imposing flux quantization through the fibres at the poles and read
\begin{equation}\label{Lambdapm}
	\Lambda_+ = \frac{(2\pi \ell_p)^3}{\sqrt{24b_1^3 \, \Vol_S(Y_7)|_{b_i^{(+)}}}}N^{1/2}\:,\qquad \Lambda_- = \sigma\, \frac{(2\pi \ell_p)^3}{\sqrt{24b_1^3 \, \Vol_S(Y_7)|_{b_i^{(-)}}}}N^{1/2}\:.
\end{equation}
Note that these are the same functional expressions as \eqref{Lambda}, though evaluated with the shifted vectors $b_i^{(\pm)}$. This is a characteristic feature of physical quantities in these fibred 
geometries: most  get contributions from the two poles, taking the same form we had in the Riemann surface case but evaluated with the orthogonal projections of the R-symmetry vector onto the fibres \eqref{xipm}. Indeed, the off-shell entropy is given by
\begin{equation}\label{entropy_block}
	\mathcal{S} = \frac{4}{b_0} \left(F_\text{grav}[\xi_+] - \sigma\, F_\text{grav}[\xi_-]\right)\Big|_{b_1 = 1}\:.
\end{equation}
Moreover, as anticipated we have now two sets of geometric R-charges associated to supersymmetric submanifolds of the fibres $S_a \subset Y_7$ 
\begin{equation}\label{Rpm}
	R_a^\pm = \frac{2\pi}{3b_1}\, \frac{\Vol_S (S_a)}{\Vol_S(Y_7)} \bigg|_{b_i^{(\pm)}} \:, 
\end{equation}
with $R_a^+ >0$ and $\sigma R_a^- >0$. The corresponding integer fluxes $M_a$ through the codimension two fibrations $S_a \hookrightarrow \Sigma_a \rightarrow \Sigma$ inside $Y_9$ are given in terms of these R-charges by
\begin{equation}\label{fluxes_block}
	M_a = \frac{b_1}{2b_0} \, (R_a^+ - R_a^-) N \:.
\end{equation}
Note that all of the above expressions reduce to those in section \ref{sec:flavtwist} upon setting $\sigma = 1$, $m_+=m_- = 1$, and sending $b_0 \to 0$, hence recovering the result for $\Sigma_g = S^2$.

\subsection{Shifted eigenvalue densities}
Besides an investigation of the localized index on $\Sigma\times S^1$ at finite $N$ \cite{Inglese:2023wky}, there are currently no systematic studies of the field theory dual to the accelerating black holes whose near-horizon geometry we presented above. This renders the task of providing a geometric characterization of the corresponding matrix models harder, especially given that  part of the 
picture we presented in section \ref{sec:matching} was motivated heuristically, 
albeit with concrete computational checks.
Nevertheless, we can exploit the lessons learned so far to describe some of the features of said matrix models, leaving a more thorough analysis to future work. 

First of all, let us notice that at the fixed points of $V_\pm$, that is at the respective poles of $\Sigma$, the geometry of the internal space becomes equivalent to that described in section \ref{sec:fibration} (with $\Sigma_g$ replacing the spindle), upon using the shifted R-symmetry vectors $b_i \to b_i^{(\pm)}$.  Moreover, the vector field $\zeta_M$ generating the $U(1)_M$ action is a characteristic feature of the original geometric engineering set-up,  and as such it only rotates directions inside the fibres $Y_7$, and does not 
act on the spindle itself -- that is, it does not have any component along $\partial_{\varphi_0}$. Hence it is clear that at the two poles of the spindle the geometric set-up is much the same as what we had before, and thus we can envisage having a doubling of all the quantities associated to the matrix model, with a copy at the ``$+$" pole and another at the ``$-$" pole.

We begin by noting that the $U(1)_M$ action is Hamiltonian on $Y_\pm$, and therefore we can define two associated moment maps $h_M^\pm$  at the poles of the spindle to be such that
\begin{equation}
	\diff h_M^{\pm} = -\, \zeta_M \lrcorner\, J|_{Y_\pm} \:.
\end{equation}
From here, everything goes through as in section \ref{sec:matching}, although with the shifted R-symmetry vector $b_i^{(\pm)}$ and consequently the shifted K\adots hler class parameters $\Lambda_\pm$ and the shifted R-charges $R_a^\pm$. Parametrizing with $\tau$ the images of $h_M^\pm$, we can define two quotients of the fibres $Y_\pm$
\begin{equation}
	Y_{\tau}^\pm \equiv (h_M^\pm)^{-1}(\tau)/U(1)_M \:.
\end{equation}
The volumes of $Y_{\tau}^\pm$ can be obtained from $\Vol(Y_\tau)$, as computed in the Riemann surface case, by shifting the R-symmetry vectors and dividing by a factor $m_\pm$ which accounts for the $\Z_{m_\pm}$ quotient in \eqref{Ypm}
\begin{equation}
	\Vol(Y_{\tau}^\pm) = \frac{1}{m_\pm} \, \Vol(Y_{\tau}) \Big|_{b_i^{(\pm)}} \:.
\end{equation}
Note that $\tau$ is just a label at this stage, and there is no reason why it should be the same at the two poles, hence we will rename it to $\tau^\pm$. We can correspondingly define two ``shifted densities" $\rho_\pm(t^\pm)$ as in \eqref{densitygeo}. In particular, denoting with $\rho_{\,\Sigma_g}(t)$ the eigenvalue density of a certain field theory on $\Sigma_g$, the same field theory on the spindle will have
\begin{equation}\label{rho_spindle}
	\rho_\pm(t^\pm) = \rho_{\,\Sigma_g}(t^\pm) \big|_{b_i^{(\pm)}} \:.
\end{equation}
Moreover, the derivatives $\rho_\pm'(t)$ are discontinuous at the critical points of $h_M^\pm$, namely where $\zeta_M$ is aligned with $\xi_\pm$ and are simply given by
\begin{equation}\label{tpm}
	t_n^\pm = t_n |_{b_i^{(\pm)}} \:,
\end{equation}
where $t_n$ are the points of discontinuity of $\rho'_{\,\Sigma_g}(t)$.

This last point can be corroborated by looking at probe M2-branes in this background. The discussion about supersymmetry we carried out in section \ref{sec:M2} goes through without any change as it relies only on the fact that $Y_9$ is a GK manifold. There we showed that the M2-branes are BPS when $\zeta_M$ is aligned with $\xi$, and from \eqref{xipm} we can see that in the loci where $V_\pm$ are zero, namely the respective poles of the spindle, this is equivalent to requiring $\zeta_M$ to be aligned with $\xi_\pm$, as stated above. With suitable tweaks, also the gauge fixing in section \ref{sec:M2match} goes through, albeit it has to be done separately at the two poles, and we find that BPS M2-branes wrapping EAdS$_2 \times S^1_M$ can sit at either poles of $\Sigma$, and their action matches \eqref{tpm}
\begin{equation}\label{IM2spindle}
	I_\text{M2}|_{\p_n^\pm} = N^{1/2} t_n^\pm \:.
\end{equation}
Here $\p_n^\pm$ are points in the $n$-th locus where $\zeta_M \propto \xi_\pm$. Correspondingly, in field theory a Wilson loop in the fundamental representation wrapping the Euclidean time circle is supersymmetric when it sits at either of the two poles, and its expectation value at leading order in $N$ will be
\begin{equation}
	\log\,  \langle W_\mathrm{fund}^\pm\rangle =  -I_\mathrm{M2}|_{\p_1^\pm}\,.
\end{equation}

All the quantities introduced up to now should then match with the corresponding quantities derived from a purely field theoretic computation. In particular, for 3d $\mathcal{N} = 2$ SCFTs on $\Sigma \times S^1$ it is convenient to introduce the shifted chemical potentials
\begin{align}
	\Delta_A^+ &= \Delta_A +  \frac{\epsilon}{2} \left(\mathfrak{n}_A - \frac{r_A}{2}\,\frac{m_- - \sigma m_+}{m_+ m_-}\right) , \label{deltap}\\
	\Delta_A^- &= \Delta_A -  \frac{\epsilon}{2} \left(\mathfrak{n}_A + \frac{r_A}{2}\,\frac{m_- - \sigma m_+}{m_+ m_-}\right) , \label{deltam}
\end{align}
where $\epsilon$ is an extra parameter which will enter the extremization procedure and the $r_A$ are subject to the constraint $\sum_{A\in W} r_A = 2$ but are otherwise free.
We expect all physical quantities to have natural expressions in terms of these, and the holographic matching to be achieved by suitably identifying $\Delta_A^\pm$ with linear combinations of $R_a^\pm$ and $\epsilon$ with $2b_0$.

To summarize,  the matrix model arising from considering the large $N$ limit of the localized partition function on $\Sigma \times S^1$ is expected to factorize into two different matrix models, which can be studied independently and result in the densities $\rho_\pm(t^\pm)$ introduced above. We expect that the information contained in said ``shifted" densities can be used to recover the total $\rho_{\,\Sigma}(t)$, as a function of a single variable $t$ obtained from $t^\pm$.
This picture is also motivated by several results in the literature where in various different contexts a factorization of physical observables have been observed, ultimately 
related to a fixed point formula.  Of these, the refined twisted index discussed in \cite{Hosseini:2022vho} is directly relevant, and will be discussed in more detail in the next subsection.

Since we have left most of the formulas in this section rather implicit, we conclude 
by presenting the explicit BPS M2-brane actions for the ABJM model
on a spindle. There are 4 such BPS M2-branes, wrapping an aligned M-theory 
circle and R-symmetry vector over each $\pm$ pole of the spindle, giving 
8 formulas in total.
From \eqref{IM2spindle} and the results in Appendix~\ref{app:summary}
we can simply write down
\begin{align}
I_\text{M2}|_{\p_a^\pm} = N^{1/2} t_a^\pm \:, \quad a=1,2,3,4\, ,
\end{align}
where 
\begin{align}\label{tpmABJM}
t_1^\pm = -\frac{\mu^\pm}{kR_2^\pm}\, , \quad t_2^\pm = -\frac{\mu^\pm}{kR_4^\pm}\, ,  \quad t_3^\pm = \frac{\mu^\pm}{kR_3^\pm}\, , \quad 
t_4^\pm = \frac{\mu^\pm}{kR_1^\pm}\, , 
\end{align}
and
\begin{align}
\mu^\pm = \pi\sqrt{2kR_1^\pm R_2^\pm R_3^\pm R_4^\pm}\, .
\end{align}
Here the R-charges $R_a^\pm$ satisfy the constraints \cite{Boido:2022mbe}
\begin{align}
R_a^+ - R_a^- = 2b_0 \frac{M_a}{N} = 2b_0 \mathfrak{n}_a\, , \quad 
\sum_{a=1}^4 R_a^+ = 2-\frac{2b_0}{m_+}\, , \quad \sum_{a=1}^4 R_a^- = 2+\frac{2b_0}{\sigma m_-}\, ,
\end{align}
which should be matched to the chemical potential variables 
$\Delta_a^\pm$ in \eqref{deltap}, \eqref{deltam}, 
where the fluxes $\mathfrak{n}_a$ for the twist and anti-twist cases are constrained to obey
\begin{align}
\sum_{a=1}^4 \mathfrak{n}_a = -\frac{\sigma m_+ + m_-}{m_+ m_-}\, .
\end{align}
Here we identify $\epsilon=2b_0$, as in the next subsection.
This gives a very precise prediction for the large $N$ matrix model 
behaviour of the ABJM theory on a spindle, with 
\eqref{tpmABJM} giving the points at which the expected eigenvalue density 
blocks have discontinuities in their derivatives. More physically, 
as we have explained the least such M2-brane action computes 
a BPS Wilson loop VEV in the field theory on the spindle.

\subsection{Connecting to the refined twisted index}

To conclude the discussion on the spindle index, we will make contact with the recent field theory results of \cite{Hosseini:2022vho}. There the authors studied the matrix models arising from the large $N$ limit of the \textit{refined} twisted index $Z_{S^2_\epsilon \times S^1}$, which is defined for the genus $g = 0$ case as a further refinement of the topologically twisted index with respect to the axial $U(1)$ isometry of the two-sphere (see section 4 of \cite{Benini:2015noa}). This index depends on an extra parameter $\epsilon$, 
which ``rotates'' the $S^1$ relative to the $S^2$. 
Holographically, its logarithm has been matched with the entropy of supersymmetric, magnetically charged, and also rotating black holes in AdS$_4$ \cite{Hristov:2018spe}, with $\epsilon$ identified with the chemical potential conjugate to the angular momentum of the black hole \cite{Hosseini:2019iad}. 

We shall see in this section that the field theory results of \cite{Hosseini:2022vho} 
may 
\emph{also} be matched to GK geometries for which $Y_7=S^7$ is fibred 
over $S^2$, and where the R-symmetry vector also rotates the $S^2$. 
Specifically, we shall see that the field theory parameter $\epsilon=2b_0$,
 in terms of the component $b_0$ of $\xi$ that rotates the $S^2$. 
A similar matching was pointed out in \cite{Boido:2022iye}, 
and also relies on some observations in \cite{Cassani:2021dwa}. 
In order to explain this, one first observes 
for the twisted $S^2_\epsilon\times S^1$,
the R-symmetry Killing vector field bilinear for this rigid supersymmetric
background is \cite{Benini:2015noa, Cassani:2021dwa}\footnote{The fugacity was denoted $\ex^{\ii\epsilon}=\ex^{\ii\varsigma/2}$ in \cite{Benini:2015noa}, which explains the factor of $\tfrac{1}{2}$ in \eqref{Krotate}.} 
\begin{align}\label{Krotate}
K = \bar{\zeta}\gamma^\mu\zeta\partial_\mu = \partial_{\theta} +\tfrac{1}{2} \epsilon 
\, \partial_{\varphi_0}\, .
\end{align}
Here $\zeta$ is the Killing spinor of the background, 
 $\theta$ has period $2\pi$ and parametrizes the $S^1$, while 
$\varphi_0$ parametrizes the axial direction on the $S^2$ and is also 
normalized to have canonical period $2\pi$. The expression 
\eqref{Krotate} is a more invariant geometric way to 
characterize the parameter $\epsilon$, in terms of specifying the R-symmetry 
vector field of the background on which one is localizing.  On the other hand, the bulk black hole solution
inherits the same Killing vector field, and when lifted to 
11d M-theory this explains why $\epsilon=2b_0$, since 
by definition $b_0$ parameterizes the component of 
the R-symmetry vector of the full solution along the horizon direction 
$\Sigma$, precisely as in equation \eqref{fullxi}. This is all detailed quite explicitly for the  class 
of dyonically charged, accelerating and rotating black holes 
studied in \cite{Ferrero:2020twa, Cassani:2021dwa} (which lie in the universal 
truncation of \cite{Gauntlett:2007ma}), but the same arguments should go through
also for the more general class of purely magnetically charged black holes in this paper, with 
general choice of $Y_7$ and magnetic fluxes, but with a trial R-symmetry 
vector field that also rotates the horizon $S^2$. We then 
expect the entropy function in \eqref{entropy_block} to match $\log Z_{S^2_\epsilon \times S^1}$ in the twist case $\sigma = 1$ and in the limit $m_\pm = 1$, which we find 
is indeed the case. This is then consistent with the conjecture in 
\cite{Boido:2022iye}: that $b_0$ may be regarded as a chemical potential 
for angular momentum $J$, with the entropy of \emph{rotating}
versions of the black holes in this paper obtained via a Legendre transform. 
Moreover, 
given the results earlier in the paper, we may go further and 
also match matrix model quantities in the refined field theory calculation, and GK geometry
with $b_0\neq0$. 

When considering the refined index, the summation over the gauge magnetic fluxes $\mathfrak{m}$ in \eqref{locZ} cannot be performed directly and instead one should keep them as variables when taking the large $N$ limit. In particular, in \cite{Hosseini:2022vho} the authors supplement the ansatz \eqref{uansatz} with\footnote{Our notation differs from that of \cite{Hosseini:2022vho} in the following way: $\mathfrak{s}^\text{there} = -\mathfrak{n}$, $\mathfrak{n}^\text{there} = \mathfrak{s}$, $\epsilon^\text{there} = \pi \epsilon$, $\Delta^\text{there} = \pi \Delta$, $T^{(\pm)}_\text{there} = t^\pm$, $\widetilde{\mathcal{W}}^\text{there} = - \mathcal{U}$.}
\begin{equation}
	\mathfrak{m}_I^\alpha = \ii N^{1/2}\,\mathfrak{s}^\alpha + \mathfrak{p}_I^\alpha \quad \longrightarrow \quad \mathfrak{m}_I(t) = \ii N^{1/2}\,\mathfrak{s}(t) + \mathfrak{p}_I(t)\:.
\end{equation}
Moreover, the partition function takes the form of a product of holomorphic blocks \cite{Beem:2012mb}, which at large $N$ can be expressed in terms of shifted Bethe potentials 
\begin{equation}\label{logZrefined}
	\log Z_{S^2_\epsilon \times S^1} = -\frac{\ii}{\pi\epsilon}\, \Big[\mathcal{U}(\rho_+(t^+),v_I^+(t^+),\Delta_A^+) - \mathcal{U}(\rho_-(t^-),v_I^-(t^-),\Delta_A^-)\Big] \:,
\end{equation}
where $\mathcal{U}$ is the same Bethe potential we introduced in section \ref{sec:localize} and the various quantities appearing are defined as
\begin{equation}\label{shifted_quantities}
	\begin{split}
		t^\pm &= t \pm \frac{\pi\epsilon}{2} \, \mathfrak{s}(t) \:, \qquad \rho_\pm(t^\pm) =\frac{ \rho(t)}{1\pm \frac{\pi\epsilon}{2}\,\mathfrak{s}'(t)} \:, \qquad v_I^\pm(t^\pm) = v_I(t) \pm \frac{\pi\epsilon}{2} \, \mathfrak{p}_I(t) \:, \\
		\Delta_A^\pm &= \Delta_A \pm \frac{\epsilon}{2}\, \mathfrak{n}_A \:,
	\end{split}
\end{equation}
and the shifted chemical potentials satisfy the constraints
\begin{equation}
	\sum_{A \in W} \Delta_A^\pm = 2\mp \epsilon \:.
\end{equation}
Note that the $\Delta_A^\pm$ introduced for the spindle in \eqref{deltap} and \eqref{deltam} precisely reduce to those in \eqref{shifted_quantities} when $\sigma=1$ and $m_+=m_- = 1$. 

At this point, one should extremize \eqref{logZrefined} with respect to $\rho(t)$, $v_I(t)$, $\mathfrak{s}(t)$, and $\mathfrak{p}_I(t)$, but in \cite{Hosseini:2022vho} the authors observed that this is actually equivalent to extremizing independently the two Bethe potentials with respect to $\rho_\pm(t^\pm)$ and $v_I^\pm(t^\pm)$. With a suitable map between the variables this is equivalent to the unrefined case and therefore the results are the same
\begin{equation}\label{rhopm_vpm}
	\rho_\pm(t^\pm) = \rho_{\Sigma_g}(t^\pm)\big|_{\Delta_A^\pm} \:, \qquad  v_I^\pm(t^\pm) = v_I^{\Sigma_g}(t^\pm)\big|_{\Delta_A^\pm} \:.
\end{equation} 
The physical $\rho(t)$ and the other functions of $t$ of the matrix model are then obtained by going backwards through the definitions of the shifted quantities in \eqref{shifted_quantities} and imposing some consistency conditions which will not be detailed here. 
Substituting \eqref{rhopm_vpm} into \eqref{logZrefined} one obtains
\begin{equation}
	\log Z_{S^2_\epsilon \times S^1} =  \frac{1}{\pi\epsilon}\,\Big[\bar{\mathcal{U}}(\Delta_A^+) -  \bar{\mathcal{U}}(\Delta_A^-)\Big] =  \frac{1}{2\epsilon}\,\Big[F_{S^3}(\Delta_A^+) -  F_{S^3}(\Delta_A^-)\Big] \:.
\end{equation}
As anticipated, this matches perfectly with the entropy  function \eqref{entropy_block} when $\sigma = 1$ and $m_+=m_-= 1$ if we identify $\epsilon = 2b_0$ and $\Delta_A^\pm$ with suitable linear combinations of $R_a^\pm(\xi_\pm)$.\footnote{Recall that $F_{S^3}=F_\text{grav}|_{b_1 = 4} = 16 F_\text{grav}|_{b_1 = 1}$ upon identifying $\Delta_A$ with the same linear combinations of $R_a(\xi)$.} Furthermore, notice that the densities \eqref{rhopm_vpm} are exactly the same as those we defined from GK geometry, for the spindle matrix model, in \eqref{rho_spindle}! 

\section{Further toric examples}\label{sec:examples}
In this final section we illustrate our results with 
some further examples. We will consider other toric choices of Calabi-Yau cones $X_4$, which are dual to various flavoured variants of the two quivers we already discussed. These are somewhat more involved computationally, but there are no extra ingredients entering the picture besides some toric geometry machinery, which we will use as a tool for computing the relevant quantities.

We first revisit briefly the toric version of the GK geometry described in section \ref{sec:GK} and explain how the extra structure provides an algorithmic way to compute both the eigenvalue density and the discontinuity points of its derivative. We then apply the above to the examples where $X_4 = \C \times C(T^{1,1})$ and $X_4 = C(Q^{1,1,1}/\Z_n)$.  

\subsection{Generalities on toric cones}
\subsubsection{GK geometry with toric fibres}\label{sec:toric}
In the examples we will consider, the cone $X_4=C(Y_7)$ is a complex toric cone. 
The advantage of dealing with toric spaces is that geometric quantities can be expressed in terms of sums over combinatorial data, the \textit{toric data}, including integrals over the internal space or its cycles. For ease of exposition in this section 
we generally take $Y_7$ to be the simply-connected covering space, 
re-establishing quotients by $\Z_k$ or $\Z_n$ in final formulas as appropriate, 
much as we did in section~\ref{sec:S7example}. 

The complex cone \eqref{gX4} is toric if $s=4$ {\it  i.e.}\ $Y_7$ has $U(1)^4$ isometry, generated by the Killing vector fields $\del_{\varphi_i}$ as before. At each fixed radius $r>0$ sits a homothetic copy of $Y_7$, with the original $Y_7$ embedded naturally at $r=1$. A toric cone is fully described by a set of $d \geq 4$ primitive integer vectors $v_a \in \Z^4$, $a=1, \dots, d$. In particular, we introduce the polyhedral cone $\mathcal{C}$ as the cone whose $d$ facets have $v_a$ as inward-pointing normal vectors
\begin{equation}\label{toric_cone}
	\mathcal{C} \equiv \left\{y\in \R^{4} \; | \; \sum_{i=1}^4 y^i\,v_{ai} \geq 0 \:, \; a=1,\dots, d \right\}.
\end{equation}
This polyhedral cone is also the image of the moment map $\mu \colon C(Y_7) \to \R^4$ associated to the $U(1)^4$ action, given explicitly by\footnote{This should not be confused with the moment map associated to the $U(1)_M$ action we introduced in \eqref{mmap}. The one given in \eqref{mmap_toric} is defined to satisfy $\diff \mu_i =- \partial_{\varphi_i} \lrcorner\, \mathcal{J}$.}
\begin{equation}\label{mmap_toric}
	\mu_i = y^i= \tfrac{1}{2}\, r^2 \, \del_{\varphi_i} \lrcorner\, \eta \:.
\end{equation}
The set $\{y^i,\varphi_i\}$, $i=1,2,3,4$ constitutes a system of coordinates for $C(Y_7)$ and the latter can be built as a $U(1)^4$ fibration over the polyhedral cone $\mathcal{C}$ itself in the following way: in the interior of the cone $\mathcal{C}_{\text{int}}$ the fibration is trivial, while on each facets at the boundary $\del \mathcal{C}$ a single $U(1) \subset U(1)^4$ copy shrinks to a point. In particular, at the $a^{\mathrm{th}}$ facet the $U(1)$ copy that shrinks is identified by the vector field
\begin{equation}\label{shrinkU1}
	\nu_a =\sum_{i=1}^{4} v_{ai} \,\partial_{\varphi_i} \:.
\end{equation}
Complex codimension one submanifolds of $C(Y_7)$ which are invariant under the $U(1)^4$ action are called toric divisors and their images under the moment map $\mu$ are precisely the $d$ facets of the cone $\mathcal{C}$. Note that the freedom of changing basis for the $U(1)^4$ action results in the freedom of making an $SL(4,\Z)$ transformation of the toric data $v_{ai}$ and the toric map coordinates $\mu_i$. The basis we use is such that the condition \eqref{hol_charge} holds and $v_{a} = (1, \vec{w}_a)$, with $\vec{w}_a \in \Z^3$. 

Notice that any consistent choice of toric data for $d=4$ is related
to that in \eqref{toricdata_S7} by an $SL(4,\Z)$ transformation, and indeed $\C^4 = C(S^7)$ is the simplest example of toric cone. In that case the toric divisors coincide with the cones over the submanifolds obtained setting the complex coordinates $z_a =0$ whose volumes are \eqref{volsubS7}, and the vectors $\nu_a$ are nothing but those rotating said coordinates, \textit{i.e.}\ $\nu_a = \partial_{\phi_a}$.
When $d>4$ instead, the $v_a$ cannot be all linearly independent, and indeed there exists a $d \times (d-4)$ constant matrix $q_A^a$ (the \emph{kernel matrix}), such that
\begin{equation}
	\sum_{a=1}^{d} q_A^a\, v_a = 0 \:, \qquad \forall A= 1,\dots, d-4 \:.
\end{equation}
This relation can also be seen as the fact that the toric divisors are not independent cycles because $\dim H_5(Y_7,\R) = d-4$.\footnote{\label{footnoteGIT}An alternative way to construct the toric cone $X_4$ is to start with the space $\C^d$ with coordinates $z_a$, and then take a (GIT) quotient by a $(\C^*)^{d-4}$ action with weights $q_A^a$.} It is then clear that the vectors $\nu_a$ do not constitute a basis for the $U(1)^4$ action as they are not linearly independent, hence we will stick with the $\partial_{\varphi_i}$ basis, unlike what we did for $Y_7 = S^7$ where $\partial_{\phi_a}$ is a more natural (or 
certainly democratic) choice.

The image of $Y_7$ under the moment map is a subset of $\mathcal{C}$ which we call the \emph{Sasakian polytope}. Introducing the vector $b=(b_1,b_2,b_3,b_4)$ as the coordinate vector for the Reeb vector field \eqref{Reebvec}, the Sasakian polytope is given by
\begin{equation}\label{saspoly}
	P(b) = \mathcal{C}\; \cap \; H(b) \:,
\end{equation}
where  $H(b)$ is the Reeb hyperplane, defined as
\begin{equation}
	H(b)  \equiv \left\{y\in \R^{4} \: | \, \sum_{i=1}^4 y^i\,b_{i} = \tfrac{1}{2} \right\}.
\end{equation}
One can show that the Sasakian volume \eqref{sasakvol} can be obtained from the Euclidean volume of the polytope $P(b)$ as
\begin{equation}\label{sasvol}
	\Vol_S (Y_7) = \frac{(2\pi)^4}{|b|} \, \Vol_E(P) \:,
\end{equation}
where $|b|$ denotes the Euclidean norm of the vector $b$ and the factor $(2\pi)^4$ comes from integrating over the angles $\varphi_i$. This is useful because computing $\Vol_E(P)$ is just a matter of basic linear algebra involving the toric data $v_a$ and the Reeb vector $b$. Note that the cones over the submanifolds $S_a$ we introduced above \eqref{Rcharges} are precisely the toric divisors and are mapped to the facets of $\mathcal{C}$ via the moment map $\mu$. Hence, the Sasakian volume $\Vol_S(S_a)$ is similarly obtained from the Euclidean volume of the facets $P_a$ of the polytope $P$
\begin{equation}
	\Vol_S (S_a) = \frac{(2\pi)^3}{|b|} \, \Vol_E(P_a) \:.
\end{equation}

All the discussion so far was true for Sasakian $Y_7$. However, in order to implement the GK extremization procedure, we want to depart from this assumption and consider a general K\"ahler form $\tildeJ$ (transverse to the foliation $\mathcal{F}_\xi$ generated by $\xi$). In particular, the transverse cohomology class of $\tildeJ$ can be parametrized by $d$ real parameters $\lambda_a\in\R$ as
\begin{equation}\label{lambdadef}
	[\tildeJ] = -2\pi \sum_{a=1}^{d} \lambda_a c_a \:,
\end{equation}
where $c_a \in H^2_B(\mathcal{F}_\xi)$ are the Poincaré duals of the toric divisors. The Sasakian case with $\tildeJ_S$ is recovered when $\lambda_a = -\frac{1}{2b_1}$ for all $a=1,\dots,d$. Likewise, the flavour twist is achieved by setting $\lambda_a = -\Lambda$ as in \eqref{Lambda} for all $a=1,\dots,d$.\footnote{There is a subtlety here that we are slightly glossing over. Since $\dim H^2_B(\mathcal{F}_\xi)=d-3$, 
	the $d$ parameters $\lambda_a$ are not uniquely specified in \eqref{lambdadef}. There is then a related
	``gauge invariance", acting as the shifts $\lambda_a \to \lambda_a + \sum_{i=1}^{4}\gamma^i (v_{ai} b_1-b_i)$, $\gamma^i$ being the arbitrary gauge parameters. More precisely then, we recover the flavour twist when there exist a gauge in which $\lambda_a = - \Lambda$.} For now it will be convenient to stick with the general case and we will impose the flavour twist at a later stage.
Allowing for a general K\"ahler class amounts to shifting the facets of the polytope \eqref{saspoly}
\begin{equation}\label{polytope}
	P(b) \; \to \;  \mathcal{P} (b, \lambda_a) \equiv \left\{y\in H(b) \; | \; \sum_{i=1}^4 (y^i-y_*^i)\,v_{ai} \geq \lambda_a \:, \; a=1,\dots, d \right\},
\end{equation}
where $y_*^i$ is an arbitrary point in the Reeb hyperplane. For the following, it will be convenient to take
\begin{equation}
	y_* = \left(\frac{1}{2b_1}\,,0,0,0\right).
\end{equation}
Note that saturating the $a^{\mathrm{th}}$ inequality in \eqref{polytope} defines the $a^{\mathrm{th}}$ facet $\mathcal{P}_a$.
The corresponding generalization of the Sasakian volume \eqref{sasvol} is the so-called \textit{master volume}
\begin{equation}
	\mathcal{V}_7 = \Vol(Y_7) \equiv \int_{Y_7} \tildeeta \wedge \frac{\tildeJ^3}{3!} = \frac{(2\pi)^4}{|b|} \, \Vol_E(\mathcal{P}) \:,
\end{equation}
which itself is a function of the Reeb vector $b$ and the K\"ahler class parameters $\lambda_a$. Concretely, $\mathcal{V}_7$ can be computed again using basic linear algebra notions (see \cite{Gauntlett:2019roi}). The supersymmetric action \eqref{Ssusy}, the constraint equation \eqref{constraint}, and the quantized fluxes \eqref{flux_quant2} can all be expressed in terms of $\mathcal{V}_7$ and its derivatives. We refer to \cite{Gauntlett:2018dpc,Gauntlett:2019roi} for more details as we will not need these expressions.

Next, note that whenever the cone $X_4$ is toric, the geometric R-charges \eqref{Rcharges2} satisfy the constraints
\begin{equation}\label{Rconstraint}
	\sum_{a=1}^{d} R_a v_{ai} = \frac{2}{b_1} \, b_i \:.
\end{equation}
The $i=1$ component tells us that the $R_a$ sum to 2, which in turn imposes a constraint on the fluxes
\begin{equation}\label{fluxconstraint}
	\sum_{a=1}^{d} M_a = p_1 N =(2g-2) N \: ,
\end{equation}
which is a manifestation of the fact that supersymmetry for these geometries 
is being implemented via a topological twist over the Riemann surface $\Sigma_g$.
Moreover, from \eqref{Rconstraint} it is possible to express the R-symmetry vector \eqref{Reebvec} in terms of the vectors $\nu_a$ in \eqref{shrinkU1} 
\begin{equation}\label{xinu}
	\xi = \frac{b_1}{2} \, \sum_{a=1}^{d} R_a \, \nu_a\:,
\end{equation}
which agrees with \eqref{xiS7}, as it should.

Finally, to connect back to the flavour twist case of interest, we note that the Sasakian volumes can be recovered from $\mathcal{V}_7$ in the following way 
\begin{align}
	\Vol_S(Y_7) = \mathcal{V}_7 \big|_{\lambda_a = - \frac{1}{2b_1}} \:,  \qquad 
	\Vol_S(S_a) = -\frac{1}{2\pi} \, \frac{\partial \mathcal{V}_7}{\partial \lambda_a} \bigg|_{\lambda_a = - \frac{1}{2b_1}} \:.
\end{align}
Substituting these expressions into the relevant quantities of section \ref{sec:flavtwist} yields the physical observables in the toric flavour twist case.\footnote{In practice we imposed the flavour twist condition before the toric condition. Alternatively, we could have started from the toric expressions for the supersymmetric action, constraint, and flux quantization conditions and imposed the flavour twist by setting $\lambda_a$ all equal. The two procedures are completely equivalent.}

\subsubsection{Critical points and M2-branes}\label{sec:toricM2}
Recall that the moment map $h_M$ associated to the $U(1)_M$ action defined in \eqref{hM} maps $Y_7$ to the interval $[\tau_\text{min}, \tau_\text{max}]$ which gives the support of the eigenvalue density. Let us compute $h_M$ explicitly. Firstly, note that for general toric $Y_7$ we have \cite{Gauntlett:2018dpc,Gauntlett:2019pqg}
\begin{equation}
	\partial_{\varphi_i}\lrcorner \,\omega = -\,\diff x^i \:,
\end{equation}
where $x^i = (y^i -y^i_*)$. Defining $\zeta_i$ to be the coefficient of the M-theory killing vector $\zeta_M$ in the toric basis, \textit{i.e.}\ $\zeta_M = \sum_{i=1}^{4}\zeta_i\,\partial_{\varphi_i}$, it follows that
\begin{equation}\label{mmap_toric2}
	h_M(y^i) = \sum_{i=1}^{4}  \zeta_i\,(y^i -y^i_*) \:.
\end{equation}
The coefficients $\zeta_i$ depend on the particular physical model we consider, and will be specified in each example. 

Recall also that the image of $h_M$ is divided in subintervals whose extrema correspond to its value at the critical points, \textit{i.e.}\ where $\diff h_M=0$. In turn such points are those where $\zeta_M$ is proportional to $\xi$, and we can find them explicitly by employing the following strategy. We firstly find all the linear combinations of the form 
\begin{equation}\label{betatoric}
	\zeta_M = \beta\, \xi + \sum_{a\in F} \alpha_a \nu_a \:,
\end{equation}
where $F$ is any set of toric indices $\{a\}$ such that all the corresponding facets $\{\mathcal{P}_a\}$ have at least one point in common.
By definition, in the intersection $\bigcap_{a \in F}\mathcal{P}_a$ all the vectors $\nu_a$ with $a\in F$ collapse, and therefore $\zeta_M = \beta\, \xi$ which is precisely the condition for criticality of $h_M$. Note that here $\beta$ could also be vanishing, which signals the presence of D6-branes. In all the examples we look at, the intersection above is either a single point in the toric polytope or one of its edges. Ultimately, the extrema of the subintervals in the image of $h_M$ are obtained by evaluating
\begin{equation}
	h_M(y_F) \:, \qquad y_F \in \bigcap_{a \in F}\mathcal{P}_a \:,
\end{equation}
for each set $F$ defined as above, where note that different sets $F$ do not always lead to different points in the image.

Furthermore, we can directly make contact with the M2-brane discussion of section~\ref{sec:M2match}. In particular, from \eqref{Mlength} we see that $\beta$ determines precisely the value of the action of the probe M2-brane, which becomes BPS when it sits at $y_F$ 
\begin{equation}
	I_{\text{M2}}
	= \frac{\Lambda}{\ell_p^3}\,\beta\:.
\end{equation}

\subsubsection{Eigenvalue density}\label{sec:toriceigenvalue}

We now want to compute the eigenvalue density as the volume of the quotient space \eqref{Ytau} for any choice of toric $X_4$. In fact, recall that in section \ref{sec:density} we stated that
\begin{equation}\label{density}
	\rho(t) \propto \frac{\diff}{\diff \Lambda}\,\Vol(Y_\tau) \,\bigg|_{\tau = \ell_p^3 N^{1/2}t} \:,
\end{equation}
where $Y_\tau$ is defined as a quotient of the $\tau$ level set of $h_M$.
In the toric setting we can always follow the strategy we exploited for the two examples considered earlier in section~\ref{sec:density}, and decouple the coordinates $y^i \in \R^4$ with Euclidean metric\footnote{Differently from section \ref{sec:density}, here the K\adots hler class parameter is absorbed into the coordinates $y^i$. This has some repercussions on the normalization of quantities. In particular, this is why we distinguish $\mathcal{Q}_\tau$ from $Q_\tau$ (see \eqref{YpiQ} versus \eqref{volYtau}) which are otherwise conceptually the same quantity.} from the angles $\varphi_i\in U(1)^4$, which  simply give a contribution of $2\pi$ each.
Firstly, we note that the polytope \eqref{polytope} is the generalization of the tetrahedron defined by $r^2 =1$ in section \ref{sec:density}. The shape of such a polytope  will depend on the toric data $v_a$. Saturating the $a^{\mathrm{th}}$ inequality, \textit{i.e.}\
\begin{equation}\label{facet}
	(y^i - y^i_*)\, v_{ai} = \lambda_a \:,
\end{equation}
defines the plane where the $a^{\mathrm{th}}$ facet $\mathcal{P}_a$ lies, and changing the corresponding K\adots hler class parameter $\lambda_a$ amounts to shifting the facet parallel to itself. Upon setting the flavour twist condition $\lambda_a = -\Lambda$, all the facets move together when changing $\Lambda$. We then restrict to the level set of the moment map \eqref{mmap_toric2} imposing
\begin{equation}\label{mmap_plane}
	(y^i -y^i_*)\,\zeta_i = \tau \:.
\end{equation}
This equation describes a hyperplane in $\R^4$ which slices the polytope $\mathcal{P}$. The intersection of the two will be a polygon $\mathcal{Q}_\tau$ whose shape depends on the value of $\tau$ and whose Euclidean area determines directly the volume we are interested in
\begin{equation}\label{YpiQ}
	\Vol(Y_\tau) = (2\pi)^3\, \Vol_E(\mathcal{Q}_\tau) \:.
\end{equation}
Every time the plane \eqref{mmap_plane} crosses a vertex or an edge of the polytope, the expression for $\Vol_E(\mathcal{Q}_\tau)$ changes, hence it is clear that these correspond to the discontinuity points in the derivative of the density, \textit{i.e.}\ the critical points of the previous subsection. Note that here we have a clear geometric interpretation of what it means taking the derivative with respect to $\Lambda$ in \eqref{density}: given what we said below \eqref{facet}, it is clear that $\frac{\diff}{\diff \Lambda}\,\Vol_E(\mathcal{Q}_\tau)$ is (proportional to) the length of the perimeter of the polygon $\mathcal{Q}_\tau$.

In general, for a fixed $\tau$ a vertex $q_{ab}\in\R^4$ of $\mathcal{Q}_\tau$ is computed by imposing simultaneously two equations like \eqref{facet} for adjacent facets $\mathcal{P}_{a}$ and $\mathcal{P}_{b}$, the Reeb hyperplane condition $y \in H(b)$, and the moment map level set in \eqref{mmap_plane}
\begin{align}\label{systemtoric}
		(q^i_{ab} - y^i_*)\, v_{ai} & = -\Lambda\, , \nonumber\\
		(q^i_{ab} - y^i_*)\, v_{bi} & = -\Lambda\, , \nonumber\\
		(q^i_{ab} -y^i_*)\,b_i & = 0\, , \nonumber\\
		(q^i_{ab} -y^i_*)\,\zeta_i & = \tau \:.
\end{align}
These are four linear equations for four variables $q^i_{ab}$, $i=1,\dots,4$, hence the vertices of $\mathcal{Q}_\tau$ are uniquely determined in terms of $\Lambda$, $\tau$ and $b$ (for fixed examples $v_a$ and $\zeta$ are fixed)\footnote{Note that in the examples we will express $b$ in terms of geometrical R-charges $R$, as in \eqref{xinu}, for better comparison with the dual field theory results.}, and we pick $y_*=(1/2b_1,0,0,0)$ for convenience.
This system should be solved for any pair of adjacent facets. Note that it may not admit any solution, for certain pairs, indicating that the moment map hyperplane cuts the polytope parallel to the edge between these two facets. Further imposing that the solution indeed lies inside the polytope
\begin{equation}\label{toricineq}
	(q^i_{a b}-y_*^i)v_{ci}\geq-\Lambda \,, \quad \forall c=1,\dots, d
\end{equation}
gives the range of $\tau$ for which $q_{ab}$ is a vertex of $\mathcal{Q}_\tau$. Therefore the set of vertices of $\mathcal{Q}_\tau$, and consequently the expression for its perimeter (which recall is the quantity identified with the eigenvalue density $\rho$), changes on each such interval of $\tau$. 
Once the sets of vertices are found, the surface area of $\mathcal{Q}_\tau$ on each interval can be computed by splitting it into triangles and computing the area of each triangle as a determinant as we did in the examples in section \ref{sec:density}. More concretely, let us call $\zetavec$ the vector in $\R^4$ with components $\zeta_i$ and introduce
\begin{equation}
	\zetavec_\perp \equiv \zetavec - \frac{\zetavec \cdot b}{|b|^2} \, b \:.
\end{equation}
This is the equivalent of \eqref{zetaperp} but all the vectors here are expressed in toric coordinates. Then, the area of a triangle with vertices $q_{a_1 a_2}, q_{a_3 a_4}, q_{a_5 a_6} \in \R^4$ can be computed as
\begin{equation}\label{determinant}
	\frac{1}{2} \,\left|\det\left(\frac{b}{|b|},\frac{\zetavec_\perp}{|\zetavec_\perp|}, q_{a_3 a_4}- q_{a_1 a_2}, q_{a_5 a_6} - q_{a_1 a_2}\right)\right| ,
\end{equation}
and $\Vol_E(\mathcal{Q}_\tau)$ will be a finite sum of such determinants. Note that it is crucial to have $\zetavec_\perp$ rather then $\zetavec$ in \eqref{determinant} because only $\zetavec_\perp$ is orthogonal to both the triangle and to $b$, so that the volume of the parallelotope computed by the determinant is simply the area of the triangle times the length of the two ``heights", which are both 1.
We refer the reader to the examples in the next subsections to see how this procedure works more concretely. 

Finally, let us highlight that we are now able to justify the normalization in \eqref{normalized_rho}. Let us compute the integral of the RHS of \eqref{density}
\begin{equation}\label{integral}
	\mathcal{I}\equiv\int_{t_{\text{min}}}^{t_\text{max}} \diff t\, \frac{\diff}{\diff\Lambda}\,\Vol(Y_\tau) = \frac{(2\pi)^3}{\ell_p^3 N^{1/2}}\,\frac{\diff}{\diff\Lambda} \int_{\tau_{\text{min}}}^{\tau_\text{max}} \diff \tau\, \Vol_E(\mathcal{Q}_\tau)\:,
\end{equation}
where in the last step we exchanged the integral and the derivative and changed the variable of integration from $t$ to $\tau$. Now the integral above with a suitable measure is nothing but the volume of the entire polytope $\mathcal{P}$, which can be expressed in terms of the master volume $\mathcal{V}_7$ 
\begin{equation}
	\int_{\tau_{\text{min}}}^{\tau_\text{max}} \frac{\diff \tau}{|\zetavec_\perp|}\, \Vol_E(\mathcal{Q}_\tau) = \Vol_E(\mathcal{P}) = \frac{|b|}{(2\pi)^4}\,\mathcal{V}_7 \:.
\end{equation}
Substituting this back into \eqref{integral} we obtain
\begin{equation}
	\mathcal{I} = \frac{|b||\zetavec_\perp|}{2\pi\ell_p^3 N^{1/2}}\,\frac{\diff\mathcal{V}_7}{\diff\Lambda} = \frac{|b||\zetavec_\perp|}{2\pi\ell_p^3 N^{1/2}}\,(2\pi \ell_p)^6 N \,, 
\end{equation}
where the second equality is the toric equivalent of the flux quantization \eqref{flux_quant3} \cite{Gauntlett:2019roi}.
Hence, the correct normalization for \eqref{density} is
\begin{equation}
	\rho(t) = \frac{1}{\mathcal{I}}\, \frac{\diff}{\diff \Lambda}\,\Vol(Y_\tau) \,\bigg|_{\tau = \ell_p^3 N^{1/2}t} = \frac{1}{32  \pi^5 \ell_p^3  N^{1/2}|b||\zetavec_\perp|}\,\frac{\diff}{\diff \Lambda} \Vol(Y_\tau)\,\bigg|_{\tau = \ell_p^3  N^{1/2}\, t} \:,
\end{equation}
which is exactly what we had in \eqref{normalized_rho}. If $Y_7$ is a quotient space by a $\Z_n$ action, similarly to section \ref{sec:matching}, there is an additional factor of $n$ relating the volumes of $Y_\tau$ and $\mathcal{Q_\tau}$ (the unquotiented case is $n=1$), and the final expression for $\rho$ reads
\begin{equation}\label{rhonormalization}
	\rho(t) = \frac{1}{4 \pi^2 \ell_p^3  N^{1/2}|b||\zetavec_\perp|n}\,\frac{\diff}{\diff \Lambda} \Vol(\mathcal{Q}_\tau)\,\bigg|_{\tau = \ell_p^3  N^{1/2}\, t} \:.
\end{equation}

\subsection{Example: $C(Y_7) = \C \times C(T^{1,1})$}\label{sec:Cxconifold}

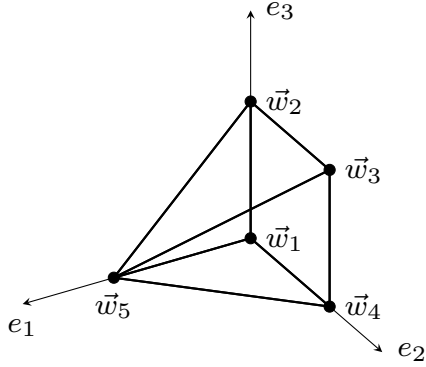
\begin{figure}
	\begin{center}
		\tdplotsetmaincoords{60}{150}
		\begin{tikzpicture}
			[tdplot_main_coords,scale=1.4,every node/.style={scale=1.4},font=\scriptsize]
			
			\draw[-stealth] (0,0,0) -- (2.5,0,0) node[anchor=north]{$e_1$};
			\draw[-stealth] (0,0,0) -- (0,2.5,0) node[anchor=west]{$e_2$};
			\draw[-stealth] (0,0,0) -- (0,0,2.5) node[anchor=west]{$e_3$};
			
			\filldraw[black] (0,0,0) circle (1.5pt) node[anchor=west]{$\vec{w}_1$};
			\filldraw[black] (0,0,1.5) circle (1.5pt) node[anchor=west]{$\vec{w}_2$};
			\filldraw[black] (0,1.5,1.5) circle (1.5pt) node[anchor=west]{$\vec{w}_3$};
			\filldraw[black] (0,1.5,0) circle (1.5pt) node[anchor=west]{$\vec{w}_4$};
			\filldraw[black] (1.5,0,0) circle (1.5pt) node[anchor=north]{$\vec{w}_5$};
			
			\draw[thick] (0,0,0) -- (0,0,1.5) -- (0,1.5,1.5) -- (0,1.5,0) -- cycle;
			\draw[thick] (0,0,0) -- (0,0,1.5) -- (1.5,0,0) -- cycle;
			\draw[thick] (0,0,0) -- (0,1.5,0) -- (1.5,0,0) -- cycle;
			\draw[thick] (0,1.5,1.5) -- (0,1.5,0) -- (1.5,0,0) -- cycle;
			\draw[thick] (0,0,1.5) -- (0,1.5,1.5) -- (1.5,0,0) -- cycle;
			
		\end{tikzpicture}
	\end{center}
	\vspace{-.5cm}
	\caption{Toric diagram for $\C\times C(T^{1,1})$.} 
	\label{fig:CxC}
\end{figure}

We now illustrate the procedure outlined in the previous subsection on the toric cone $X_4 = \C \times C(T^{1,1})$, which admits two field theory duals depending on the choice of the M-theory circle action. First let us describe some properties of this geometry. The toric vectors are
\begin{equation}\label{toricCxC}
	\begin{split}
		v_1&=\left(1,0,0,0\right),\quad v_2=\left(1,0,0,1\right),\quad v_3=\left(1,0,1,1\right),\\ v_4&=\left(1,0,1,0\right),\quad v_5=\left(1,1,0,0\right),
	\end{split}
\end{equation}
and the corresponding toric diagram is shown in Figure~\ref{fig:CxC}.
A subtle point to address is that the space $Y_7$ has a singularity and therefore computing the master volume naively leads to the wrong result. As discussed in  \cite{Gauntlett:2019roi} this is fixed by imposing a restriction on the Kähler class parameters
\begin{equation}
	\lambda_1 - \lambda_2 + \lambda_3 - \lambda_4 = 0 \:.
\end{equation}
Incidentally, this constraint also enforces the flavour twist condition \eqref{flavour_twist}, which is precisely what we need to compare with field theory results.
The Sasakian volume of the link $Y_7$ of $\C \times C(T^{1,1})$ is  
\begin{equation}
	\Vol_S(Y_7) = \frac{\pi^4 (b_1 - b_2)}{3 b_2 b_3 b_4 (b_1 - b_2 - b_3)(b_1 - b_2 - b_4)}\,.
\end{equation}
The geometric R-charges are
\begin{equation}\label{RCxC}
	\begin{split}
		& R_1 = \frac{2(b_1 -b_2 - b_3)(b_1 - b_2 - b_4)}{b_1(b_1 - b_2)} \:, \qquad 
		R_2 = \frac{2b_4(b_1 - b_2 - b_3)}{b_1 (b_1 - b_2)} \:, \\ \qquad &R_3 = \frac{2b_3b_4}{b_1 (b_1 - b_2)}  \:, \qquad R_4 = \frac{2b_3(b_1 - b_2 - b_4)}{b_1 (b_1 - b_2)}  \:, \qquad R_5 = \frac{2b_2}{b_1} \:,
	\end{split}
\end{equation}
so they satisfy
\begin{align}
	\sum_{a=1}^{5} R_a = 2 \:, \qquad	R_1 R_3 - R_2 R_4 = 0 \:.
\end{align}
Note that the second constraint implements the partial extremization over the baryonic directions in field theory. Finally, the Kähler class parameter defined in \eqref{Lambda} can be expressed in terms of the R-charges as
\begin{equation}\label{lambdaCxC}
	\Lambda=\frac{\ell_p^3N^{1/2}\pi}{2}\sqrt{\frac{2(R_1+R_2)(R_3+R_4)(R_1+R_4)(R_2+R_3)R_5}{R_1+R_2+R_3+R_4}}\, .
\end{equation}
Notice that the denominator may also be expressed as $2-R_5$. 

\subsubsection{Flavoured $U(N)$}

\paragraph{Geometry}

\begin{figure}
	\centering
	\begin{tikzpicture}
		\filldraw[color=red, fill=red!10, very thick] (2,.9)--(2.25,2.2)-- (2.75,-.9)--(2,.9) node[midway, left] {$\mathcal{Q}_{\tau<0}$};	
		\filldraw[color=blue, fill=blue!10, very thick] (4,1.8)--(4.5,4.45)--(5.5,-1.9)--(4,1.8) node[midway, left] {$\mathcal{Q}_{\tau=0}$};	
		\filldraw[color=Green, fill=Green!10, very thick] (6,1.3)--(6.25,2.6)-- (6.72,-.5)--(6,1.3) node[midway, left] {$\mathcal{Q}_{\tau>0}$};	
		
		\draw[thick,dashed] (0,0) -- (4,1.8) -- (8,.8);
		\draw[thick,dashed] (4,1.8) -- (4.5,4.5);
		\draw[thick] (4.5,4.5)--(5.5,-2);
		\draw[thick] (0,0)--(4.5,4.5)--(8,.8)--(5.5,-2)--(0,0);
		\filldraw[black] (0,0) circle (2pt) node[anchor=east]{$p_{145}$};
		\filldraw[black] (8,.8) circle (2pt) node[anchor=west]{$p_{235}$};
		\filldraw[black] (4.5,4.5) circle (2pt) node[anchor=south ]{$p_{1234}$};
		\filldraw[black] (5.5,-2) circle (2pt) node[anchor=north ]{$p_{125}$};	
		\filldraw[black] (4,1.8) circle (2pt) node[anchor=south west]{$p_{345}$};	
		
		\filldraw[red] (2,.9) circle (2pt) node[anchor=south east]{$q_{45}$};
		\filldraw[red] (2.25,2.25) circle (2pt) node[anchor=south east]{$q_{14}$};		
		\filldraw[red] (2.75,-1) circle (2pt) node[anchor=north]{$q_{15}$};		
		\filldraw[blue] (4,1.8) circle (2pt) node[anchor=south east]{};
		\filldraw[blue] (4.5,4.5) circle (2pt) node[anchor=south west]{};
		\filldraw[blue] (5.5,-2) circle (2pt) node[anchor=north west]{};
		\filldraw[Green] (6,1.3) circle (2pt) node[anchor=south east]{$q_{35}$};
		\filldraw[Green] (6.25,2.65) circle (2pt) node[anchor=west]{$q_{23}$};	
		\filldraw[Green] (6.75,-.6) circle (2pt) node[anchor= west ]{$q_{25}$};	
	\end{tikzpicture}
	\caption{Pyramid describing $\C\times C(T^{1,1})$ and its slicing corresponding to flavoured $U(N)$. The red slice is $\mathcal{Q}_\tau$ for $\tau_{\text{min}} < \tau <0$,  the blue slice for $\tau = 0$, and the green slice for $0 <\tau < \tau_{\text{max}} $. Each vertex $p_{abc}$ denotes the intersection of the facets $\mathcal{P}_a$, $\mathcal{P}_b$, $\mathcal{P}_c$, and the Reeb hyperplane. The perimeter of $\mathcal{Q}_\tau$ varies with $\tau$ qualitatively as $\rho$ in Figure~\ref{fig:rhoADHM}.}
	\label{fig:rhoUNgeom} 
\end{figure}
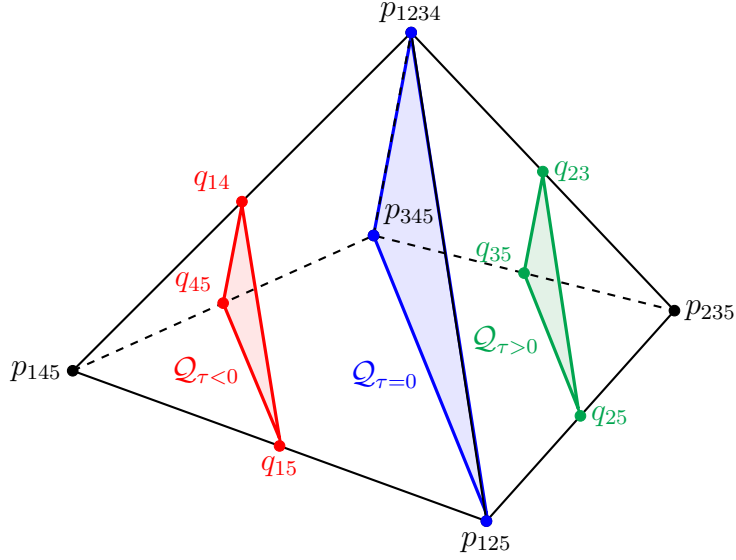

The first choice of $U(1)_M$ action we consider has weights $(1,-1,0,0,0)$ on the complex coordinates $z_a$ in the ambient space of the GIT quotient defining $X_4$ (see footnote \ref{footnoteGIT}), which are rotated by $\del_{\phi_a}\equiv\nu_a=\sum_i v_{ai}\del_{\varphi_i}$ \cite{Benini:2009qs}. Using the toric data \eqref{toricCxC}, the M-theory circle action on $X_4$ therefore reads
\begin{equation}
	\zeta_M = - \partial_{\varphi_4}\,.
\end{equation}
For this choice of $\zeta_M$ and toric data \eqref{toricCxC}, \eqref{betatoric} admits two non-trivial solutions for $\beta$,\footnote{In practice we solve the system in terms of $b_i$ and then convert the results in the $R_a$ variables using \eqref{RCxC}.} which we multiply by $\Lambda$ to obtain $\tau$
\begin{equation}\label{tauUN}
	\tau_\tmin = -\frac{2\Lambda}{R_2+R_3}\,, \qquad \tau_\tmax = \frac{2\Lambda}{R_1+R_4}\, .
\end{equation}
The non-zero $\alpha_a$ coefficients for these solutions are $\{\alpha_1,\alpha_4,\alpha_5\}$ and $\{\alpha_2,\alpha_3,\alpha_5\}$ respectively, which is reflected by the vertices in Figure~\ref{fig:rhoUNgeom}. Note that $\beta=\tau/\Lambda=0$ is also a solution  for either $\{\alpha_1,\alpha_2\}$ or $\{\alpha_3,\alpha_4\}$ non-vanishing. Recall that these results also correspond to the action of BPS M2-branes via \eqref{ttaubetaM2}.

Next we want to obtain the full distribution $\rho$ by solving the system \eqref{systemtoric}. We find solutions for $\{q_{14},q_{15},q_{45},q_{23},q_{25},q_{35}\}$. We do not report the explicit expressions of the vertices as they are not particularly insightful. They have a similar structure to the ones presented in the examples of section~\ref{sec:density}. Further imposing the inequalities \eqref{toricineq}, we find that the first three solutions are the vertices of $\mathcal{Q}_\tau$ for $\tau\in[\tau_\tmin,0]$, while the last three define $\mathcal{Q}_\tau$ for $\tau\in[0,\tau_\tmax]$, where we reproduce the expression of $\tau_\tmin$, $\tau_\tmax$ in \eqref{tauUN}. $\mathcal{Q}_\tau$ is therefore a triangle in both regions, and its area can be computed via the determinant \eqref{determinant} first for $\{q_{14},q_{15},q_{45}\}$ and then for $\{q_{23},q_{25},q_{35}\}$, see Figure~\ref{fig:rhoUNgeom}. Finally taking the $\Lambda$ derivative to obtain the perimeter and normalizing by the appropriate factor \eqref{rhonormalization} gives the geometrical density 
\begin{equation}
	\rho(t) = \frac{1}{\ell_p^3N^{1/2}}\left.\begin{cases}
		\dfrac{2\Lambda+(R_2+R_3)\tau}{\pi^2(R_1+R_2)(R_3+R_4)R_5}\,, \quad \tau_\tmin<\tau<0\\[10pt]
		\dfrac{2\Lambda-(R_1+R_4)\tau}{\pi^2(R_1+R_2)(R_3+R_4)R_5}\,, \quad 0<\tau<\tau_\tmax
	\end{cases}\right|_{\tau=\ell_p^3N^{1/2}t}.
\end{equation}

\paragraph{Field theory}
\begin{figure}
	\begin{center}
		\begin{tikzpicture}[scale=1.2,every node/.style={scale=1.2},font=\scriptsize]
			\node[gauge] (t0) at (0,0) {$\, N\,$};
			\node[flavor] (t1) at (2.2,1.8) {$1$};
			\node[flavor] (t2) at (2.2,-1.8) {$1$};
			\draw[->] (t0) edge [out=80,in=180-35,bend left] node[anchor=east,yshift=3pt] {$Q^{(1)}$} (t1);
			\draw[<-] (t0) edge [out=35,in=180+35,bend right] 
			node[anchor=west,yshift=-3pt] {$\tilde Q^{(1)}$}(t1);
			\draw[->] (t0) edge [out=-35,in=180-35,bend left] node[anchor=west,yshift=3pt] {$Q^{(2)}$} (t2);
			\draw[<-] (t0) edge [out=-80,in=180+35,bend right] 
			node[anchor=east,yshift=-3pt] {$\tilde Q^{(2)}$}(t2);
			\draw[<<<-] (t0) edge [out=180-35,in=180+35,loop,looseness=9.5] node[anchor=north,xshift=12pt,yshift=27pt] {$\Phi_{1,2,3}$} (t0);
		\end{tikzpicture}
	\end{center}
	\vspace{-.5cm}
	\caption{Quiver diagram of the flavoured U$(N)$ theory.}
	\label{fig:flavUN}
\end{figure}
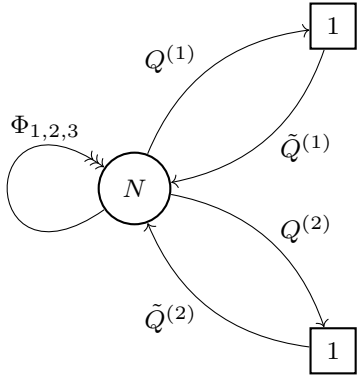

The dual field theory to this geometry is a flavoured $U(N)$ theory which contains two pairs of fundamental chiral fields $\{Q^{(1)},\tilde Q^{(1)}\}$, $\{Q^{(2)},\tilde Q^{(2)}\}$ and three adjoint chiral fields $\Phi_i$, $i=1,2,3$, see Figure \ref{fig:flavUN}, and has superpotential
\begin{equation}
	W = \mathrm{Tr}\left[\Phi_1[\Phi_2,\Phi_3]+\tilde Q^{(1)} \Phi_1 Q^{(1)}+\tilde Q^{(2)} \Phi_2 Q^{(2)}\right] \, ,
\end{equation}
such that
\begin{equation}
	T\tilde T = \Phi_1\Phi_2 \implies \Delta_T+\Delta_{\tilde T}=\Delta_{\Phi_1}+\Delta_{\Phi_2}\,.
\end{equation}
The results can be derived from the Bethe potential given in Appendix~\ref{app:fieldth}, and read \cite{Hosseini:2016ume}
\begin{equation}
	t_\tmin = -\frac{\mu}{\Delta_{\tilde T}}\, , \qquad 
	t_\tmax = \frac{\mu}{\Delta_T}\, ,
\end{equation}
\begin{equation}
	\rho(t)=\begin{cases}
		\dfrac{\mu+\Delta_{\tilde T} t}{\pi^2\Delta_{\Phi_1}\Delta_{\Phi_2}\Delta_{\Phi_3}}\,, \qquad t_\tmin<t<0\,, \\[10pt]
		\dfrac{\mu-\Delta_T t}{\pi^2\Delta_{\Phi_1}\Delta_{\Phi_2}\Delta_{\Phi_3}}\,, \qquad 0<t<t_\tmax\,,
	\end{cases},
\end{equation}
with 
\begin{equation}
	\mu = \pi\sqrt{\frac{2\Delta_{\Phi_1}\Delta_{\Phi_2}\Delta_{\Phi_3}\Delta_T\Delta_{\tilde T}}{\Delta_{\Phi_T}+\Delta_{\Phi_{\tilde T}}}}\, .
\end{equation}

\paragraph{Matching} The field theory and geometrical parameters are related as  \cite{Gauntlett:2019roi}
\begin{equation}
	\begin{aligned}
	\Delta_{\Phi_1}=R_1+R_2\,,\quad \Delta_{\Phi_2}=R_3+R_4\,, \quad \Delta_{\Phi_3} = R_5\,, \\ \Delta_T = R_1+R_4\,, \quad \Delta_{\tilde T}=R_2+R_3\,.
	\end{aligned}
\end{equation}
As expected, under this identification, $\mu=2\Lambda/(\ell_p^3N^{1/2})$.
It is then easy to see that the discontinuity points of $\rho'(t)$ match, and also that we reproduce $\rho(t)$.

\subsubsection{Flavoured ABJM}

\begin{figure}
	\centering
	\begin{tikzpicture}
		
		\begin{scope}[fill opacity=.5]
			\filldraw[color=Green, fill=Green!20, very thick] (4.5,1.66)--(4.5+.58*5.5-.2*.1625,1.66-.58*2+.2*3.25)--(4.5+.58*5.5,1.66-.58*2)-- (4.5,1.66) node[midway, left] {};	
			\filldraw[color=blue, fill=blue!20, very thick] (2.2,1)--(2.2+.82*5.5,1-.82*2)--(2.2+.82*5.5-.9*.1625,1-.82*2+.9*3.25)--(2.2+.82*5.5-.9*.1625-.825*2.865,1-.82*2+.9*3.25+.825*.5)--(2.2,1) node[midway, left] {};
			\filldraw[color=Red, fill=Red!20, very thick] (.7,.7)--(.7+.845*5.5,.7-.845*2)--(.7+.845*5.5-1.48*.1625,.7-.845*2+1.48*3.25)-- (.7+.845*5.5-1.48*.1625-.24*2.865,.7-.845*2+1.48*3.25+.24*.5)--(.7,.7) node[midway, left] {};			
		\end{scope}
		
		\draw[thick,dashed] (0,0) -- (4,1.8) -- (8,.8);
		\draw[thick,dashed] (4,1.8) -- (4.5,4.5);
		\draw[thick] (4.5,4.5)--(5.5,-2);
		\draw[thick] (0,0)--(4.5,4.5)--(8,.8)--(5.5,-2)--(0,0);
		\filldraw[black] (0,0) circle (2pt) node[anchor=east]{$p_{145}$};
		\filldraw[black] (8,.8) circle (2pt) node[anchor=west]{$p_{235}$};
		\filldraw[black] (4.5,4.5) circle (2pt) node[anchor=south ]{$p_{1234}$};
		\filldraw[black] (5.5,-2) circle (2pt) node[anchor=north ]{$p_{125}$};	
		\filldraw[black] (4,1.8) circle (2pt) node[anchor=south west]{$p_{345}$};	
		
		\filldraw[Green] (4.5,1.66) circle (2pt) node[anchor=north]{$q_{35}$};
		\filldraw[Green](4.5+.58*5.5-.2*.1625,1.66-.58*2+.2*3.25) circle (2pt) node[anchor=south west]{$q_{23}$};		
		\filldraw[Green]  (4.5+.58*5.5,1.66-.58*2) circle (2pt) node[anchor=north west]{$q_{25}$};		
		\filldraw[blue] (2.2,1) circle (2pt) node[anchor=south east]{$q_{45}$};
		\filldraw[blue]
		(2.2+.82*5.5,1-.82*2) circle (2pt) node[anchor=north west]{$q_{25}$};
		\filldraw[blue] (2.2+.82*5.5-.9*.1625,1-.82*2+.9*3.25) circle (2pt) node[anchor=south west]{$q_{23}$};
		\filldraw[blue] (2.2+.82*5.5-.9*.1625-.825*2.865,1-.82*2+.9*3.25+.825*.5) circle (2pt) node[anchor= east]{$q_{34}$};
		\filldraw[red] (.7,.7) circle (2pt) node[anchor=south east]{$q_{14}$};
		\filldraw[red] (.7+.845*5.5,.7-.845*2) circle (2pt) node[anchor=west]{$q_{12}$};	
		\filldraw[red] (.7+.845*5.5-1.48*.1625,.7-.845*2+1.48*3.25) circle (2pt) node[anchor= south west ]{$q_{23}$};	
		\filldraw[white]  (.7+.845*5.5-1.48*.1625-.24*2.865-.5,.7-.845*2+1.48*3.25+.24*.5)  circle (5pt);
		\filldraw[red]  (.7+.845*5.5-1.48*.1625-.24*2.865,.7-.845*2+1.48*3.25+.24*.5)  circle (2pt) node[anchor=  east]{$q_{34}$};		
		
	\end{tikzpicture}
	\caption{Pyramid describing $\C\times C(T^{1,1})$ and its slicing corresponding to flavoured ABJM. The red slice is $\mathcal{Q}_\tau$ for $\tau_1 < \tau < \tau_2$, the blue slice for $\tau_2 < \tau < \tau_3$, and the green slice for $\tau_3 <\tau < \tau_4$. Each vertex $p_{abc}$ denotes the intersection of the facets $\mathcal{P}_a$, $\mathcal{P}_b$, $\mathcal{P}_c$, and the Reeb hyperplane. Note that we do not draw the slice at $\tau=\tau_2=0$, where $q_{14}=q_{45}=p_{145}$ and $q_{12}=q_{25}=p_{125}$. The perimeter of $\mathcal{Q}_\tau$ varies with $\tau$ qualitatively as $\rho$ in Figure~\ref{fig:rhoflavABJM}.}
	\label{fig:rhoflavABJMgeom} 
\end{figure}
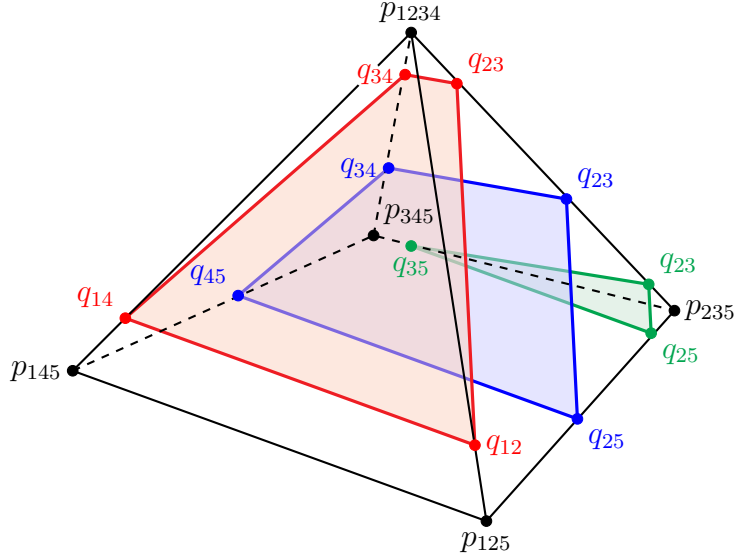

\paragraph{Geometry} 

A second possible choice of $U(1)_M$ is to have weights $(1,0,0,0,-1)$ on $z_a$, which using the toric data \eqref{toricCxC} reads on the cone
\begin{equation}\label{zetaMflavABJM}
	\zeta_M = - \partial_{\varphi_2}\,.
\end{equation}
In the following we assume $R_4<R_2$. Solving \eqref{betatoric} for this choice of M-theory circle gives four solutions (recall $\tau=\Lambda\beta$)
\begin{equation}\label{tauflavABJM}
	\tau_1 = -\frac{2\Lambda}{R_5} \,, \quad \tau_2=0\,, \quad \tau_3 = \frac{2\Lambda}{R_1+R_2}\,, \quad  \tau_4 = \frac{2\Lambda}{R_1+R_4}\,,
\end{equation}
and associated non-zero $\alpha_a$, $\{\alpha_1,\alpha_2,\alpha_3,\alpha_4\}$, $\{\alpha_1,\alpha_5\}$,
$\{\alpha_3,\alpha_4,\alpha_5\}$ $\{\alpha_2,\alpha_3,\alpha_5\}$,  respectively, as clearly appears in Figure \ref{fig:rhoflavABJMgeom}. Physically these results correspond to the value of the action of BPS probe M2-branes.  

To reproduce $\rho$ we turn to the system \eqref{systemtoric}, which in this case admits solutions for any pair of neighbouring $v_a$, except for $q_{15}$. The M-theory hyperplane therefore slices the toric polytope parallel to the edge between $\mathcal{P}_1$ and $\mathcal{P}_5$ as shown in Figure~\ref{fig:rhoflavABJMgeom}. Then from the inequalities \eqref{toricineq}, we find that $\mathcal{Q}_\tau$ has vertices $\{q_{12},q_{23},q_{34},q_{14}\}$, $\{q_{25},q_{23},q_{34},q_{45}\}$, $\{q_{25},q_{23},q_{35}\}$ in the regions $[\tau_1,\tau_2]$, $[\tau_2,\tau_3]$, $[\tau_3,\tau_4]$ respectively, where we recover the values of $\tau$ from \eqref{tauflavABJM}. In the regions where $\mathcal{Q}_\tau$ is quadrilateral the area is computed by splitting into two triangles, and summing the two areas computed via the determinant \eqref{determinant}. After taking the derivative and normalizing properly, see \eqref{rhonormalization}, we deduce that the density is given by
\begin{equation}
	\rho(t) = \frac{1}{ \ell_p^3  N^{1/2}}\left.\begin{cases}
			\frac{4\Lambda + 2 R_5 \tau}{\pi^2(R_2+R_3)(R_3+R_4)(R_1+R_5+R_2)(R_1+R_5+R_4)}\,, \quad {\scriptstyle \tau_1 < \tau <\tau_2}\\[6pt]
			\frac{4\Lambda - (2R_1+R_2R_4 - R_3(R_1+R_5)) \tau}{\pi^2(R_2+R_3)(R_3+R_4)(R_1+R_5+R_2)(R_1+R_5+R_4)}\,, \quad {\scriptstyle\tau_2 < \tau <\tau_3}\\[6pt]
			\frac{2\Lambda - (R_1+R_4) \tau}{\pi^2(R_2-R_4)(R_4+R_3)(R_1+R_5+R_4)}\,, \qquad\qquad\quad\:\: {\scriptstyle\tau_3 < \tau < \tau_4}
		\end{cases}\hspace{-.2cm}\right|_{\tau = \ell_p^3  N^{1/2} \, t}.
\end{equation}

\paragraph{Field theory}

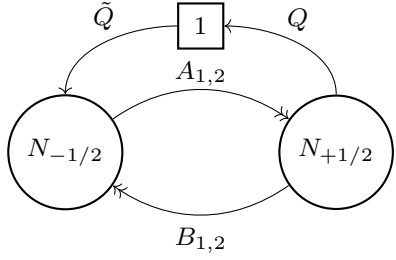
\begin{figure}
	\begin{center}
		\begin{tikzpicture}[scale=1.2,every node/.style={scale=1.2},font=\scriptsize]
			\node[gauge] (t0) at (0,0) {$N_{-1/2}$};
			\node[gauge] (t1) at (3,0) {$N_{+1/2}$};
			\node[flavor] (t2) at (1.5,1.4) {$1$};
			\draw[->>] (t0) edge [out=35,in=180-35,loop,looseness=1] node[anchor=north,yshift=13pt] {$A_{1,2}$} (t1);
			\draw[<<-] (t0) edge [out=-35,in=180+35,loop,looseness=1] 
			node[anchor=south,yshift=-16pt] {$B_{1,2}$}(t1);
			\draw[<-] (t0) edge [out=90,in=180,loop,looseness=.9] node[anchor=north,yshift=17pt] {$\tilde Q$} (t2);
			\draw[<-] (t2) edge [out=0,in=90,loop,looseness=.9] 
			node[anchor=north,yshift=15pt] {$Q$}(t1);
		\end{tikzpicture}
	\end{center}
	\vspace{-.5cm}
	\caption{Quiver diagram for the flavoured ABJM theory.}
	\label{fig:flavABJM}
\end{figure}

The field theory dual to $X_4 = \C \times C(T^{1,1})$ equipped with the M-theory circle \eqref{zetaMflavABJM} is an instance of the flavoured ABJM theory. That is the \MII\, with an additional pair of fundamental chiral fields $(Q,\tilde Q)$ and with Chern-Simons level $(-1/2,1/2)$ \cite{Benini:2009qs}.\footnote{The half-integer level is needed to cancel the parity anomaly due to the additional pair of chiral fields. Note that we exchange $k$ and $-k$ compared to \cite{Benini:2009qs}, according to the prescription in table 1 of \cite{Hosseini:2016tor}.} The quiver diagram is shown in Figure \ref{fig:flavABJM}. The superpotential is the ABJM superpotential, plus a deformation term
\begin{equation}\label{flavABJMsuper}
	W = \mathrm{Tr}\left[A_1B_1A_2B_2-A_1B_2A_2B_1 + Q A_1 \tilde{Q}\right] ,
\end{equation}
\begin{equation}
	T\tilde T = A_1 \implies \Delta_T+\Delta_{\tilde T}=\Delta_{A_1}\,.
\end{equation}
The results are derived in Appendix~\ref{app:fieldth}, assuming $\Delta_{B_1}<\Delta_{B_2}$:
\begin{equation}
	\begin{aligned}
		t_1 &= -\frac{\mu}{\Delta_{\tilde T} }\:, \qquad\qquad\: t_2 = 0 \:, \\
		t_3 &= \frac{\mu}{\Delta_{T}  + \Delta_{B_2}} \:, \qquad t_4 = \frac{\mu}{\Delta_{T} + \Delta_{B_1}} \:. 
	\end{aligned}
\end{equation}
\begin{equation}
	\rho(t)=\resizebox{0.8\hsize}{!}{$\begin{dcases} 
			\frac{2\mu +2\Delta_{\tilde T}t }{\pi^2 (\Delta_{A_1} + \Delta_{B_1})(\Delta_{A_1} + \Delta_{B_2})(\Delta_{A_2} + \Delta_{B_1})(\Delta_{A_2} + \Delta_{B_2})}
			 \:, \quad t_1< t < t_2 \:, \\[4pt]
			\frac{2\mu -(2\Delta_{T}+\Delta_{B_1}\Delta_{B_2} - \Delta_{A_1}\Delta_{A_2})t}{\pi^2 (\Delta_{A_1} + \Delta_{B_1})(\Delta_{A_1} + \Delta_{B_2})(\Delta_{A_2} + \Delta_{B_1})(\Delta_{A_2} + \Delta_{B_2})} \:, \quad t_2< t < t_3\,,  \\[4pt] \frac{ \mu -(\Delta_{T}+\Delta_{B_1})t}{\pi^2 (\Delta_{A_1}+\Delta_{B_1}) (\Delta_{A_2}+\Delta_{B_1}) (\Delta_{B_2}-\Delta_{B_1})}
			 \:, \qquad\qquad\qquad t_3< t < t_4 \:.
		\end{dcases}
		$ }
\end{equation}
\begin{equation}
	\mu=\pi\sqrt{\frac{2(\Delta_{A_2}+\Delta_{B_1})(\Delta_{A_2}+\Delta_{B_2})(\Delta_{T}+\Delta_{B_1})(\Delta_{T}+\Delta_{B_2})\Delta_{\tilde T}}{2-\Delta_{\tilde T} }}\, .
\end{equation}

\paragraph{Matching} Finally, the parameters on both sides are related by \cite{Hosseini:2019ddy}\footnote{Note we used a different toric basis from \cite{Hosseini:2019ddy}. In particular, we have $R_1 = \Delta_1^\text{there}$, $R_2 = \Delta_3^\text{there}$, $R_3 = \Delta_5^\text{there}$, $R_4 = \Delta_2^\text{there}$, $R_5 = \Delta_4^\text{there}$.} 
\begin{equation}
	\begin{aligned}
	\Delta_{A_1}=R_1+R_5 \:, \quad \Delta_{A_2}=R_3 \:, \quad \Delta_{B_1}= R_4 \:, \quad \Delta_{B_2}=R_2 \:, \\
	\Delta_T = R_1\:, \quad \Delta_{\tilde T}=R_5 \:,
	\end{aligned}
\end{equation}
such that again, $\mu=2\Lambda/(\ell_p^3 N^{1/2})$, $t$ and $\tau$ match and we exactly reproduce $\rho$ in the geometry. Note that the orderings we assumed on each side, namely $R_4<R_2$ and $\Delta_{B_1}<\Delta_{B_2}$, are equivalent.

\subsection{Example: $Y_7=Q^{1,1,1}/\Z_n$}

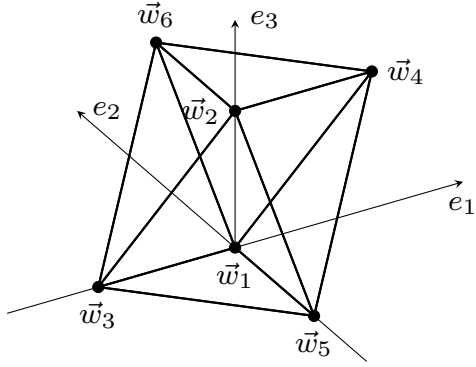
\begin{figure}
	\begin{center}
		\tdplotsetmaincoords{60}{-30}
		\begin{tikzpicture}
			[tdplot_main_coords,scale=1.4,every node/.style={scale=1.4},font=\scriptsize]
			
			\draw[-stealth] (-2.5,0,0) -- (2.5,0,0) node[anchor=north]{$e_1$};
			\draw[-stealth] (0,-2.5,0) -- (0,3,0) node[anchor=west]{$e_2$};
			\draw[-stealth] (0,0,0) -- (0,0,2.5) node[anchor=west]{$e_3$};
			
			\filldraw[black] (0,0,0) circle (1.5pt) node[anchor=north]{$\vec{w}_1$};
			\filldraw[black] (0,0,1.5) circle (1.5pt) node[anchor=east]{$\vec{w}_2$};
			\filldraw[black] (-1.5,0,0) circle (1.5pt) node[anchor=north]{$\vec{w}_3$};
			\filldraw[black] (1.5,0,1.5) circle (1.5pt) node[anchor=west]{$\vec{w}_4$};
			\filldraw[black] (0,-1.5,0) circle (1.5pt) node[anchor=north]{$\vec{w}_5$};
			\filldraw[black] (0,1.5,1.5) circle (1.5pt) node[anchor=south]{$\vec{w}_6$};
			
			\draw[thick] (0,0,0) -- (-1.5,0,0) -- (0,1.5,1.5) -- cycle;
			\draw[thick] (0,0,0) -- (1.5,0,1.5) -- (0,1.5,1.5) -- cycle;
			\draw[thick] (0,0,0) -- (-1.5,0,0) -- (0,-1.5,0) -- cycle;
			\draw[thick] (0,0,0) -- (1.5,0,1.5) -- (0,-1.5,0) -- cycle;
			\draw[thick] (0,0,1.5) -- (-1.5,0,0) -- (0,1.5,1.5) -- cycle;
			\draw[thick] (0,0,1.5) -- (1.5,0,1.5) -- (0,1.5,1.5) -- cycle;
			\draw[thick] (0,0,1.5) -- (-1.5,0,0) -- (0,-1.5,0) -- cycle;
			\draw[thick] (0,0,1.5) -- (1.5,0,1.5) -- (0,-1.5,0) -- cycle;
			
		\end{tikzpicture}
	\end{center}
	\vspace{-.5cm}
	\caption{Toric diagram for $Q^{1,1,1}$.}
	\label{fig:Q111}
\end{figure}

A second example of a toric cone is 
$Y_7=Q^{1,1,1}/\Z_n$. Here we denote the quotient by $\Z_n$, rather than 
$\Z_k$, since in this case the positive integer $n$ is the number of flavours 
(D6-branes), rather than a Chern-Simons level (RR two-form flux).
The toric data are given by the following, with the toric diagram shown in Figure~\ref{fig:Q111}:
\begin{equation}\label{toricQ111}
	\begin{split}
		v_1 &= (1,0,0,0) \:, \quad v_2 = (1,0,0,1) \:,  \quad 
		v_3 = (1,-1,0,0) \:, \quad \\ v_4 &= (1,1,0,1) \:, \quad v_5 = (1,0,-1,0) \:, \quad v_6 = (1,0,1,1)\:.
	\end{split} 
\end{equation}
We will consider the simplifying ansatz of \cite{Gauntlett:2019roi} for the Reeb vector
\begin{align}\label{bansatzQ111}
	b=(1,0,b_3,1/2+b_3)\:.
\end{align} 
In this case we have
\begin{equation}\label{volQ111}
	\Vol_S(Q^{1,1,1}/\Z_n) = \frac{32\pi^4}{3 n}\frac{3-4b_3^2}{(1-4b_3^2)^2} \:.
\end{equation}
The expressions for the geometrical R-charges in terms of $b_3$ can be found in \cite{Gauntlett:2019roi}. We do not need them here and only notice that their positivity imposes $|b_3|\leq1/2$. Finally we record that 
\begin{equation}
	\Lambda=\frac{\ell_p^3N^{1/2}\pi}{2}\sqrt{n}\,\frac{1-4b_3^2}{\sqrt{3-4b_3^2}}\,.
\end{equation}

\subsubsection{Flavoured ABJM}

\begin{figure}
	\centering
	\begin{tikzpicture}
		\filldraw[color=red, fill=red!10, very thick] (2.5,0)--(2.5,4)-- (2.5-.4*2.5,4+.4*2)--(2.5-.4*2.5,0+.4*2)--(2.5,0) node[midway, left] {};	
		\filldraw[color=blue, fill=blue!10, very thick] (3.5,2)--(3.5,6)--(6,4)--(6,0)--(3.5,2) node[midway, left] {};	
		\filldraw[color=Green, fill=Green!10, very thick] (7.,2)--(7.,6)-- (7.+.4*2.5,6-.4*2)--(7.+.4*2.5,2-.4*2)--(7.,2) node[midway, left] {};	
		
		\draw[thick] (0,0) -- (0,4) -- (6,4)--(6,0)--(0,0);
		\draw[thick,dashed] (0,0)-- (3.5,2) -- (3.5,6)--(3.5,2)--(9.5,2);
		\draw[thick] (3.5,6) -- (9.5,6)--(9.5,2);
		\draw[thick] (0,4) -- (3.5,6); 
		\draw[thick] (6,4) -- (9.5,6); 
		\draw[thick] (6,0) -- (9.5,2); 		
		\filldraw[black] (0,0) circle (2pt) node[anchor=east]{$p_{236}$};
		\filldraw[black] (0,4) circle (2pt) node[anchor=east]{$p_{246}$};
		\filldraw[black] (6,4) circle (2pt) node[anchor= north west]{$p_{245}$};
		\filldraw[black] (6,0) circle (2pt) node[anchor= north west]{$p_{235}$};	
		\filldraw[black] (3.5,2) circle (2pt) node[anchor=south west]{$p_{136}$};	
		\filldraw[black] (3.5,6) circle (2pt) node[anchor=south west]{$p_{146}$};			
		\filldraw[black] (9.5,6) circle (2pt) node[anchor=south west]{$p_{145}$};			
		\filldraw[black] (9.5,2) circle (2pt) node[anchor=south west]{$p_{135}$};			
		
		\filldraw[red] (2.5,0) circle (2pt) node[anchor=north west]{$q_{23}$};
		\filldraw[red] (2.5,4) circle (2pt) node[anchor=south west]{$q_{24}$};		
		\filldraw[red] (2.5-.4*2.5,4+.4*2) circle (2pt) node[anchor=south east]{$q_{46}$};	
		\filldraw[red] (2.5-.4*2.5,0+.4*2) circle (2pt) node[anchor=south east]{$q_{36}$};	
		\filldraw[blue] (3.5,2) circle (2pt) node[anchor=south east]{};
		\filldraw[blue] (3.5,6) circle (2pt) node[anchor=south east]{};
		\filldraw[blue] (6,4) circle (2pt) node[anchor=north west]{};
		\filldraw[blue] (6,0) circle (2pt) node[anchor=north west]{};
		\filldraw[Green] (7,2) circle (2pt) node[anchor=south east]{$q_{13}$};
		\filldraw[Green] (7,6) circle (2pt) node[anchor=south]{$q_{14}$};	
		\filldraw[Green] (7+.4*2.5,6-.4*2) circle (2pt) node[anchor=north west ]{$q_{45}$};	
		\filldraw[Green] (7+.4*2.5,2-.4*2) circle (2pt) node[anchor=north west ]{$q_{35}$};	
	\end{tikzpicture}
	\caption{Polytope describing $Q^{1,1,1}$ and its slicing corresponding to flavoured ABJM. The red slice is $\mathcal{Q}_\tau$ for $\tau_{\text{min}} < \tau <0$, the blue slice for $\tau = 0$, and the green slice for $0 <\tau < \tau_{\text{max}} $.  Each vertex $p_{abc}$ denotes the intersection of the facets $\mathcal{P}_a$, $\mathcal{P}_b$, $\mathcal{P}_c$, and the Reeb hyperplane. Note that $\mathcal{Q}_\tau$ is not 0 at $\tau_{\tmin,\tmax}$ but an edge of the polytope.}
	\label{fig:rhoflavABJM2geom} 
\end{figure}
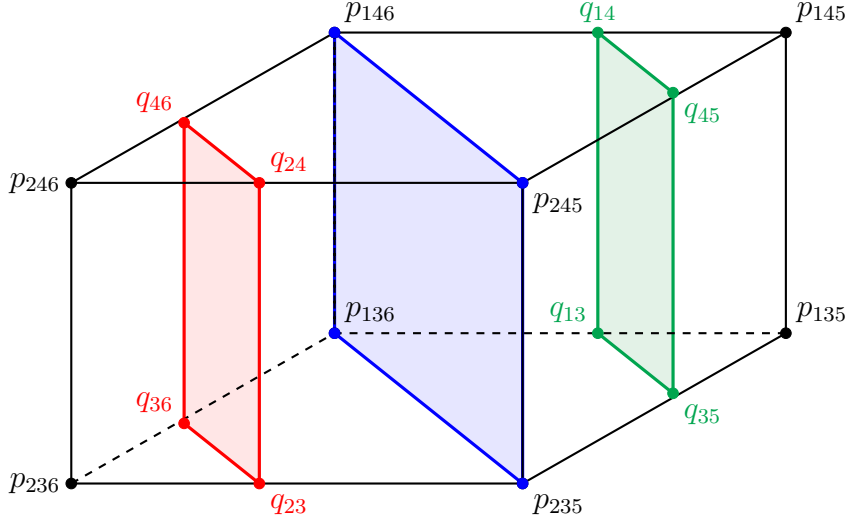

\paragraph{Geometry} We pick a $U(1)_M$ action with weights $(-1,0,0,0,0,1)$ on the coordinates of the ambient space, which using the toric data \eqref{toricQ111} reads on $Q^{1,1,1}/\Z_n$ 
\begin{equation}
	\zeta_M=\frac{1}{n}\,(\del_{\varphi_3}+\del_{\varphi_4})\,.
\end{equation}
Then the equation \eqref{betatoric} admits two\footnote{Actually for general $b$ there are four solutions, which degenerate to these two under the ansatz \eqref{bansatzQ111}.} non-trivial solutions for $\beta=\tau/\Lambda$ (corresponding to BPS M2-brane actions) with non-vanishing $\{\alpha_2,\alpha_6\}$ and $\{\alpha_1,\alpha_5\}$ respectively, which read
\begin{equation}
	\tau_\tmin = -\frac{2\Lambda}{n-2nb_3}\,,\quad 	\tau_\tmax = \frac{2\Lambda}{n+2nb_3}\,.
\end{equation}
It also admits $\beta=0$ as a solution with either $\{\alpha_2,\alpha_5\}$ or $\{\alpha_1,\alpha_6\}$ non-zero, as can be observed in Figure~\ref{fig:rhoflavABJM2geom}.

The system \eqref{systemtoric} has solutions $\{q_{23},q_{24},q_{46},q_{36}\}$ and $\{q_{13},q_{35},q_{45},q_{14}\}$ which, according to the inequalities \eqref{toricineq} are the vertices of $\mathcal{Q}_\tau$ for $\tau$ in $[\tau_\tmin,0]$ and $[0,\tau_\tmax]$ respectively (see Figure \ref{fig:rhoflavABJM2geom}), reproducing $\tau_{\tmin,\tmax}$ given above. The volume of $\mathcal{Q}_\tau$ is computed on each region by splitting the quadrilateral into two triangles whose areas are the determinant \eqref{determinant}. Taking the derivative with respect to $\Lambda$ and normalizing as in \eqref{rhonormalization}, we eventually obtain
\begin{equation}
	\rho(t) = \frac{1}{\ell_p^3N^{1/2}}\left.\begin{cases}
		\dfrac{4\Lambda+(n-4nb_3)\tau}{\pi^2}\,, \quad \tau_\tmin<\tau<0 \\[8pt]
		\dfrac{4\Lambda-(n+4nb_3)\tau}{\pi^2} \,,\quad 0<\tau<\tau_\tmax
	\end{cases}\right|_{\tau=\ell_p^3N^{1/2}t}.
\end{equation}

\paragraph{Field Theory}

\begin{figure}
	\begin{center}
		\begin{tikzpicture}[scale=1.2,every node/.style={scale=1.2},font=\scriptsize]
			\node[gauge] (t0) at (0,0) {$N_{}$};
			\node[gauge] (t1) at (3,0) {$N_{}$};
			\node[flavor] (t2) at (1.5,1.4) {$n$};
			\node[flavor] (t3) at (1.5,2.2) {$n$};
			\draw[->>] (t0) edge [out=35,in=180-35,loop,looseness=1] node[anchor=north,yshift=13pt] {$A_{1,2}$} (t1);
			\draw[<<-] (t0) edge [out=-35,in=180+35,loop,looseness=1] 
			node[anchor=south,yshift=-16pt] {$B_{1,2}$}(t1);
			\draw[<-] (t0) edge [out=90-20,in=180,loop,looseness=.9] node[anchor=north,yshift=16pt] {$\tilde Q^{(1)}$} (t2);
			\draw[<-] (t2) edge [out=0,in=90+20,loop,looseness=.9] 
			node[anchor=north,yshift=19pt] {$Q^{(1)}$}(t1);
			\draw[<-] (t0) edge [out=90,in=180,loop,looseness=1] node[anchor=north,yshift=20pt] {$\tilde Q^{(2)}$} (t3);
			\draw[<-] (t3) edge [out=0,in=90,loop,looseness=1] 
			node[anchor=north,yshift=24pt] {$Q^{(2)}$}(t1);
		\end{tikzpicture}
	\end{center}
	\vspace{-.5cm}
	\caption{Quiver diagram for the flavoured ABJM theory.}
	\label{fig:flavABJM2}
\end{figure}
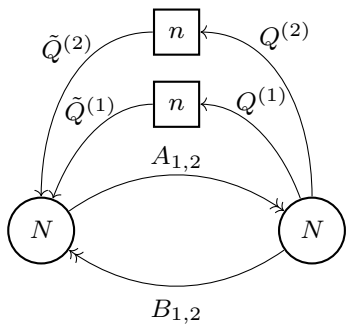

The field theory dual is a flavoured ABJM theory with two sets of pairs of fundamental chiral fields $\{Q^{(1)}_j,\tilde Q^{(1)}_j\}$, $\{Q^{(2)}_j,\tilde Q^{(2)}_j\}$, $j=1,\dots,n$, and Chern-Simons level 0, see Figure \ref{fig:flavABJM2}. It has the superpotential
\begin{equation}
	W = \mathrm{Tr}\left[A_1B_1A_2B_2-A_1B_2A_2B_1 + \sum_{j=1}^n Q^{(1)}_j A_1 \tilde{Q}^{(1)}_j+ \sum_{j=1}^n Q^{(2)}_j A_2 \tilde{Q}^{(2)}_j\right] ,
\end{equation}
and the monopole operators obey  the relations
\begin{equation}
	T\tilde T = A_1^n A_2^n \implies \Delta_T+\Delta_{\tilde T}=n(\Delta_{A_1}+\Delta_{A_2})\,.
\end{equation}
In terms of these variables the chemical potential for the topological symmetry is 
\begin{equation}
	\Delta_m
	=- \frac{1}{2}(\Delta_T-\Delta_{\tilde T})\,.
\end{equation}
The results of interest have been obtained in \cite{Hosseini:2016ume}, as explained in Appendix~\ref{app:fieldth}, and read\footnote{Fixing a typo in the sign of  $\Delta_m$.} 
\begin{equation}
	t_\tmin = - \frac{\mu}{n+\Delta_m}\, , \quad 
	t_\tmax =  \frac{\mu}{n-\Delta_m}\, ,
\end{equation}
\begin{equation}
	\rho(t)=\begin{cases}
		\dfrac{2\mu+(n+2\Delta_m) t}{\pi^2}\,, \qquad t_\tmin<t<0\,, \\[10pt]
		\dfrac{2\mu-(n-2\Delta_m) t}{\pi^2}\,, \qquad 0<t<t_\tmax\,,
	\end{cases}
\end{equation}
with
\begin{equation}
	\mu = \frac{\pi}{\sqrt{n}}\frac{|n^2-\Delta_m^2|}{\sqrt{3n^2-\Delta_m^2}}\, .
\end{equation}
Note that $\rho(t_{\tmin,\tmax})\neq 0$, in accordance with the geometric picture.

\paragraph{Matching} There is one parameter on each side that are related via \cite{Gauntlett:2019roi}
\begin{equation}
	-\Delta_m = 2 n b_3\, .
\end{equation}
Again under this identification $\mu=2\Lambda/(\ell_p^3 N^{1/2})$ and $\rho(t)$ is correctly reproduced in the geometry. An interesting observation is that the postivity of the geometrical R-charges, which implies that $|b_3|\leq 1/2$ gives the constraint $|\Delta_m|\leq n$ (such that we could drop the absolute value in $\mu$). 

\section{Discussion}\label{sec:discussion}

In this paper we have further studied the relationship 
between a class of supersymmetric, magnetically charged
(and potentially accelerating) black holes in M-theory, and 
their dual matrix model description. The matching 
between entropy functions 
on the two sides (for classes of examples) had already been established 
in  \cite{Gauntlett:2019roi, Hosseini:2019ddy}. More recently,
 a gravitational block formula for the entropy function of accelerating black holes
\eqref{entropy_block} was proven in \cite{Boido:2022iye, Boido:2022mbe} 
(see also \cite{Martelli:2023oqk}). One observation we have made in this 
paper is that the gravity formula \eqref{entropy_block} in GK geometry may already 
be matched to the refined twisted index field theory results in 
\cite{Hosseini:2022vho}, on setting $m_+=m_-=1$ so that the 
spindle horizon becomes simply $S^2$. Similarly, 
the actions of BPS probe M2-branes at the poles of this $S^2$
then also match with points where the eigenvalue density derivatives $\rho'_\pm$ for the 
two blocks in the matrix model are discontinuous. These results extend straightforwardly 
to the accelerating case with a spindle on the gravity side, and it will be fascinating to 
match the large $N$ limit of the spindle index of \cite{Inglese:2023wky} to the gravitational 
results in GK geometry we have presented here.

Perhaps the most interesting open question is to further 
match quantities in the matrix model to quantities in gravity. 
For example, it seems likely that there is also a simple geometric 
characterization of  the fluxes 
$\mathfrak{s}(t)$. But one also expects a more direct holographic 
interpretation of these variables, and of the density function $\rho(t)$. 
The analogous quantity on $S^3$ was interpreted in 
 \cite{Gulotta:2011aa} in terms of counting 
operators in the chiral ring of the gauge theory. 
However, the present paper perhaps suggests a 
more direct physical interpretation on the gravity side. 
Specifically, the near-horizon solutions are sourced by $N$ M2-branes, 
where in AdS/CFT the brane sources disappear and are replaced 
by $N$ units of dissolved flux over the internal space. 
On the other hand, 
in type IIA we have
$N$ D2-branes, where $h_M=\tau=\ell_p N^{1/2} t$ is the location of the D2-branes 
(before the back-reaction and near-horizon limit). This 
suggests that the eigenvalue density $\rho(t)$ is also playing the role 
of a density function directly in the gravity dual, describing how the dissolved 
D2/M2-branes are distributed in the internal space. 
We leave this speculation, and other questions, for future work.

\section*{Acknowledgements}

\noindent 
We thank Nikolay Bobev, Pieter Bomans, Matteo Sacchi, and Seyed Morteza Hosseini for helpful discussions. 
This work was supported in part by STFC grant 
ST/T000864/1. 
AL is supported by the Palmer scholarship in Mathematical Physics of Merton college. 

 \appendix

\section{Bethe potential at large $N$}\label{app:fieldth}

In this appendix we review the expression for the Bethe potential of topologically twisted Chern-Simons-matter theories at large $N$, derived in \cite{Hosseini:2016tor}. We focus on flavoured $U(N)$ and ABJM theories, reproducing and completing the discussion of \cite{Hosseini:2016ume} (and fixing some minor typographical errors). 

At leading order in $N$ the Bethe potential consists of contributions coming from the various gauge and matter content of the theory\footnote{There are additional bifundamental contributions to the following equation that are subleading in $N$. Their role is discussed in \cite{Benini:2015eyy}, but we do not need to consider them in this work.}
\begin{equation}
	\mathcal{U}=\mathcal{U}^\textrm{Gauge}+
	\mathcal{U}^\textrm{Bifund}+\mathcal{U}^\textrm{Flavour}\,.
\end{equation}
For theories with gauge group $\mathcal{G}=\prod_{a=1}^\gauge U(N)_{k_a}$ with $\sum_{a=1}^\gauge k_a=0$ such that they have an M-theory dual, 
\begin{equation}
	\mathcal{U}^\textrm{Gauge}=-\ii N^{3/2}\int\dd t \, t\, \rho(t)\sum_{a=1}^{\gauge} \left(k_a v_a(t)+\pi\Delta_m^{(a)}\right)\,,
\end{equation}
where $\Delta_m^{(a)}$ is the chemical potential of the topological symmetry of $U(N)_{k_a}$. 
We write $\Delta_m=\sum_a \Delta_m^{(a)}$ in terms of  variables $\Delta_T$, $\Delta_{\tilde{T}}$, identified with the R-charges on $S^3$
of the diagonal monopole operators, via
\begin{align}\label{RT}
	\Delta_m =-\frac{1}{2}(\Delta_{T} - \Delta_{\tilde{T}} )\, .
\end{align}
The general expressions of $\mathcal{U}^\textrm{Bifund}$ and $\mathcal{U}^\textrm{Flavour}$ are given in \cite{Hosseini:2016tor}. Here we want to state their simplification (together with that for $\mathcal{U}^\textrm{Gauge}$) in the special cases of $\mathcal{G}=U(N)$ and $\mathcal{G}=U(N)_k\times U(N)_{-k}$, from which all the examples of the present paper can be recovered.

\subsection{Flavoured $U(N)$ theories}\label{app:flavUN}

\begin{figure}
	\begin{center}
		\begin{tikzpicture}[scale=1.2,every node/.style={scale=1.2},font=\scriptsize]
			\node[gauge] (t0) at (0,0) {$\, N\,$};
			\node[flavor] (t1) at (0,2.5) {$n_1$};
			\node[flavor] (t2) at (2.5,0) {$n_2$};
			\node[flavor] (t3) at (0,-2.5){$n_3$};
			\draw[->] (t0) edge [out=80,in=180-35,bend left] node[anchor=east,yshift=0pt] {$Q^{(1)}_j$} (t1);
			\draw[<-] (t0) edge [out=35,in=180+35,bend right] 
			node[anchor=west,yshift=0pt] {$\tilde Q^{(1)}_j$}(t1);
			\draw[->] (t0) edge [out=-35,in=180-35,bend left] node[anchor=south,yshift=0pt] {$Q^{(2)}_j$} (t2);
			\draw[<-] (t0) edge [out=-80,in=180+35,bend right] 
			node[anchor=north,yshift=0pt] {$\tilde Q^{(2)}_j$}(t2);
			\draw[->] (t0) edge [out=-35,in=180-35,bend left] node[anchor=west,yshift=0pt] {$Q^{(3)}_j$} (t3);
			\draw[<-] (t0) edge [out=-80,in=180+35,bend right] 
			node[anchor=east,yshift=0pt] {$\tilde Q^{(3)}_j$}(t3);
			\draw[<<<-] (t0) edge [out=180-35,in=180+35,loop,looseness=9.5] node[anchor=north,xshift=12pt,yshift=27pt] {$\Phi_{1,2,3}$} (t0);
		\end{tikzpicture}
	\end{center}
	\vspace{-.5cm}
	\caption{Quiver diagram for the flavoured $U(N)$ theory.}
	\label{fig:flavUNgen}
\end{figure}
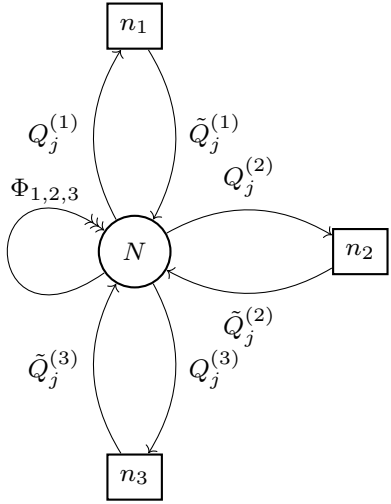

A general flavoured $U(N)$ theory contains three sets of pairs of fundamental chiral fields $\{Q_j^{(i)},\tilde Q_j^{(i)}\}_{j=1}^{n_i}$, and three adjoint chiral fields $\Phi_i$, $i=1,2,3$, see Figure \ref{fig:flavUNgen}, and has superpotential
\begin{equation}\label{superpotentialUN}
	W = \textrm{Tr} \left[ \Phi_1\left[\Phi_2,\Phi_3\right]
	+ \sum_{j=1}^{n_1} Q_j^{(1)} \Phi_1 \tilde Q_{j}^{(1)}
	+ \sum_{j=1}^{n_2} Q_j^{(2)} \Phi_2 \tilde Q_{j}^{(2)}
	+ \sum_{j=1}^{n_3} Q_j^{(3)} \Phi_3 \tilde Q_{j}^{(3)}\right]\, .
\end{equation}
The chemical potentials satisfy  constraints coming from the superpotential
\begin{align}
	\sum_{i=1}^3 \Delta_{\Phi_i} = 2\, , \quad \Delta_{Q^{(i)}_j}+\Delta_{\tilde{Q}^{(i)}_j}+\Delta_{\Phi_i}=2\,,
\end{align}
while the monopole operators obey
\begin{equation}\label{deltaTUN}
	T\tilde T = \Phi_1^{n_1}\Phi_2^{n_2}\Phi_3^{n_3} \quad \implies \quad \Delta_{T}+\Delta_{\tilde{T}} = \sum_{i=1}^3 n_i\Delta_{\Phi_i}\, .
\end{equation}

We have 
\begin{equation}
	\begin{aligned}
	\mathcal{U}^\textrm{Gauge}&=-\ii N^{3/2}\int\dd t \, t\, \rho(t)\, \pi\Delta_m\, , \\
	\mathcal{U}^\textrm{Bifund}&=\ii N^{3/2}\int\dd t\, \rho(t)^2 \sum_{i=1}^3 g_+(\pi\Delta_{\Phi_i})\, , \\
	\mathcal{U}^\textrm{Flavour}&=\ii N^{3/2}\int\dd t\,\frac{1}{2}\, |t|\, \rho(t) \sum_{i=1}^3 n_i\,\pi\Delta_{\Phi_i}\,,
\end{aligned}
\end{equation}
where
\begin{equation}
	g_+(v) = \frac{v^3}6 - \frac\pi2 v^2 + \frac{\pi^2}3 v \, .
\end{equation}

Extremizing this potential with respect to $\rho$, adding a Lagrange multiplier guarantying its normalization 
\begin{equation}
	\frac{\mathcal{U}}{\ii N^{3/2}}-\pi\mu\left(\int\dd t\,\rho(t)-1\right)\,,
\end{equation}
gives
\begin{equation}
	t_\tmin = -\frac{\mu}{\Delta_{\tilde T}}\, , \qquad 
	t_\tmax = \frac{\mu}{\Delta_T}\, ,
\end{equation}
\begin{equation}
	\rho(t)=\begin{cases}
		\dfrac{\mu+\Delta_{\tilde T} t}{\pi^2\Delta_{\Phi_1}\Delta_{\Phi_2}\Delta_{\Phi_3}}\,, \qquad t_\tmin<t<0\,, \\[10pt]
		\dfrac{\mu-\Delta_T t}{\pi^2\Delta_{\Phi_1}\Delta_{\Phi_2}\Delta_{\Phi_3}}\,, \qquad 0<t<t_\tmax\,,
	\end{cases},
\end{equation}
with 
\begin{equation}
	\mu = \pi\sqrt{\frac{2\Delta_{\Phi_1}\Delta_{\Phi_2}\Delta_{\Phi_3}\Delta_T\Delta_{\tilde T}}{\Delta_{T}+\Delta_{\tilde T}}}\, ,
\end{equation}
where $t_{\tmin,\tmax}$ are obtained as the solutions of $\rho(t)=0$, and $\mu$ imposing $\int_{t_\tmin}^{t_\tmax} \dd t\, \rho(t)=1$.
Note that these results only depend on the flavours through the constraint \eqref{deltaTUN}, while the density takes the exact same form for any flavoured $U(N)$ theory. 
This is one motivation for introducing the variables $\Delta_{T,\tilde T}$.

In this paper we consider the ADHM theory ({\it i.e.} Model I, which is dual to $S^7/\Z_k$) corresponding to $n_1=n_2=0$, $n_3=k$, and the flavoured $U(N)$ dual to $\C\times C(T^{1,1})$ corresponding to $n_1=n_2=1$, $n_3=0$.

\subsection{Flavoured ABJM theories}\label{app:flavABJM}

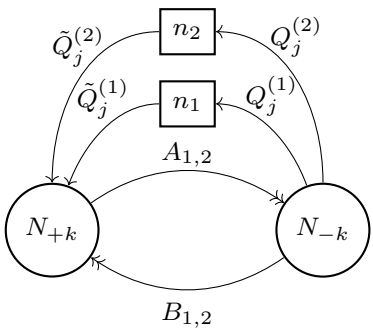
\begin{figure}
	\begin{center}
		\begin{tikzpicture}[scale=1.2,every node/.style={scale=1.2},font=\scriptsize]
			\node[gauge] (t0) at (0,0) {$N_{+k}$};
			\node[gauge] (t1) at (3,0) {$N_{-k}$};
			\node[flavor] (t2) at (1.5,1.4) {$n_1$};
			\node[flavor] (t3) at (1.5,2.2) {$n_2$};
			\draw[->>] (t0) edge [out=35,in=180-35,loop,looseness=1] node[anchor=north,yshift=13pt] {$A_{1,2}$} (t1);
			\draw[<<-] (t0) edge [out=-35,in=180+35,loop,looseness=1] 
			node[anchor=south,yshift=-16pt] {$B_{1,2}$}(t1);
			\draw[<-] (t0) edge [out=90-20,in=180,loop,looseness=.9] node[anchor=north,yshift=20pt] {$\tilde Q^{(1)}_j$} (t2);
			\draw[<-] (t2) edge [out=0,in=90+20,loop,looseness=.9] 
			node[anchor=north,yshift=21pt] {$Q^{(1)}_j$}(t1);
			\draw[<-] (t0) edge [out=90,in=180,loop,looseness=1] node[anchor=north,yshift=21pt] {$\tilde Q^{(2)}_j$} (t3);
			\draw[<-] (t3) edge [out=0,in=90,loop,looseness=1] 
			node[anchor=north,yshift=24pt] {$Q^{(2)}_j$}(t1);
		\end{tikzpicture}
	\end{center}
	\vspace{-.5cm}
	\caption{Quiver diagram for the flavoured ABJM theory.}
	\label{fig:flavABJMgen}
\end{figure}

Flavoured ABJM theories have $U(N)_{+k}\times U(N)_{-k}$ gauge group, 
contain two sets of pairs of fundamental chiral fields $\{Q_j^{(i)},\tilde Q_j^{(i)}\}_{j=1}^{n_i}$ and bifundemental chiral fields $A_i$, $B_i$ ($i=1,2$), transforming in the  $(\mathbf{N},\overline{\mathbf{N}})$, $(\overline{\mathbf{N}},\mathbf{N})$ representations respectively, and have superpotential
\begin{equation}
	\begin{aligned}
	W = \mathrm{Tr}\left[A_1B_1A_2B_2-A_1B_2A_2B_1 +  \sum_{j=1}^{n_1} Q^{(1)}_j A_1 \tilde{Q}^{(1)}_j+ \sum_{j=1}^{n_2} Q^{(2)}_j A_2 \tilde{Q}^{(2)}_j\right] \,.
	\end{aligned}
\end{equation}
We could consider flavouring $B_1$ and $B_2$ as well and a similar analysis follows, but we are not interested in spelling out this more general case here.
The chemical potentials
satisfy the superpotential constraints
\begin{equation}
	\begin{aligned}
		\Delta_{A_1}+\Delta_{B_1}+\Delta_{A_2}+\Delta_{B_2} = 2 \, , \quad \Delta_{Q^{(a)}_j}+\Delta_{\tilde{Q}^{(a)}_j}+\Delta_{A_a}=2\,,
	\end{aligned}
\end{equation}
while the monopole operators obey
\begin{equation}
	T\tilde T = A_1^{n_1}A_2^{n_2}\quad \implies \quad \Delta_T+\Delta_{\tilde T}= n_1\Delta_{A_1}+n_2\Delta_{A_2}\,. 
\end{equation}

The two gauge groups have opposite Chern-Simons levels
$(k_1,k_2)=(k,-k)$, and we define the following combinations
\begin{equation}
	\delta v(t)\equiv v_2(t)-v_1(t) \,,\quad  \Delta_m = \Delta_m^{(1)}+\Delta_m^{(2)}\,, 
\end{equation}
such that\footnote{Compared to \cite{Hosseini:2016ume}, we add the $k$ term and define $\Delta_m$ slightly differently.} 
\begin{equation}\label{UflavABJM}
	\begin{aligned}
		\mathcal{U}^\textrm{Gauge}&=\ii N^{3/2}\int\dd t \, t \,\rho(t)\,\big( k\, \delta v(t)-\pi\Delta_m\big) \: , \\
		\mathcal{U}^\textrm{Bifund}&=\ii N^{3/2}\int\dd t\, \rho(t)^2\sum_{a=1,2}\Big[g_+(\delta v(t)+\pi\Delta_{B_a}) -g_-(\delta v(t)-\pi\Delta_{A_a})\Big]\, , \\
		\mathcal{U}^\textrm{Flavour}&=\ii N^{3/2}\int\dd t\,\frac{1}{2}\, |t|\, \rho(t)\big[n_1\left(\pi\Delta_{A_1}-\delta v(t))+n_2(\pi\Delta_{A_2}-\delta v(t)\right) \big]\, ,
	\end{aligned}
\end{equation}
where
\begin{equation}
	g_\pm(v) = \frac{v^3}6 \mp \frac\pi2 v^2 + \frac{\pi^2}3 v \, .
\end{equation}
We do not present the solution for extremizing this potential for general flavours. Instead in the next subsection we turn to a specific example, which has not been treated in the literature. 

The cases of interest for this paper are the pure ABJM theory (Model II), which has no flavours turned on ($n_1=n_2=0$) and Chern-Simons level $k$. In this case, the symmetries of the theories allow to set $\Delta_m^{(1)}=\Delta_m^{(2)}=0$ without loss of generality \cite{Benini:2015eyy}. Then the instance of the flavoured ABJM dual to $\C\times C(T^{1,1})$ has $n_1=1$, $n_2=0$, and $k=-1/2$. Finally, the flavoured ABJM dual to $Q^{1,1,1}/\Z_n$ has $n_1=n_2=n$, and $k=0$.

\subsubsection{Flavoured ABJM dual to $\C\times C(T^{1,1})$}

In the following we present the computations to obtain the eigenvalue density of the flavoured ABJM field theory dual to the $\C\times C(T^{1,1})$ geometry. This computation is interesting to present both because it produces new results, and because it is representative of how the results have been obtained in all of the examples. Indeed, this theory includes both flavours and non-zero Chern-Simons levels. We therefore expect a behaviour that combines the features from Model I and II, {\it i.e.}\ both a discontinuity point of $\rho'(t)$ at $t=0$ coming from the D6-branes, and some $t\neq0$ and $\rho\neq0$ discontinuity points. 

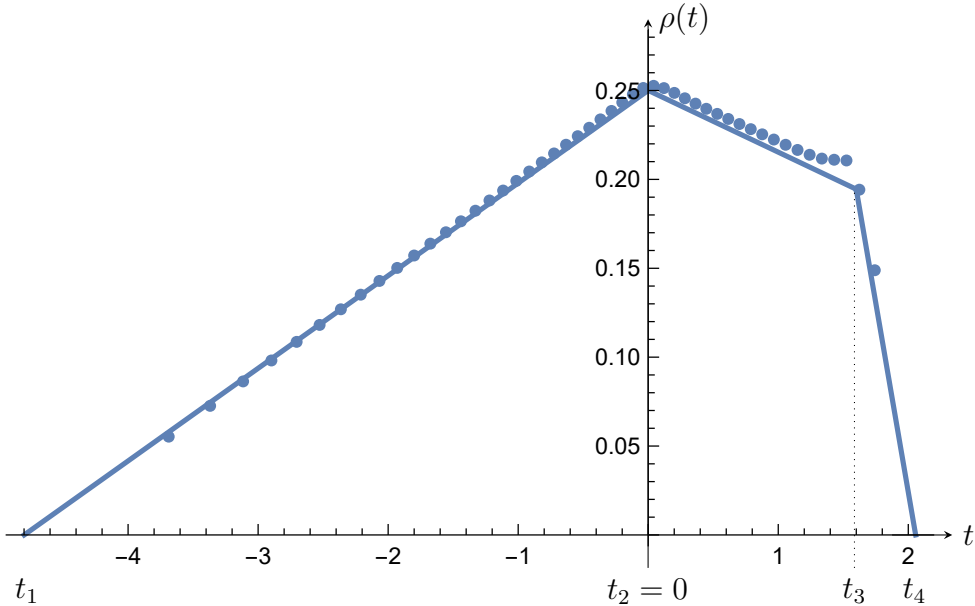
\begin{figure}
\begin{tikzpicture}
	\centering
	\node[anchor=south west,inner sep=0] at (-0.3,0) {\includegraphics[width=.8\linewidth]{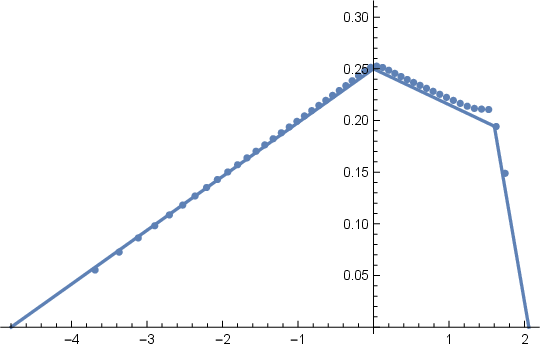}};
	\filldraw[white] (7.95,7.6) circle (15pt);
	\filldraw[white] (-1.,7.5) circle (15pt);
	\draw[dotted] (11,0.) -- (11,5.);
	\draw[-stealth] (8.2565,0) -- (8.2565,7.3) node[anchor=west] {$\rho(t)$};
	\draw[thin,-stealth] (11.5,.438) -- (12.3,.438) node[anchor=west] {$t$};
	\draw	(0,0) node[anchor=north] {$t_1$}
	(8.25,0) node[anchor=north] {$t_2=0$}
	(11,0) node[anchor=north] {$t_3$}
	(11.8,0) node[anchor=north] {$t_4$};
\end{tikzpicture}
\caption{Comparison of the analytical solution in \eqref{rhoflavABJM} (solid line) with the numerics (dots), for $\Delta_{A_1}=1/2$, $\Delta_{A_2}=2/3$, $\Delta_{B_1}=1/3$, $\Delta_{B_2}=1/2$, $\Delta_m=0$.}
\label{fig:rhoflavABJM}
\end{figure}

As explained in section~\ref{sec:largeN} the analytical expression for $\rho(t)$ is obtained by extremizing the Bethe potential, adding a Lagrange multiplier term
\begin{equation}
	\frac{\mathcal{U}}{\ii N^{3/2}}-\pi\mu\left(\int\dd t\,\rho(t)-1\right)\,.
\end{equation}
For the model we consider, the Bethe potential is given by \eqref{UflavABJM}, with $k=-1/2$, $n_1=1$, $n_2=0$. Extremizing this potential gives solutions for $\rho$ and $\delta v$, which are valid for\footnote{These conditions come from the derivation of $\mathcal{U}^\textrm{Bifund}$, see \cite{Hosseini:2016tor}.} 
\begin{align}
		0 &< \delta v(t) + \pi\Delta_{B_a} < 2\pi \:, \qquad   -2\pi < \delta v(t) - \pi\Delta_{A_a} < 0 \:, \qquad  a=1,2\:.
	\end{align}
As $\sum\Delta=2$, the 0 bounds are always more restrictive than the $2\pi$ bounds, and we only need to consider the former. When $\delta v$ hits one of these bounds, it stays frozen at that value, creating some ``tails'' where $\rho$ is obtained again by extremizing $\mathcal{U}$, but with $\delta v$ set at its frozen value. For $\delta v$ extremizing \eqref{UflavABJM}, the bounds are saturated at
\begin{align}
		& \delta v(t_1) = \pi\Delta_{A_2}  \:, \quad\qquad  \delta v(t_\star) = \pi\Delta_{A_1}  \:,\nonumber \\
		& \delta v(t_3) = -\pi\Delta_{B_1} \:, \:\qquad\delta v(t_4) = -\pi\Delta_{B_2} \:,
	\end{align}
with
\begin{equation}
	\begin{aligned}
		t_1 &= -\frac{2\mu}{\Delta_{A_1}  + 2\Delta_m}\:, \quad\qquad\qquad t_\star = \frac{2\mu}{\Delta_{A_1}  - 2\Delta_m- 2\Delta_{A_2}} \:, \\
		t_3 &= \frac{2\mu}{\Delta_{A_1}  - 2\Delta_m+ 2\Delta_{B_2}} \:, \:\qquad t_4 = \frac{2\mu}{\Delta_{A_1}  - 2\Delta_m+ 2\Delta_{B_1}}  \:. 
	\end{aligned}
\end{equation}
By symmetry of the quiver (see Figure~\ref{fig:flavABJM}), we can assume without loss of generality that $\Delta_{B_1}<\Delta_{B_2}$, such that $t_1<0<t_3<t_4$ (where we have also assumed that $\Delta_{A_1}+2\Delta_m>0$). Therefore there is a tail on the right, on which $\delta v(t)=-\pi \Delta_{B_1}$. The extremal values of $t$ are determined by $\rho(t_\tmin)=\rho(t_\tmax)=0$, which correspond to $t_\tmin=t_1$ and $t_\tmax=t_4$.  
We discard the $t_\star$ solution, which lies outside of $[t_1,t_4]$ for $\Delta_{A_2}<\Delta_{A_1}$. For $\Delta_{A_2}>\Delta_{A_1}$, we do not have an {\it a priori} reason to discard $t_\star$, but we note that if it were a discontinuity point of $\rho'(t)$, it would give a tail on the left, which is not what we expect either from the numerics (see Figure~\ref{fig:rhoflavABJM}) or the gravity dual analysis.
We know that there is also a discontinuity point of $\rho'(t)$ at $t_2=0$ as D6-branes are present. This is confirmed by the presence of a $|t|$ term in the solution for $\rho(t)$, which in summary is
\begin{equation}\label{rhoflavABJM}
		\rho(t)=\resizebox{0.8\hsize}{!}{$\begin{dcases} 
\frac{4\mu -(-4\Delta_m+\Delta_{B_1}\Delta_{B_2} - \Delta_{A_1}\Delta_{A_2})t - (\Delta_{A_1} + \Delta_{B_1})(\Delta_{A_1} + \Delta_{B_2})|t|}{2\pi^2 (\Delta_{A_1} + \Delta_{B_1})(\Delta_{A_1} + \Delta_{B_2})(\Delta_{A_2} + \Delta_{B_1})(\Delta_{A_2} + \Delta_{B_2})} \:, \quad t_1< t < t_3 \:, \\
\frac{2 \mu -(\Delta_{A_1}-2 \Delta_{m}+2\Delta_{B_1})t}{2\pi^2 (\Delta_{A_1}+\Delta_{B_1}) (\Delta_{A_2}+\Delta_{B_1}) (\Delta_{B_2}-\Delta_{B_1})} \:, \qquad\qquad\qquad\qquad\qquad\quad t_3< t < t_4 \:.
\end{dcases}
$ }
\end{equation}
Finally $\mu$ is determined by imposing the normalization of $\rho$ and reads
\begin{equation}
	\resizebox{0.9\hsize}{!}{$\mu=\pi\sqrt{\frac{(\Delta_{A_2}+\Delta_{B_1})(\Delta_{A_2}+\Delta_{B_2})(\Delta_{A_1}+2\Delta_m)(\Delta_{A_1}-2\Delta_m+2\Delta_{B_1})(\Delta_{A_1}-2\Delta_m+2\Delta_{B_2})}{8-4\Delta_m-2\Delta_{A_1} }}\,$ .}
\end{equation}
Note that both $\Delta_1$ and $\Delta_m$ are related to the monopole variables as $\Delta_1=\Delta_T+\Delta_{\tilde T}$ and $-\Delta_m=\frac{1}{2}(\Delta_T+\Delta_{\tilde T})$, such that 
\begin{equation}
	\Delta_{A_1}-2\Delta_m=2\Delta_T\,,\quad \Delta_{A_1}+2\Delta_m=2\Delta_{\tilde T}\,.
\end{equation}
In the main text, we quote the above results in terms of $\Delta_T$ and $\Delta_{\tilde T}$. This simplifies both the expressions themselves and their comparison with the geometry results.

We also checked our results numerically following the method of \cite{Benini:2015eyy}. A near-perfect agreement can be observed in Figure \ref{fig:rhoflavABJM}.

\section{Example: $Y_7=S^7/\Z_k$ -- Summary of results}\label{app:summary}

The field theories dual to $S^7/\Z_k$, namely the ADHM and ABJM theories, were used to illustrate the concepts from section~\ref{sec:fieldth} to section~\ref{sec:matching}. The aim of this appendix is to gather the results that have appeared scattered in the main text, and present them in the same fashion as in the examples of section~\ref{sec:examples}. Note that $C(S^7)=\C^4$ is also an example of a toric cone, but the geometry in this case is 
so elementary that the results could be obtained as in section~\ref{sec:matching}, without resorting to the toric machinery introduced in section~\ref{sec:examples}. Here instead we use the general toric framework to present the results. 

\begin{figure}
	\begin{center}
		\tdplotsetmaincoords{60}{150}
		\begin{tikzpicture}
			[tdplot_main_coords,scale=1.4,every node/.style={scale=1.4},font=\scriptsize]
			
			\draw[-stealth] (0,0,0) -- (2.5,0,0) node[anchor=north]{$e_1$};
			\draw[-stealth] (0,0,0) -- (0,2.5,0) node[anchor=west]{$e_2$};
			\draw[-stealth] (0,0,0) -- (0,0,2.5) node[anchor=west]{$e_3$};
			
			\filldraw[black] (0,0,0) circle (1.5pt) node[anchor=north]{$\vec{w}_1$};
			\filldraw[black] (1.5,0,0) circle (1.5pt) node[anchor=north]{$\vec{w}_2$};
			\filldraw[black] (0,1.5,0) circle (1.5pt) node[anchor=west]{$\vec{w}_3$};
			\filldraw[black] (0,0,1.5) circle (1.5pt) node[anchor=west]{$\vec{w}_4$};

			\draw[thick] (0,0,0) -- (0,0,1.5) -- (1.5,0,0) -- cycle;
			\draw[thick] (0,0,0) -- (0,1.5,0) -- (1.5,0,0) -- cycle;
			\draw[thick] (0,0,0) -- (0,0,1.5) -- (0,1.5,0) -- cycle;
			
		\end{tikzpicture}
	\end{center}
	\vspace{-.5cm}
	\caption{Toric diagram for $S^7$.} 
	\label{fig:toricS7}
\end{figure}
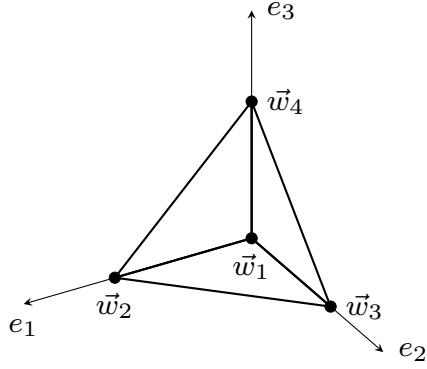
$S^7/\Z_k$ has the following toric data,
\begin{equation}
	\begin{split}
		v_1=\left(1,0,0,0\right),\quad v_2=\left(1,1,0,0\right),\quad v_3=\left(1,0,1,0\right),\quad v_4=\left(1,0,0,1\right)\, ,
	\end{split}
\end{equation}
with toric diagram shown in Figure~\ref{fig:toricS7}. We have
\begin{equation}
	\Vol_S(S^7/\Z_k) = \frac{1}{k}\, \frac{\pi^4}{3b_2\,b_3\,b_4\,(b_1-b_2-b_3-b_4)}\: ,
\end{equation}
\begin{equation}
R_1=\frac{2}{b_1}\left(b_1-b_2-b_3-b_4\right), \qquad\qquad  R_a=\frac{2 b_a}{b_1}\:, \quad a=2,3,4 \:,
\end{equation}
such that
\begin{equation}
	\Lambda = \frac{\pi \ell_p^3\, N^{1/2}}{2} \, \sqrt{2k R_1 R_2 R_3 R_4} \:.
\end{equation}

\subsection{ADHM -- Model I}

\begin{figure}
	\centering
	\begin{tikzpicture}
		\filldraw[color=red, fill=red!10, very thick] (2,.2)--(2.25,2.2)-- (2.75,-1)--(2,.2) node[midway, left] {};	
		\filldraw[color=blue, fill=blue!10, very thick] (4,.4)--(4.5,4.45)--(5.5,-2)--(4,.4) node[near end, left] {};	
		\filldraw[color=Green, fill=Green!10, very thick] (6,.6)--(6.25,2.6)-- (6.75,-.6)--(6,.6) node[midway, left] {};	
		
		\draw[thick,dashed] (0,0) -- (8,.8);
		\draw[thick] (4.5,4.5)--(5.5,-2);
		\draw[thick] (0,0)--(4.5,4.5)--(8,.8)--(5.5,-2)--(0,0);
		\filldraw[black] (0,0) circle (2pt) node[anchor=east]{$p_{134}$};
		\filldraw[black] (8,.8) circle (2pt) node[anchor=west]{$p_{234}$};
		\filldraw[black] (4.5,4.5) circle (2pt) node[anchor=south]{$p_{124}$};
		\filldraw[black] (5.5,-2) circle (2pt) node[anchor=north]{$p_{123}$};	
		
		\filldraw[red] (2,.2) circle (2pt) node[anchor=south east]{$q_{34}$};
		\filldraw[red] (2.25,2.25) circle (2pt) node[anchor=east]{$q_{14}$};		
		\filldraw[red] (2.75,-1) circle (2pt) node[anchor=north]{$q_{13}$};		
		\filldraw[blue] (4,.4) circle (2pt) node[anchor=south east]{};
		\filldraw[blue] (4.5,4.5) circle (2pt) node[anchor=south west]{};
		\filldraw[blue] (5.5,-2) circle (2pt) node[anchor=north west]{};
		\filldraw[Green] (6,.6) circle (2pt) node[anchor=south east]{$q_{34}$};
		\filldraw[Green] (6.25,2.65) circle (2pt) node[anchor=west]{$q_{24}$};	
		\filldraw[Green] (6.75,-.6) circle (2pt) node[anchor= west ]{$q_{23}$};	
	\end{tikzpicture}
	\caption{Tetrahedron describing $S^7/\Z_k$ and its slicing corresponding to the ADHM theory. The red slice is $\mathcal{Q}_\tau$ for $\tau_{\text{min}} < \tau <0$,  the blue slice for $\tau = 0$, and the green slice for $0 <\tau < \tau_{\text{max}} $. The perimeter of $\mathcal{Q}_\tau$ varies with $\tau$ qualitatively as $\rho$ in Figure~\ref{fig:rhoADHM}.}
	\label{fig:rhoADHMtoric} 
\end{figure}

\paragraph{Geometry}
We first consider
\begin{equation}
	\zeta_M =  \frac{1}{k}\,(\del_{\phi_1}-\del_{\phi_2})= - \frac{1}{k}\,\partial_{\varphi_2}\,,
\end{equation}
leading to 
\begin{equation}
	\tau_\tmin = -\frac{2\Lambda}{R_2+R_3}\,, \qquad \tau_\tmax = \frac{2\Lambda}{R_1+R_4}\,,
\end{equation}
and
\begin{equation}
	\rho(t) = \frac{1}{\ell_p^3N^{1/2}}\left.\begin{cases}
		\dfrac{2\Lambda+k R_2\tau}{\pi^2R_3R_4(R_1+R_2)}\,, \quad \tau_\tmin<\tau<0\\[10pt]
		\dfrac{2\Lambda-kR_1\tau}{\pi^2R_3R_4(R_1+R_2)}\,, \quad 0<\tau<\tau_\tmax
	\end{cases}\right|_{\tau=\ell_p^3N^{1/2}t},
\end{equation}
as shown in Figure \ref{fig:rhoADHMtoric}.

\paragraph{Field theory}

The field theory dual to this geometry is the ADHM  theory, which corresponds to the instance $n_1=n_2=0$, $n_3=k$ of the flavoured $U(N)$ theory described in Appendix~\ref{app:flavUN}. The results read
\begin{equation}
	t_\tmin = -\frac{\mu}{\Delta_{\tilde T}}\, , \qquad 
	t_\tmax = \frac{\mu}{\Delta_T}\, ,
\end{equation}
\begin{equation}
	\rho(t)=\begin{cases}
		\dfrac{\mu+\Delta_{\tilde T} t}{\pi^2\Delta_{\Phi_1}\Delta_{\Phi_2}\Delta_{\Phi_3}}\,, \qquad t_\tmin<t<0\,, \\[10pt]
		\dfrac{\mu-\Delta_T t}{\pi^2\Delta_{\Phi_1}\Delta_{\Phi_2}\Delta_{\Phi_3}}\,, \qquad 0<t<t_\tmax\,,
	\end{cases},
\end{equation}
with 
\begin{align}
	\mu  = \pi\sqrt{\frac{2}{k}\Delta_{\Phi_1}\Delta_{\Phi_2}\Delta_T\Delta_{\tilde{T}}}\, .
\end{align}

\paragraph{Matching} 	
\begin{align}
	\Delta_{T} = kR_1\, , \quad \Delta_{\tilde{T}} = kR_2\, , \quad 
	\Delta_{\Phi_1} = R_3\, , \quad \Delta_{\Phi_2} = R_4\, ,
\end{align}
such that, $\mu=2\Lambda/(\ell_p^3N^{1/2})$, and $\rho(t)$ match.

\subsection{ABJM -- Model II}

\begin{figure}
	\centering
	\begin{tikzpicture}
		
		\filldraw[color=red, fill=red!10, very thick] (2,.2)--(2-2*0.2,.2+2*0.7)-- (2+.8*2.7,.2-.8*2.1)--(2,.2) node[near end, below] {};	
		\filldraw[color=blue, fill=blue!10, very thick] (4,.4)--(4-4*0.2,.4+4*0.7)--(4-4*0.2+.73*2.5,.4+4*0.7-.73*2.8)--(4+.8*2.7,.4-.8*2.1)--(4,.4) node[near end, below] {};	
		\filldraw[color=Green, fill=Green!10, very thick] (6,.6)--(6-4.75*0.2,.6+4.75*0.7)-- (6+.4*2.7,.6-.4*2.1)--(6,.6) node[near end, below] {};	
		
		\draw[thick,dashed] (0,0) -- (8,.8);
		\draw[thick] (4.5,4.5)--(5.5,-2);
		\draw[thick] (0,0)--(4.5,4.5)--(8,.8)--(5.5,-2)--(0,0);
		\filldraw[black] (0,0) circle (2pt) node[anchor=east]{$p_{134}$};
		\filldraw[black] (8,.8) circle (2pt) node[anchor=west]{$p_{234}$};
		\filldraw[black] (4.5,4.5) circle (2pt) node[anchor=south]{$p_{124}$};
		\filldraw[black] (5.5,-2) circle (2pt) node[anchor=north]{$p_{123}$};	
		
		\filldraw[red] (2,.2) circle (2pt) node[anchor=south east]{$q_{34}$};
		\filldraw[red] (2-2*0.2,.2+2*0.7) circle (2pt) node[anchor=east]{$q_{14}$};		
		\filldraw[red] (2+.8*2.7+.05,.2-.8*2.1-.05) circle (2pt) node[anchor=north]{$q_{13}$};		
		\filldraw[blue] (4,.4) circle (2pt) node[anchor=south east]{$q_{34}$};
		\filldraw[blue] (4-4*0.2,.4+4*0.7) circle (2pt) node[anchor=east]{$q_{14}$};		
		\filldraw[blue] (4-4*0.2+.73*2.5,.4+4*0.7-.73*2.8) circle (2pt) node[anchor=east]{$q_{12}$};
		\filldraw[blue] (4+.8*2.7,.4-.8*2.1) circle (2pt) node[anchor=west]{$q_{23}$};
		\filldraw[Green] (6,.6) circle (2pt) node[anchor=south east]{$q_{34}$};
		\filldraw[Green] (6-4.75*0.2,.6+4.75*0.7) circle (2pt) node[anchor=west]{$q_{24}$};	
		\filldraw[Green] (6+.4*2.7,.6-.4*2.1) circle (2pt) node[anchor= west ]{$q_{23}$};	
	\end{tikzpicture}
		\caption{Tetrahedron describing $S^7/\Z_k$ and its slicing corresponding to the ABJM theory. The red slice is $\mathcal{Q}_\tau$ for $\tau_1 < \tau < \tau_2$, the blue slice for $\tau_2 < \tau < \tau_3$, and the green slice for $\tau_3 <\tau < \tau_4$. The perimeter of $\mathcal{Q}_\tau$ varies with $\tau$ qualitatively as $\rho$ in Figure~\ref{fig:rhoABJM}.}
	\label{fig:rhoABJMtoric} 
\end{figure}
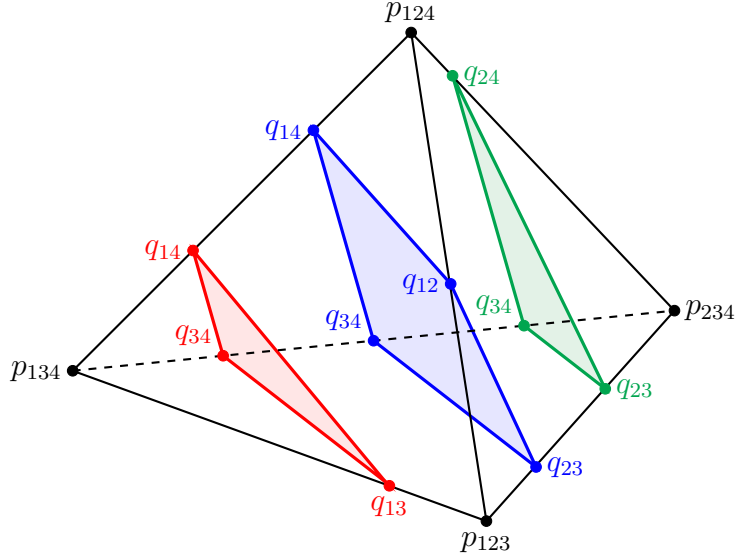

\paragraph{Geometry}
The second choice of M-theory circle is
\begin{equation}
	\zeta_M =   \frac{1}{k}\,(\del_{\phi_1}-\del_{\phi_2}+\del_{\phi_3}-\del_{\phi_4})=  \frac{1}{k}\,(-\partial_{\varphi_2}+\partial_{\varphi_3}-\partial_{\varphi_4})\,,
\end{equation}
leading to 
\begin{equation}
	\tau_1=-\frac{2\Lambda}{kR_2}\,,\quad
	\tau_2=-\frac{2\Lambda}{kR_4}\,,\quad
	\tau_3=\frac{2\Lambda}{kR_3}\,,\quad
	\tau_4=\frac{2\Lambda}{kR_1}\,,\quad
\end{equation}
and
\begin{equation}
	\rho(t) = \frac{1}{ \ell_p^3  N^{1/2}}\left.\begin{cases}
		\frac{2\Lambda + k R_2 \tau}{\pi^2(R_1+R_2)(R_2+R_3)(R_4-R_2)}\,, \,\qquad\quad {\scriptstyle\tau_1 < \tau <\tau_2}\\[4pt] 
		\frac{4\Lambda + k (R_2R_4 - R_1 R_3) \tau}{\pi^2(R_1+R_2)(R_2+R_3)(R_1+R_4)(R_3+R_4)}\,, \: {\scriptstyle\tau_2 < \tau <\tau_3}\\[4pt]
		\frac{2\Lambda - k R_1 \tau}{\pi^2(R_1+R_2)(R_1+R_4)(R_3-R_1)}\,, \qquad\quad\, {\scriptstyle\tau_3 < \tau < \tau_4}
	\end{cases}\hspace{-.2cm} \right|_{\tau = \ell_p^3  N^{1/2}t }
\end{equation}
as shown in Figure \ref{fig:rhoABJMtoric}.

\paragraph{Field theory}

The field theory dual to this geometry is the ABJM theory, which corresponds to the unflavoured instance ($n_1=n_2=0$) of the flavoured ABJM theory  described in Appendix~\ref{app:flavABJM}. The results read
\begin{align}
	t_1= -\frac{\mu}{k\Delta_{B_1}}\, , \quad 
	t_2 = -\frac{\mu}{k\Delta_{B_2}}\, , \quad t_3 = \frac{\mu}{k\Delta_{A_2}}\, , \quad \quad  t_4= \frac{\mu}{ k\Delta_{A_1}}\, ,
\end{align}
\begin{align}
	\rho(t) = \begin{cases} \displaystyle\frac{\mu +  k \Delta_{B_1}t}{\pi^2(\Delta_{A_1}+\Delta_{B_1})(\Delta_{A_2}+\Delta_{B_1})(\Delta_{B_2}-\Delta_{B_1})}\, , & t_1<t<t_2\, ,\\[12pt]
		\frac{2\mu + k (\Delta_{B_1}\Delta_{B_2}-\Delta_{A_1}\Delta_{A_2})t}{\pi^2(\Delta_{A_1}+\Delta_{B_1})(\Delta_{A_1}+\Delta_{B_2})(\Delta_{A_2}+\Delta_{B_1})(\Delta_{A_2}+\Delta_{B_2})}\, , &  t_2<t<t_3\, , \\[10pt]
		\displaystyle\frac{\mu - k \Delta_{A_1}t}{\pi^2(\Delta_{A_1}+\Delta_{B_1})(\Delta_{A_1}+\Delta_{B_2})(\Delta_{A_2}-\Delta_{A_1})}\, , & t_3<t<t_4\, .
	\end{cases}
\end{align}
with
\begin{align}\label{muII_app}
	\mu = \pi\sqrt{2 k\Delta_{A_1}\Delta_{A_2}\Delta_{B_1}\Delta_{B_2}}\, .
\end{align}

\paragraph{Matching} 	
	\begin{align}
	\Delta_{A_1} = R_1\, , \quad 
	\Delta_{B_1} = R_2\, , \quad 
	\Delta_{A_2} = R_3\, , \quad 
	\Delta_{B_2} = R_4\, ,
\end{align}
such that, $\mu=2\Lambda/(\ell_p^3N^{1/2})$, and $\rho(t)$ match.

\section{BPS condition for M2-branes}\label{app:bps}

In the following we denote with $M = (0,1,m)$ the 11d spacetime index and with $\mathsf{A} = (0,1,\mathsf{a})$ the 11d frame indices, and we introduce a local orthonormal frame $e^\mathsf{A}$ for the metric \eqref{11dmetric} such that $e^{10} = \ex^{-B/3}\,\eta$. If $\epsilon$ is the Killing spinor of the background, an M2-brane whose configuration is described by the embedding coordinates $X^M$ does not break supersymmetry if it is such that $\epsilon$ satisfies the following Euclidean projection condition \cite{Becker:1995kb} 
\begin{equation}\label{projection}
	\mathbb{P} \,\epsilon=0, \quad \text { with } \quad \mathbb{P} \equiv \frac{1}{2}\left(1-\frac{\ii}{3 !}\, \varepsilon^{i j k} \,\partial_i X^M \partial_j X^N \partial_k X^P\, \Gamma_{M N P}\right).
\end{equation}
Here $i$, $j$, $k$ are worldvolume indices, $\varepsilon^{ijk}$ is the 3d Levi-Civita tensor, and $\Gamma_{MNP}$ are antisymmetric products of the 11d gamma matrices in curved space, {\it i.e.}\ $\Gamma_M = \Gamma_\mathsf{A}\, e^\mathsf{A}_{\;\,M}$. The configuration of branes we are interested in is wrapping EAdS$_2$ ({\it i.e.}\ the (0,1) directions) and $S^1_\mathrm{M2}$, which we take to be a generic linear combination of the directions in $Y_9$, {\it i.e.}\footnote{Of course the fact that $S^1_\mathrm{M2}$ must be a circle will in principle put restrictions on the coefficients $c_\mathsf{a}$, but we will not need them as the projection condition \eqref{projection} is enough to fix all the coefficients.}
\begin{equation} \label{volS1M}
	\vol_{S^1_\mathrm{M2}} =\sum_{\mathsf{a}=2}^{10} c_\mathsf{a}\, e^\mathsf{a}\:, \qquad c_\mathsf{a} \in \R\:. 
\end{equation}
We choose a parametrization where 
\begin{equation}
	\begin{split}
		\partial_0 X^0 = \partial_1 X^1 = 1 \:, \quad (\partial_3 X^m)\, e^\mathsf{a}_{\;\,m} = c_\mathsf{a} \:, \\
		\partial_0 X^m = \partial_1 X^m = \partial_3 X^0 = \partial_3 X^1 = 0 \:.
	\end{split}
\end{equation}
Then the only non-zero contributions in the sum in $\mathbb{P}$ are those for which $(M, N, P) = (0,1,m)$ or permutations, yielding
\begin{equation}
	\mathbb{P} \,\epsilon=0, \quad \text { with } \quad	\mathbb{P} = \frac{1}{2}\left(1 - \ii\sum_{\mathsf{a}=2}^{10} c_\mathsf{a} \, \Gamma_{0}\Gamma_1 \Gamma_\mathsf{a} \right) .
\end{equation}
The next step is to decompose this projection on our AdS$_2\times Y_9$ background. Following the conventions of \cite{Gauntlett:2003di,Gauntlett:2007ts} but rotated to Euclidean signature, as those are compatible with our GK backgrounds,
we split the 11d gamma matrices as $\Gamma_\mathsf{A} = (\beta_0 \otimes \mathbb{I}_9,\, \beta_1 \otimes \mathbb{I}_9, \, \beta_*\otimes \gamma_\mathsf{a})$, where $\beta_0 = \sigma_2$ and $\beta_1 = \sigma_1$ are Pauli matrices,  $\beta_* = -\ii\beta_0 \beta_1 = -\sigma_3$ is the chiral gamma matrix in 2d, and $\gamma_\mathsf{a}$ generates the 9d Clifford algebra. This gives
\begin{equation}
	\mathbb{P} = \frac{1}{2} \left(\mathbb{I}_2 \otimes \mathbb{I}_9 -  \ii\beta_0  \beta_1 \beta_* \otimes \sum_{\mathsf{a}=2}^{10} c_\mathsf{a} \gamma_\mathsf{a}\right) = \frac{1}{2}\,\mathbb{I}_2 \otimes \left(\mathbb{I}_9 + \sum_{\mathsf{a}=2}^{10} c_\mathsf{a} \gamma_\mathsf{a}\right),
\end{equation}
where we used the fact that $ \ii\beta_0 \beta_1 \beta_* = -\mathbb{I}_2$. Notice that in order for $\mathbb{P}$ to be a well-defined projector we have to ensure that $\mathbb{P}^2 = \mathbb{P}$, which implies that $\sum_{\mathsf{a} = 2}^{10} c_\mathsf{a}^2 = 1$. Splitting also the 11d Killing spinor as $\epsilon = \theta \otimes \chi + c.c.$, where $\theta$ is a Killing spinor on AdS$_2$ and $\chi$ on $Y_9$, the condition $\mathbb{P}\, \epsilon=0$  results in the following equation involving the 9d spinor only
\begin{equation}\label{BPScond}
	\sum_{\mathsf{a}=2}^{10} c_\mathsf{a} \gamma_\mathsf{a} \chi = -\chi \:.
\end{equation}
Finally, taking $\chi$ to be normalized as $\bar{\chi} \chi = 1$, it can be shown that for a GK geometry the 1-form $\eta$ is given by the bilinear \cite{Kim:2006qu}
\begin{equation}
	\eta = \ex^{B/3}\,\bar{\chi} \gamma_\mathsf{a} \chi \, e^\mathsf{a} \:,
\end{equation}
but recall that by definition of the frame $\eta=\ex^{B/3}e^{10}$, so that $\chi$ necessarily satisfies
\begin{equation}\label{spinorprop1}
	\gamma_{10}\chi = \chi \:, \qquad\qquad \bar{\chi}\gamma_\mathsf{a} \chi = 0 \:, \quad \mathsf{a} \neq 10 \:.
\end{equation}
These equations tell us that $\chi$ is an eigenvector of $\gamma_{10}$ with unit eigenvalue and $\gamma_\mathsf{a} \chi$ is orthogonal to $\bar{\chi}$ when $\mathsf{a}\neq 10$. Similarly, the Kähler 2-form of the GK background is given by 
\begin{equation}
	J = -\frac{\ii}{2} \, \bar{\chi}\gamma_{\mathsf{a}\mathsf{b}} \chi \, e^\mathsf{a} \wedge e^\mathsf{b} \:,
\end{equation}
where $\gamma_{\mathsf{a}\mathsf{b}} = \gamma_{[\mathsf{a}} \gamma_{\mathsf{b}]} $. However, one can always choose a frame such that
\begin{equation}
	J = e^2 \wedge e^3 + e^4 \wedge e^5 + e^6 \wedge e^7 + e^8 \wedge e^9 \:,
\end{equation}
so that $\ii \gamma_{23} \chi = \ii \gamma_{45} \chi =\ii \gamma_{67} \chi = \ii \gamma_{89} \chi = \chi$
and $\bar{\chi}\gamma_{\mathsf{a}\mathsf{b}} \chi = 0$ for any other combination of indices $\mathsf{a}, \,\mathsf{b}$.
As a consequence, one may check that the two sets of spinors
\begin{equation}\label{spinorprop2}
	\{ \chi, \gamma_2\chi, \gamma_4\chi, \gamma_6\chi, \gamma_8\chi\} \quad \text{and} \quad \{ \chi, \gamma_3\chi, \gamma_5\chi, \gamma_7\chi, \gamma_9\chi\}
\end{equation}
are both linearly independent. Combining \eqref{spinorprop1} with \eqref{spinorprop2}, we find that \eqref{BPScond} is solved by setting 
\begin{equation} \label{ca}
	\begin{cases}
		c_\mathsf{a} = 0 \:,\quad \mathsf{a} \neq 10 \\
		c_{10} = -1 
	\end{cases} \quad \implies \quad \vol_{S^1_\mathrm{M2}} = - e^{10} = -\ex^{-B/3}\,\eta\:.
\end{equation}
We thus conclude  that an M2-brane wrapping EAdS$_2 \times S^1_\mathrm{M2}$ is supersymmetric precisely when the circle direction is aligned with the R-symmetry circle.

\section{Wilson loop duals and flat $C$-fields}\label{app:witten}

In section \ref{sec:Wilsonian} we have matched BPS Wilson loop VEVs in the 
large $N$ matrix model with the action of M2-branes/fundamental strings 
in the holographic dual. We have seen that this matching works 
in a variety of examples  throughout the paper, including 
the toric examples in section \ref{sec:examples}, and the extension 
to black holes with spindle horizons/the refined twisted index in 
section \ref{sec:spindle}. 

There is, however, an important subtlety in this matching, which 
we address in this appendix. The matching of 
Wilson loops VEVs with the renormalized action of fundamental 
strings, whose worldvolume ends on the Wilson loop at
the conformal boundary,  goes back to  \cite{Maldacena:1998im}. 
However, depending on the topology of the bulk 
spacetime this Wilson loop VEV in the gravity dual 
can sometimes be zero, due to integration over a certain bulk 
zero mode. In particular this can be the case for Euclidean black holes, 
as first pointed out in a non-supersymmetric context in 
\cite{Witten:1998zw}. One might then be puzzled by why 
this effect does not  lead to the Wilson loops in this paper being zero. 
They are not zero, and it is instructive to consider the mechanism in 
\cite{Witten:1998zw} more closely to see why.

The type IIA string has a coupling $\exp(\ii \int  B)$, where 
in the supergravity partition function we should be careful to sum 
over all bulk configurations with the same boundary conditions on the 
conformal boundary.  In particular, the bulk field $B$ is a two-form potential 
for the NS three-form $H=\diff B$, and one can add a closed/flat
$B$-field to a supergravity background without affecting 
the gauge-invariant flux $H$ or its equation of motion. In the present set-up, with boundary 
conditions fixed at conformal infinity, we should then sum 
over $B$-fields in the bulk spacetime that are zero at infinity,
modulo gauge transformations $B\rightarrow B+\diff\Lambda$, 
where the one-form $\Lambda$ is also zero at conformal infinity. 
Such physically distinct $B$-fields are then classified by 
the compactly supported cohomology group $H^2_{\mathrm{cpt}}(M)$,
 where $M$ is the spacetime.\footnote{Even more precisely, 
including large gauge transformations this becomes $H^2_{\mathrm{cpt}}(M,U(1))$.} 

Now for  Euclidean black holes the black hole part 
of the spacetime has  topology $\R^2\times\Sigma$, where
$\Sigma$ is the horizon and $\R^2$ is parametrized by the 
Euclidean time circle and the radial direction for the black hole. In fact the topology 
is the same whether one considers the full black hole, or as in this paper 
just the near-horizon region of an extremal black hole, where 
Euclidean AdS$_2$ also has the topology of $\R^2$. 
The fundamental strings we consider in the paper are precisely wrapping 
this $\R^2$ direction. Since $H^2_{\mathrm{cpt}}(\R^2\times\Sigma,\R)\cong \R$ 
is generated by a closed two-form that integrates to 1 over the 
$\R^2$ direction and has rapid decay towards the boundary of 
$\R^2$ (where the conformal boundary is), this 
naively leads to a moduli space of flat $B$-fields to integrate over. 
As we will explain further below, in fact this is not the case, but 
\emph{if it were the case} the supergravity saddle point approximation for the path integral with the
type IIA string inserted would give
\begin{align}\label{Wilsonzero}
\langle\, \mbox{string}\, \rangle  = \int_{\vartheta=0}^{2\pi}\exp[-I_{\mathrm{string}}+\ii 
\vartheta] = 0 \, . 
\end{align}
Here $I_{\mathrm{string}}$ is the renormalized string action, for a fundamental 
string wrapping the $\R^2$ direction, and 
 $\vartheta=\int_{\R^2}B$ parametrizes the integrals of the 
inequivalent $B$-fields. The $2\pi$-periodicity in $\vartheta$ arises here 
precisely due to the large gauge transformations mentioned above.

The above effect was explained in \cite{Witten:1998zw}, 
but the conclusion \eqref{Wilsonzero} is not correct for the set-up we are considering
here. In particular, we skipped over two important steps above: (i)
 the full spacetime $M$ is not simply the black hole part of the geometry, 
since there is also an internal space geometry, and (ii) we must 
take into account that the M-theory circle is non-trivially fibred 
over the type IIA spacetime.  In particular the latter  introduces a 
RR field in addition to the $B$-field, and these couple to each other. 
In fact it is then simpler to think of the fundamental string 
as a wrapped M2-brane, wrapping the M-theory circle, 
and consider instead the M2-brane coupling $\exp(\ii \int C)$, with 
compactly support $C$-field modes in the full 11d spacetime.  

We denote the 11d Euclidean spacetime by $E$, and note we are interested 
in the case where $E=\R^2\times Y_9$, where the $\R^2$ factor could be either Euclidean AdS$_2$ ({\it i.e.}\ 
the Poincar\'e disk), or the transverse directions to the horizon of a non-extremal deformation of the 4d black hole ({\it i.e.}\ the radial and Euclidean time circle). The topology of $E$ is the same in both cases, with the horizon geometry absorbed into $Y_9$.

We suppose also that $Y_9$ is the total space of a circle fibration over a base 
$B_8$. This could be viewed as either the M-theory circle fibration, or 
a choice of (off-shell) R-symmetry direction for the purposes of the argument that follows. 
In general $B_8$ might also have orbifold singularities (for 
example for a quasi-regular choice of R-symmetry Killing vector), 
and that also does not affect the following statements in de Rham cohomology. 
We denote the choice of circle action by $U(1)$, and write $B_8=Y_9/U(1)$ and 
$M=E/U(1)=\R^2\times B_8$. 

In general given a total space $E$ which is an $S^1$ bundle over a non-compact 
space $M$, 
we have the following associated long exact Gysin sequence in cohomology:
\begin{align}\label{gysin}
\cdots \longrightarrow H^3_{\mathrm{cpt}}(M)\stackrel{\pi^*}{\longrightarrow} H^3_{\mathrm{cpt}}(E)\stackrel{\pi_*}{\longrightarrow} H^2_{\mathrm{cpt}}(M)\stackrel{\wedge\,  c_1}{\longrightarrow} H^4_{\mathrm{cpt}}(M)\longrightarrow\cdots
\end{align}
Here $\pi:E\rightarrow M$ is the projection, $\pi^*$ the usual pull-back of forms,
 $\pi_*$ denotes ``integration over the fibre'', which means integrating the 
form (partially) along the $S^1$ fibre, and finally $c_1\in H^2(M)$ denotes the 
first Chern class of the $S^1$ bundle, {\it i.e.}\ $c_1=[\mathcal{F}/2\pi]$, where 
$\mathcal{F}=\diff \mathcal{A}$ denotes the curvature of a connection $\mathcal{A}$ on this circle bundle.\footnote{When the $S^1$ corresponds to the M-theory circle 
direction, here $\mathcal{F}$ is naturally represented by the  RR two-form field 
strength. It is precisely when this has a non-trivial cohomology class 
that the original $B$-field argument we presented is affected, and 
from a purely type IIA perspective this is due to the couplings between these fields.} 
 
Viewing $M=\R^2\times B_8$ as a trivial $\R^2$ bundle over $B_8$, the Thom isomorphism theorem 
tells us that $H^{n+2}_{\mathrm{cpt}}(M)\cong H^{n}(B_8)$ for all $n\geq 0$. 
In particular $H^2_{\mathrm{cpt}}(M)\cong H^{0}(B_8)\cong \R$, 
which is generated by a unit integral two-form along  $\R^2$. 
Notice this is the same two-form discussed in the argument 
that led to \eqref{Wilsonzero}, as in \cite{Witten:1998zw}. 
The M2-branes of interest are precisely wrapping this $\R^2$ direction, 
together with the $S^1$ fibre (whether that is the M-theory 
circle or the R-symmetry direction -- for BPS M2-branes they
are necessarily aligned for the locus the M2-brane is wrapping).

Examining 
 \eqref{gysin} further, the Thom isomorphism above gives us 
$H^3_{\mathrm{cpt}}(M)\cong H^1(B_8)\cong 0$, which follows 
since our internal spaces $Y_9$ (and hence their quotients $B_8$) all have zero first Betti number, 
{\it i.e.}\ there are no non-trivial closed one-forms. This is a crucial ingredient, 
as we comment further below. 
Thus more specifically in this case
\eqref{gysin} reads
\begin{align}\label{gysinspecific}
\cdots \longrightarrow 0\longrightarrow H^3_{\mathrm{cpt}}(E)\stackrel{\pi_*}{\longrightarrow}\R \stackrel{\wedge\,  c_1}{\longrightarrow} H^4_{\mathrm{cpt}}(M)\longrightarrow\cdots
\end{align}
As long as $c_1\neq 0$, {\it i.e.}\ the circle bundle we have is non-trivial, then
wedging the volume form on $\R^2$ (which generates the $\R$ factor) with $c_1$ is non-zero in $H^4_{\mathrm{cpt}}(M)$. In fact as already noted the Thom isomorphism tells us $H^4_{\mathrm{cpt}}(M)\cong 
H^2(B_8)$, with this isomorphism map corresponding to wedging with the above-mentioned volume 
form. So as long as $c_1\neq 0\in H^2(B_8)$, the image in $H^4_{\mathrm{cpt}}(M)$ 
will also be non-zero, and the kernel of the $\wedge\, c_1$ map is hence trivial. But then exactness 
 of \eqref{gysinspecific}  says that the generator of $\R$ is not 
the image of any class in $H^3_{\mathrm{cpt}}(E)$, when integrated over $S^1$ under 
the $\pi_*$ map. 

Coming back to the M2-branes of interest, the cohomology group $H^3_{\mathrm{cpt}}(E)$
precisely captures the flat $C$-field modes in the bulk that should be integrated over 
when evaluating the full supergravity partition function. The above argument 
shows that when the circle bundle we have is non-trivial, 
the integral of any such $C$-field mode over the circle direction
and $\R^2$ direction is necessarily zero. There are hence no modes 
that contribute to an argument similar to that in \cite{Witten:1998zw}, 
and the holographic Wilson loop we have is non-zero.
 
Let us conclude by remarking that this argument is certainly 
evaded in the case that the circle bundle is trivial: then 
$Y_9=S^1\times B_8$, and there is a flat $C$-field 
that is simply the volume form on $\R^2$ wedged with the 
volume form along the $S^1$. Such M2-branes would lead 
to zero VEVs, as in \eqref{Wilsonzero}, but in our examples 
$Y_9$ always have zero first Betti number, and in particular 
do not have $S^1\times B_8$ topology. 

\bibliographystyle{utphys} 
 \bibliography{helical.bib}{}

\end{document}